\documentclass[aps,prd,longbibliography,twocolumn,floatfix,superscriptaddress,nofootinbib]{revtex4-1}
\usepackage{hyperref,color}
\usepackage{natbib}
\setcitestyle{square, comma, numbers,sort&compress, super}

\RequirePackage{graphicx,amsmath}
\RequirePackage{mathptmx}      
\RequirePackage{flushend}
\RequirePackage{lineno}

\newcommand{\finaluno}{--0.0034$\pm$0.0042~cpd/kg/keV} 
\newcommand{\finaldos}{0.0003$\pm$0.0037~cpd/kg/keV} 

\pdfoutput=1 

\usepackage[T1]{fontenc} 
\usepackage{amsmath}
\usepackage{multirow}
\usepackage{float}

\usepackage{siunitx} 
\begin{document}

\title{\boldmath Annual modulation results from three-year exposure of ANAIS-112}

\author{J.~Amar{\'e}}
\author{S.~Cebri{\'an}}
\author{D.~Cintas}
\author{I.~Coarasa}
\author{E.~Garc\'{\i}a}
\affiliation{Centro de Astropart\'{\i}culas y F\'{\i}sica de Altas Energ\'{\i}as (CAPA), Universidad de Zaragoza, Pedro Cerbuna 12, 50009 Zaragoza, Spain}
\affiliation{Laboratorio Subterr\'aneo de Canfranc, Paseo de los Ayerbe s.n., 22880 Canfranc Estaci\'on, Huesca, Spain}
\author{M.~Mart\'{\i}nez}
\email{mariam@unizar.es}
\affiliation{Centro de Astropart\'{\i}culas y F\'{\i}sica de Altas Energ\'{\i}as (CAPA), Universidad de Zaragoza, Pedro Cerbuna 12, 50009 Zaragoza, Spain}
\affiliation{Laboratorio Subterr\'aneo de Canfranc, Paseo de los Ayerbe s.n., 22880 Canfranc Estaci\'on, Huesca, Spain}
\affiliation{Fundaci\'on ARAID, Avenida de Ranillas 1D, 50018 Zaragoza, Spain}
\author{M.A. Oliv{\'a}n}
\affiliation{Centro de Astropart\'{\i}culas y F\'{\i}sica de Altas Energ\'{\i}as (CAPA), Universidad de Zaragoza, Pedro Cerbuna 12, 50009 Zaragoza, Spain}
\affiliation{Laboratorio Subterr\'aneo de Canfranc, Paseo de los Ayerbe s.n., 22880 Canfranc Estaci\'on, Huesca, Spain}
\affiliation{Fundaci\'on CIRCE, Avenida de Ranillas 3D, 50018 Zaragoza, Spain}
\author{Y.~Ortigoza}
\author{A.~Ortiz~de~Sol{\'o}rzano}
\author{J.~Puimed{\'o}n}
\author{A.~Salinas}
\author{M.L.~Sarsa}
\affiliation{Centro de Astropart\'{\i}culas y F\'{\i}sica de Altas Energ\'{\i}as (CAPA), Universidad de Zaragoza, Pedro Cerbuna 12, 50009 Zaragoza, Spain}
\affiliation{Laboratorio Subterr\'aneo de Canfranc, Paseo de los Ayerbe s.n., 22880 Canfranc Estaci\'on, Huesca, Spain}
\author{P.~Villar}
\affiliation{Centro de Astropart\'{\i}culas y F\'{\i}sica de Altas Energ\'{\i}as (CAPA), Universidad de Zaragoza, Pedro Cerbuna 12, 50009 Zaragoza, Spain}
\begin{abstract}
ANAIS (annual modulation with NaI scintillators) is a dark matter direct
detection experiment consisting of 112.5~kg of NaI(Tl) detectors in operation at the Canfranc Underground Laboratory (LSC), in Spain, since August 2017. ANAIS' goal is to confirm or refute in a model independent way the DAMA/LIBRA positive result:
an annual modulation in the low-energy detection rate having all the features expected for the signal induced by dark matter particles in a standard galactic halo. This modulation, observed for about 20~years, is in strong tension with the negative results of other very sensitive experiments, but a model-independent comparison is still lacking. By using the same target material, NaI(Tl), such a comparison is more direct and almost independent in dark matter particle and halo models. 
Here, we present the annual modulation analysis corresponding to three years of ANAIS data (for an effective exposure of 313.95~kg$\times$y), applying a blind procedure, which updates the one developed for the 1.5~years analysis, and later applied to 2~years. The analysis also improves the background modeling in the fitting of the region of interest rates. 
We obtain for the best fit 
in the [1--6]~keV ([2--6]~keV) energy region a modulation amplitude of {\finaluno} ({\finaldos}), supporting the absence of modulation in our data, and incompatible with the DAMA/LIBRA result at 3.3 (2.6)~$\sigma$, for a sensitivity of 2.5 (2.7)~$\sigma$. 
Moreover, we include two complementary analyses: a phase-free annual modulation search and the exploration of the possible presence of a periodic signal at other frequencies. Finally, we carry out several consistency checks of our result, and we update the ANAIS-112 projected sensitivity for the scheduled 5~years of operation.
\end{abstract}
\maketitle
\section{Introduction}
\label{sec:intro}
Dark matter and dark energy are required to explain the observations of the Universe at different scales if interpreted in the framework of the cosmological standard model. In spite of their absolute preponderance in the Universe energy budget, no many hints about their nature can be drawn, beyond some general considerations. Because the standard model of particle physics does not provide convenient candidates for any of them, they are cornerstones in the search for physics beyond the standard model, in the frontier between particle physics, astrophysics and cosmology~\cite{Tanabashi:2018oca,Bertone:2016nfn}. 
\par
Focusing on dark matter particle candidates, they are attractive because many of them arise naturally in the field of particle physics to solve other puzzles; explaining the dark matter can be seen just as by-product. Among them, the weakly interacting massive particles (WIMPs), paradigmatic dark matter candidates for more than two decades, were very successful because of a plausible thermal origin and a naturally convenient relic abundance associated. WIMPs appear in well-established theories beyond the standard model, like supersymmetry, and their direct detection is possible because of their weak (but nonzero) coupling to conventional matter. Nowadays, the scene has moved strongly, both by the stringent limits on WIMP candidates and by the development of new detection strategies for lighter particles~\cite{Battaglieri:2017aum,Zyla:2020zbs}.
\par
Very sensitive experiments searching for WIMPs have increased their sensitivity by orders of magnitude in the last ten years. This was done by applying specifically developed detection techniques that profit from detector masses of the order of several tons and background discrimination abilities. However, so far they have not found any hint of dark matter particles and now they are approaching the neutrino floor~\cite{Undagoitia:2015gya,Schumann:2019eaa}. On the other hand, for about twenty years, the DAMA/LIBRA Collaboration has been claiming the observation of an annual modulation in the detection rate, which fulfills all of the requirements expected for energy depositions of WIMP dark matter distributed in a standard galactic halo~\cite{Drukier:1986tm,Freese:1987wu}. The DAMA/LIBRA detector is installed at Gran Sasso Underground Laboratory (LNGS), in Italy. It consists of highly radiopure NaI(Tl) scintillators, having a total mass of 250~kg~\cite{Bernabei:2020mon}. The current statistical significance of the DAMA/LIBRA modulation result reaches the 12~$\sigma$ level in the [2--6]~keV energy region, but it has neither been reproduced by any other experiment, nor ruled out in a fully model-independent way. Compatibility among the different experimental results in most conventional WIMP dark matter scenarios is actually disfavored~\cite{Adhikari:2018ljm,Kobayashi:2018jky,Akerib:2018zoq,Abe:2018mxq,Aprile:2017yea,Savage:2009mk,Aprile:2015ade,Herrero-Garcia:2015kga,Baum:2018ekm,Kang:2018qvz,Herrero-Garcia:2018mky}. Other experiments using the same target are crucial to ascertain whether the DAMA/LIBRA positive signal is a signature of the halo dark matter particles or due to some systematic artifact. This is the goal of the ANAIS-112 experiment~\cite{Amare:2019jul,Amare:2018sxx}, and others like COSINE~\cite{Adhikari:2018ljm,PhysRevLett.123.031302}, and planned SABRE~\cite{Antonello:2018fvx,Antonello:2020xhj} and COSINUS~\cite{Angloher:2016ooq,Angloher:2017sft}. 
\par
After two decades of accumulating annual modulation evidence, for the first time, the DAMA/LIBRA result can be tested close to the three sigma confidence level by analyzing the same distinctive feature, the annual modulation, in the same energy regions and using the same target nuclei~\cite{Amare:2018sxx,Coarasa:2018qzs}. By using the same target material as DAMA/LIBRA, NaI(Tl), the comparison is more direct and almost independent on the dark matter particle and halo models. 
However, the poor knowledge of some sodium iodide properties can introduce a model-dependent uncertainty in the comparison among different experiments using this target. We refer, in particular, to the possibility that sodium and iodine quenching factors, which allow converting nuclear recoil energies into electron equivalent ones, are dependent on specific crystal properties (growth procedure, thallium content, impurities, etc.). 
This question is still under study, and a deeper understanding of the energy conversion mechanism in inorganic scintillators should be investigated in the near future in order to reduce the uncertainties still present in the knowledge of such quenching factors in NaI~\cite{Gerbier:1998dm,dama-quench-hypo,spooner-quench,tovey-quench,Chagani:2008in,collar-quench,Xu:2015wha,joo-quench,cosine-quench,Bignell:2021bjx}. With respect to the comparison between ANAIS-112 and DAMA/LIBRA annual modulation results, the corresponding uncertainty would affect the interpretation for those particles producing nuclear recoils in the detector. Quenching factors for nuclear recoil scintillation in ANAIS crystals have  recently been measured at the Triangle Universities Nuclear Laboratory (TUNL), and results will be published soon. Unless otherwise stated, throughout this paper, all the energies shown will correspond to electron equivalent energies.
\par
ANAIS-112 is an experimental effort relying on a large expertise on NaI(Tl) scintillators operation from researchers at the University of Zaragoza and the Canfranc Underground Laboratory (LSC), in Spain~\cite{Sarsa:1996pa,Sarsa:1997hb}.
It consists of 112.5~kg of NaI(Tl), distributed in a 3$\times$3 array of modules made by Alpha Spectra (AS), Inc. (CO, US), with a mass of 12.5~kg each. The experiment is installed in the Hall B of the LSC, and data taking started in August 2017. More details on the experimental setup can be found in~\cite{Amare:2018sxx}. 
\par
We designed a blind protocol for the annual modulation analysis of ANAIS-112 data before data taking started. Energy and time distribution from single-hit events in the region of interest (ROI), from 1 to 6~keV, were kept blinded since the beginning of the data taking. 
The fine-tuning of the events rejection procedures, general background assessment and sensitivity estimate were accomplished using 10\% of the data from the first year of measurement (randomly distributed days along the data taking period)~\cite{Amare:2018sxx,Amare:2018ndh,Coarasa:2018qzs}.
The designed analysis protocol was applied after 1.5 and 2~years of data taking, unblinding the corresponding ROI events. Results, corresponding to exposures of 157.55~kg$\times$y and 220.69~kg$\times$y, respectively, were published in 2019 and 2020~\cite{Amare:2019jul, Amare:2019ncj}. These results were compatible with the absence of modulation and allowed confirmation of our sensitivity projections, while producing some tension with DAMA/LIBRA annual modulation result.
\par
After having accumulated 3~years of data by August 2020, and using the same analysis protocol as in Ref.~\cite{Amare:2019jul}, we have carried out a new unblinding of ANAIS-112 data and the corresponding annual modulation analysis in a fully comparable way. We review in Sec.~\ref{sec:anais} the most relevant features of the ANAIS-112 experimental setup and the stability of its performance in the period starting on August 2017 until August 2020. In Sec.~\ref{sec:dataAna}, we summarize briefly the ROI energy calibration, data selection protocols and updated efficiencies. We present in Sec.~\ref{sec:modulation} the results of the model independent blind analysis searching for annual modulation in the same regions for which DAMA/LIBRA has published results ([1--6]~keV and [2--6]~keV) for an exposure of 313.95~kg$\times$y and different modeling of backgrounds. In this section we also include some consistency checks in order to evaluate the possible presence of systematic effects either in our data or in our analysis procedures, and other complementary analyses: a phase-free annual modulation search, and the exploration of possible periodic signals at other frequencies that could be hidden in the data. Finally, in Sec.~\ref{sec:sensitivity}, we update the sensitivity prospects of ANAIS-112 experiment for the scheduled 5-years operation. We present our conclusions in Sec.~\ref{sec:conclusions}.   
\section{ANAIS-112 setup and updated performance}
\label{sec:anais}
\begin{figure*}[htbp]
\centering 
\includegraphics[width=0.45\textwidth]{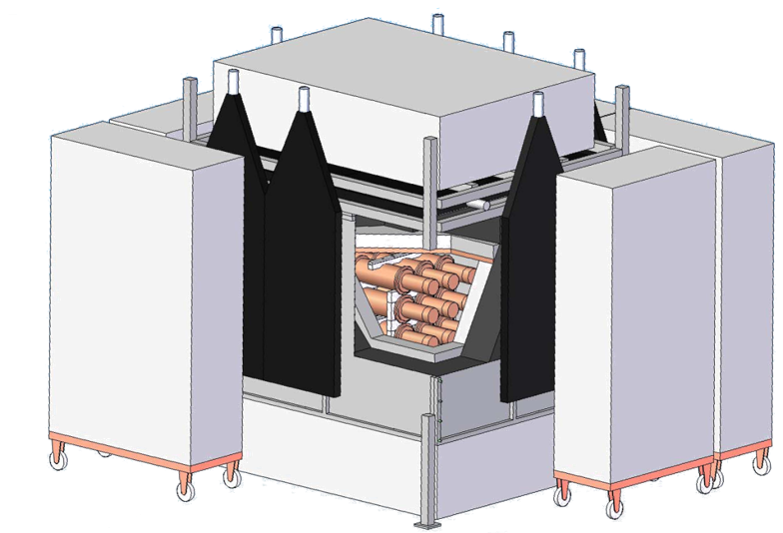}
\includegraphics[width=0.45\textwidth]{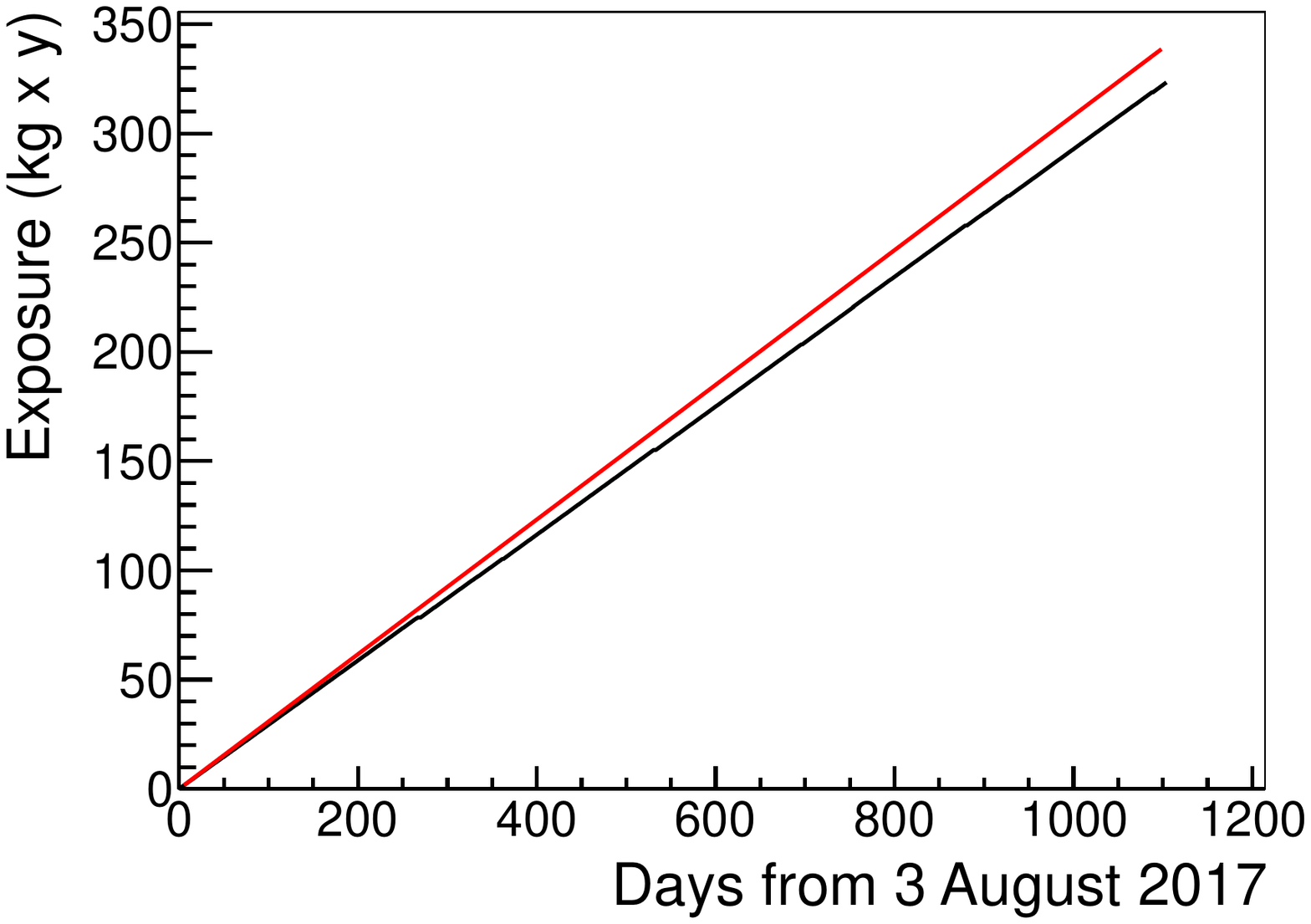}
\caption{\label{fig:duty} Left panel: Artistic view of ANAIS-112 setup showing the 9 NaI(Tl) modules inside a shielding made of lead, antiradon box, active muon vetoes, polyethylene and water tanks. Right panel: the black line represents the accumulated exposure since August 2017 with the nine ANAIS-112 modules. For comparison, the red line corresponds to a 100\% live time for 112.5~kg detection mass.
} 
\end{figure*}

ANAIS-112 setup has been described in~\cite{Amare:2018sxx}, and it is shown in Fig.~\ref{fig:duty}, left panel. We highlight in this section only those specific features of the ANAIS-112 setup which can be relevant for the comparison with the DAMA/LIBRA results, concerning both, setup and operation conditions. 
\par
First, we want to emphasize the remarkable ANAIS-112 duty cycle. Annual modulation analysis can profit from the full exposure with all the nine modules since August 2017, amounting to more than 94\% of the real time elapsed. Down time is mostly due to the periodical calibrations of the modules. Besides, only a few incidents along the three years of operation happened, e.g. power failures at the experiment. This excellent duty cycle guarantees that the data are evenly distributed along the year, reducing possible systematic contributions in the annual modulation analysis. The right panel of Fig.~\ref{fig:duty} shows the ANAIS-112 accumulated exposure since August 2017 as a function of real time, amounting to 322.83~kg$\times$y. 
Table~\ref{tab:duty} shows the percentages of live, down and dead time, as well as the total live time per operation year.
\setlength{\tabcolsep}{2em}
\begin{table*}[htbp]
\centering
\begin{tabular}{ccccc}
\hline\hline
Time period& Live time&Live time &Down time &Dead time \\
 & (days) & (\%) & (\%) & (\%) \\
\hline
08/03/2017 -- 07/31/2018 & 341.722 & 94.40 & 2.84 & 2.76\\
08/01/2018 -- 08/28/2019 & 374.302 & 95.48 & 2.44 & 2.07\\
08/29/2019 -- 08/13/2020 & 333.791 & 95.10 & 2.62 & 2.28\\
\hline\hline
\end{tabular}
\caption{\label{tab:duty} Distribution of live, dead and down time during ANAIS-112 data taking and total live time accumulated per operation year.}
\end{table*}
\par
The ANAIS-112 data acquisition system has been also described in~\cite{Amare:2018sxx}, and it has proven to be robust and not affected by electrical or mechanical disturbances in the environment, allowing for the high duty cycle commented above. The two photomultiplier tube (PMT) signals from each module are individually fully processed by the electronic chain. The conveniently preamplified low energy (LE) signals are sent to MATACQ-CAEN V1729A digitizers, which sample the waveforms at 2~GS/s in a 1.2 $\mu$s window with high resolution (14 bits). The trigger is set up at a photoelectron (p.e.) level in each PMT, and each module is triggered by the coincidence between the two PMT trigger signals in a 200~ns window. Corresponding high energy (HE) signals are also available for every module. They allow us to assess backgrounds in the different energy ranges.  
\par
\begin{figure}[htbp]
\centering 
\includegraphics[width=.48\textwidth]{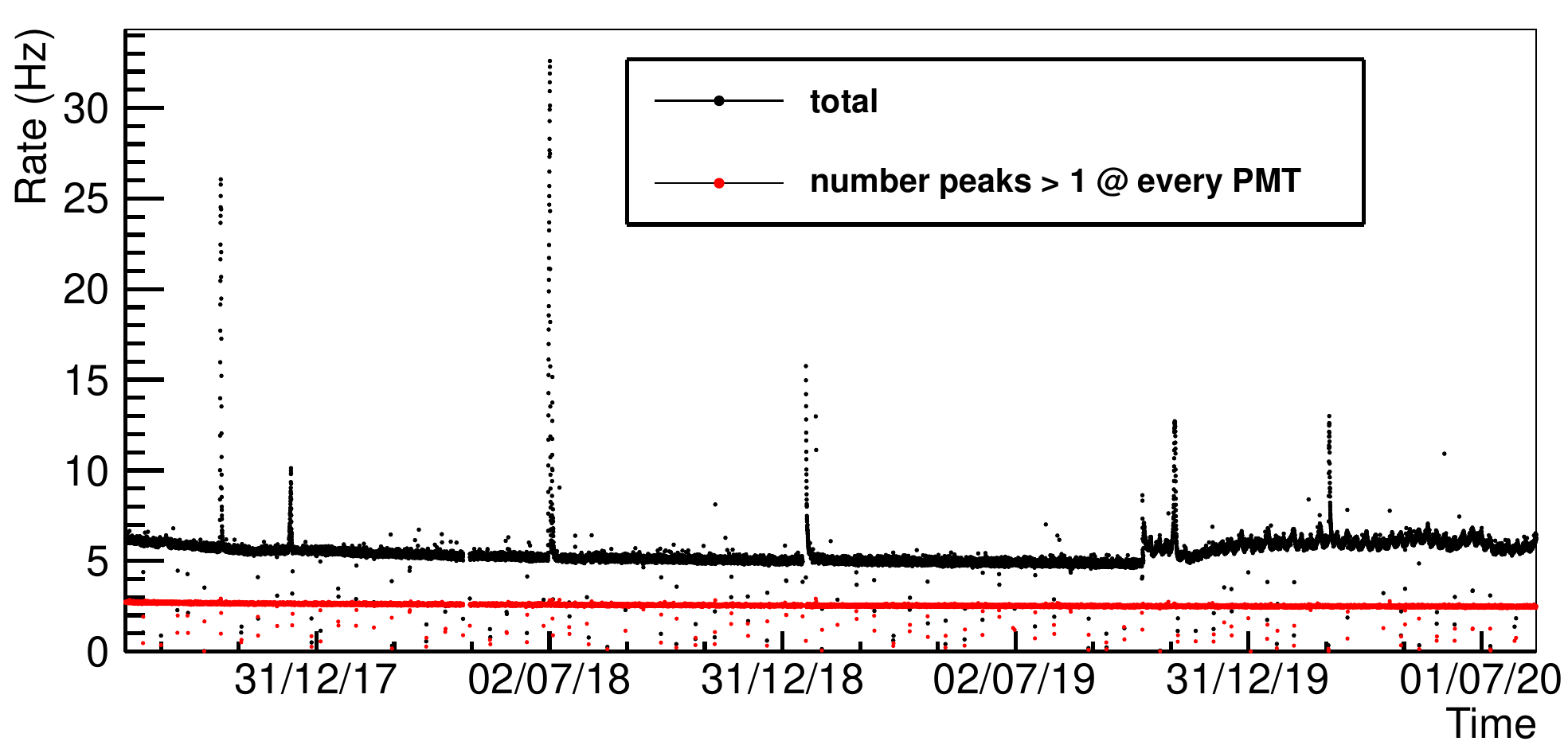}
\caption{\label{fig:trigger} The ANAIS-112 trigger rate from August 2017 until August 2020. Total rate is shown in black, and the rate of events having more than one peak in every PMT detected by the peak-finding algorithm is shown in red. 
} 
\end{figure}

Figure~\ref{fig:trigger} shows in black the total trigger rate of ANAIS-112 along the three years of operation, 
whereas in red is shown the rate of events having more than one peak in each PMT\footnote{In a first analysis step, a peak-finding algorithm is applied to both PMT signals in order to identify peaks attributable to individual photoelectrons: see next section for more details.}.
It can be observed that the trigger rate is dominated by events having only one peak at each PMT signal, which are not associated with bulk scintillation in the sodium iodide material. It is worth remarking that our event selection protocol (see next section for more details) requires more than four peaks in each PMT signal to be considered a bulk scintillation event.
We can conclude from Fig.~\ref{fig:trigger} that although a few high trigger rate periods along the three years of operation occurred, they do not affect the total rate of selected events, as shown by the remarkably stable rate of events having more than one peak at each PMT.
All of these anomalous high rate periods are associated to changes in the supply of gas flushing into the ANAIS-112 shielding used to prevent the entrance of radon. However, the abnormal increases of the trigger rate cannot be explained by the entrance of radon into the shielding: it does not correlate with any increase in background events in any energy region, and it also appears when using radon-free air provided by LSC for the flushing.
\par
The ANAIS-112 modules were specifically designed to have a Mylar window in the copper housing to allow for a low-energy calibration with external x-ray/gamma sources down to 10~keV. This particular feature of the ANAIS detectors allows periodical calibration of all the detectors using several $^{109}$Cd external sources mounted on flexible wires that are introduced into the shielding and positioned in front of the Mylar windows, irradiating simultaneously the nine modules for a period of about 4~hrs every two weeks. Lines of 88.0, 22.6 and 11.9~keV\footnote{22.6 and 11.9~keV are average energies, corresponding to the weighted average of the different x rays produced following an electron capture (EC) decay in the first of them, and photoelectric absorption in the second.} are used for the control of the modules' response stability below 100~keV. The latter is not directly produced by the $^{109}$Cd decay, but it is the result of the subsequent Br x rays produced in the flexible wire surrounding the sources.  
The modules' response can be thought of as the combination of two independent factors: the PMTs gain and the light collection, which can be studied separately. 
\par
\begin{figure}[htb]
\centering 
\includegraphics[width=.49\textwidth]{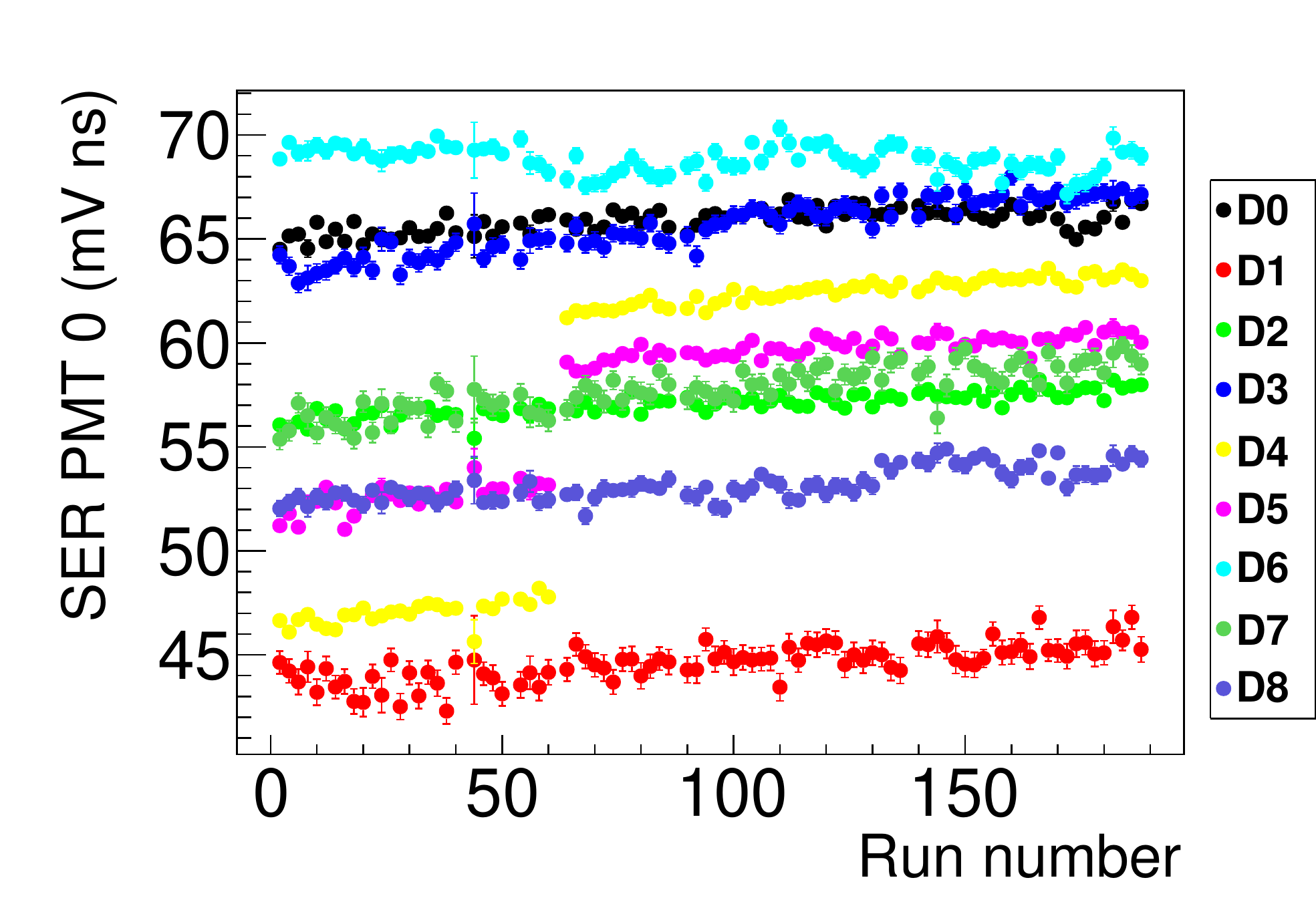}
\includegraphics[width=.49\textwidth]{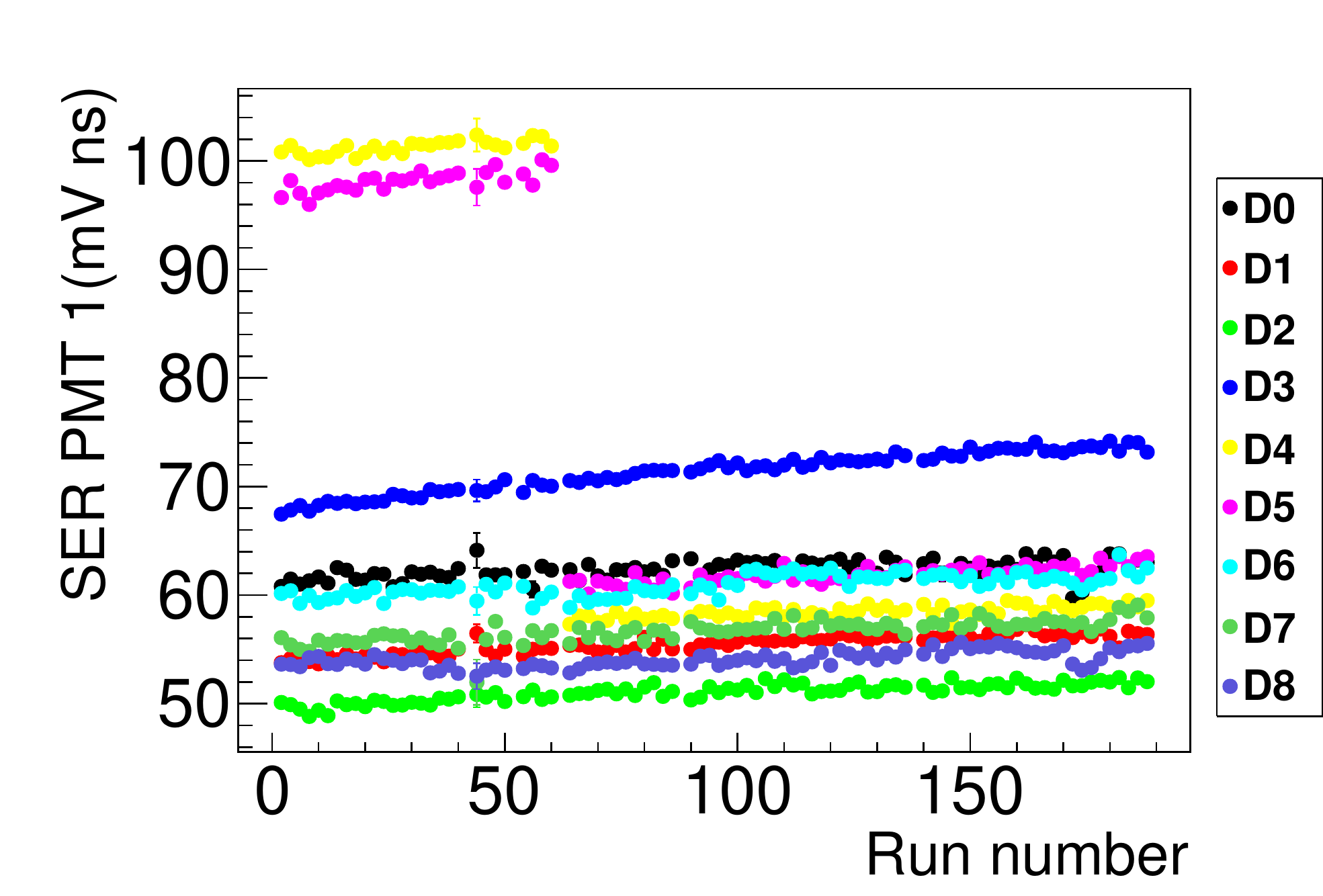}
\caption{\label{fig:SERstab} Evolution of the PMT gain provided by the SER for PMT-0 (upper panel) and PMT-1 (lower panel) coupled to each module. The HV setting of D4 and D5's PMTs was modified after the first year of data taking.
} 
\end{figure}

The PMT gain is directly monitored by the single electron response (SER) of each PMT. We calculate periodically, every two weeks approximately, this SER for each PMT (two PMTs per module, which are referred to as PMT-0 and PMT-1, depending on the PMT position with respect to the ANAIS setup: PMT-0 corresponds to the west side, PMT-1 to the east side). We use for that estimate the distribution of peaks identified by the peak searching algorithm in the tail of pulses that have a very low number of peaks. The evolution in time of the average of SER values for each PMT in the ANAIS-112 setup along the three years is shown in Fig.~\ref{fig:SERstab}.
\par
\begin{figure*}[htb]
\centering 
\includegraphics[width=.95\textwidth]{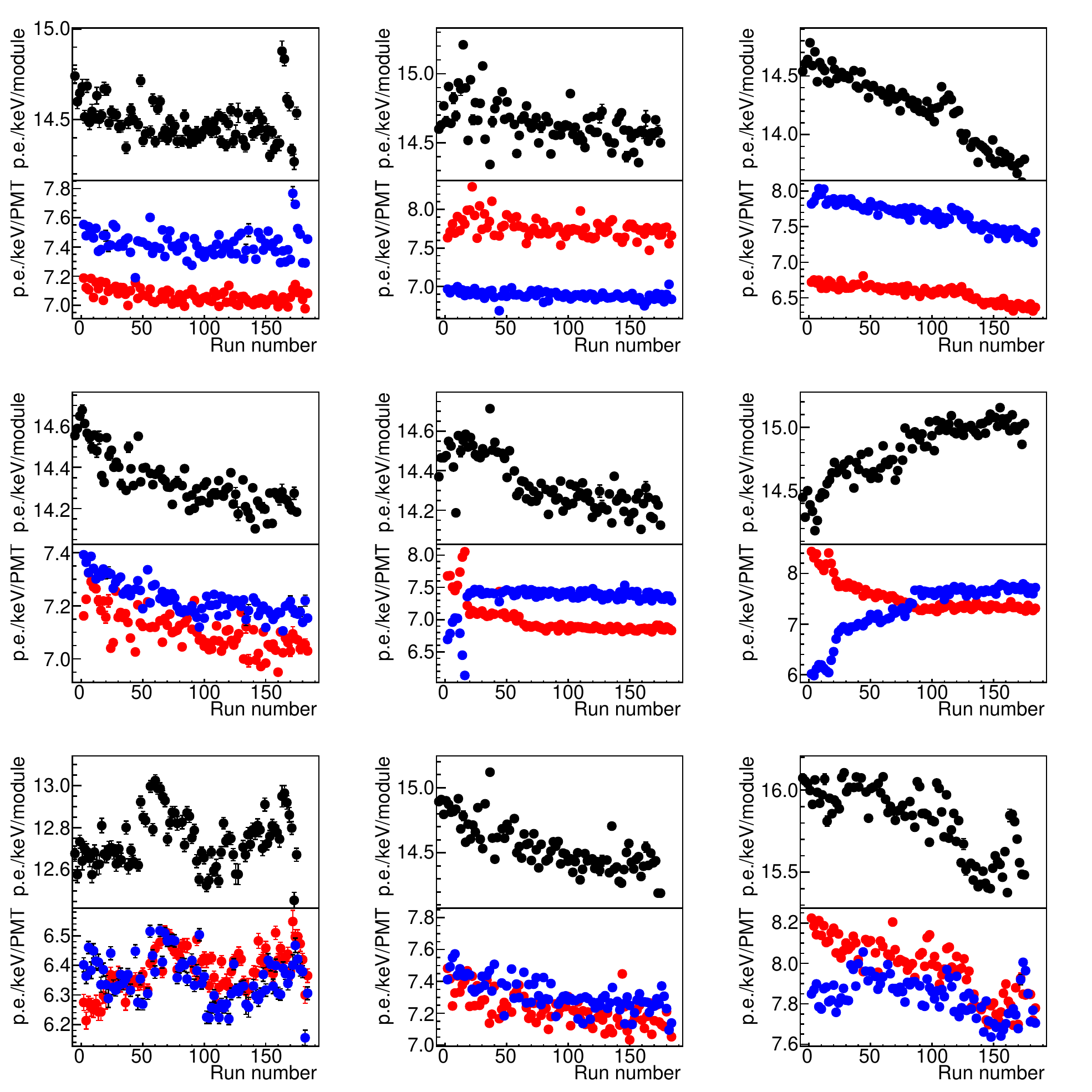}
\caption{\label{fig:lightyield} Evolution of the light collection in ANAIS-112 modules, shown separately for each PMT: PMT-0 values are shown in red and PMT-1 values in blue (lower panels); total values are shown in black (upper panels). In all the figures shown in this article, the data from the nine modules will be shown in the same way: upper panels from left to right correspond to modules D0, D1 and D2; middle row from left to right D3, D4 and D5, lower row from left to right D6, D7 and D8. 
} 
\end{figure*}

The ANAIS-112 modules show a very high light collection, at the level of 15~p.e./keV of electron equivalent energy deposited~\cite{Olivan:2017akd}. These light collection values are higher and more homogeneous than those reported of the DAMA/LIBRA modules~\cite{Bernabei:2012zzb}. 
Some of the reasons behind such a high light collection are the very good optical properties of the AS crystals, the high quantum efficiency (Q.E.) Hamamatsu PMTs used (Q.E. of the PMTs units used in ANAIS-112 is shown in Table~\ref{tab:lightyield}), and the good optical coupling between crystals and quartz windows and between quartz windows and PMTs. 
The ANAIS-112 light collection is being continuously monitored during the three years of data taking. Figure~\ref{fig:lightyield} shows the light collection per PMT and per module, and Table~\ref{tab:lightyield} compares the light collection estimates carried out in 2017, before starting the data taking, to the average values obtained after the three years of ANAIS-112 operation. The total light collection per module has been quite stable, but it shows a different behavior in the nine modules: slight decreases are observed in most of them, but module D5. Their time evolution is not correlated neither with environmental conditions, nor the PMT gain.
\par
\begin{figure*}[htb]
\centering 
\includegraphics[width=.95\textwidth]{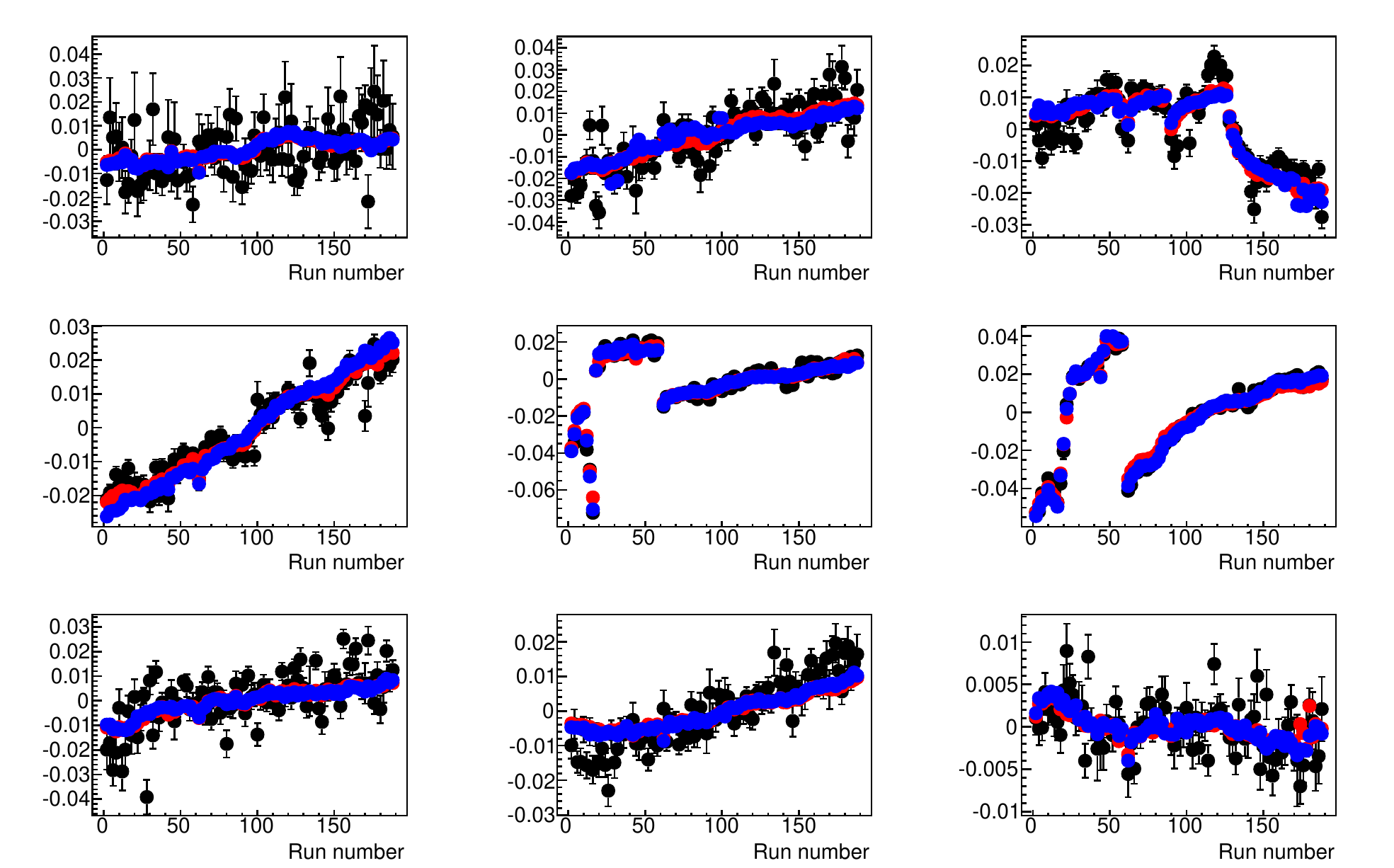}
\caption{\label{fig:calstab} Evolution during the three years of data taking of the position of the three calibration lines (determined in mV$\times$ns units) from $^{109}$Cd in the ANAIS-112 modules, shown as relative deviation with respect to the average values. The relative position of the lines 11.9, 22.6 and 88.0~keV are shown in black, red and blue solid dots, respectively. Slow drifts are observed in most of the modules, but periodical calibration allows for correcting those drifts. Only the modules D4 and D5 showed strong instability during the first year of operation. Then the PMT operation high voltage (HV) was changed just before starting the second year of measurement; because of that, for D4 and D5 in this figure, different average values have been considered for the periods before and after the HV change.
} 
\end{figure*}

Changes in the light collection and in the PMT gain combine to modify the total response of each module to a given energy deposition. This can be corrected by the calibration in energy which is explained in Sec.~\ref{sec:dataAna}. Figure~\ref{fig:calstab} shows the relative deviation in the position (in mV$\times$ns units) of the three lines observed in the calibration runs along the three years of data taking.
It can be observed that the modules D4 and D5 suffered from strong ($\sim$10\%) variability in the response during the first year of operation. Because of that, the PMT HV was reduced before the second year (run number 62). The behavior of the different modules has been quite different in the three years reported: D0, D6, D7 and D8 are stable at the level of $\pm$1\%, D1 and D3 show a similar drift, slight but continuous, as also do D4 and D5 after the HV change previously commented, whereas D2 suffered from a sudden change ($\sim$4\%) in response after the second year of operation. Concerning possible systematic effects which could affect the results presented in this work, we emphasize that our calibration procedure, carried out biweekly and independently for every detector, allows us to correct for any drift or change in the response (see Sec.~\ref{sec:controlPopulations}).
Table~\ref{tab:lightyield} also shows the energy resolution derived from the 3.2~keV line (produced by the decay of $^{40}$K in the crystal bulk) in the coincidence spectra corresponding to the full three years exposure. This resolution, which is only weakly correlated with the light collection, would get worse if response instabilities were not properly corrected by the ANAIS-112 calibration protocols (explained in Sec.~\ref{sec:dataAna}). 
\setlength{\tabcolsep}{1.5em}
\begin{table*}[tb]
\centering
\begin{tabular}{cccccc}
\hline\hline
\multirow{3}{1.5cm}{\centering Module} & Q.E. & \multicolumn{3}{c}{Total light collection (p.e./keV)} & Energy resolution\\
 & PMT0/PMT1 & 2017 results & \multicolumn{2}{c}{3 years results} & FWHM @ 3.2 keV\\
 & (\%) & \cite{Olivan:2017akd} & average & std. deviation & (keV)\\
\hline
D0 &38.2/37.2 & 14.6 $\pm$ 0.1 & 14.49 & 0.11 & 1.26$\pm$0.03\\
D1 & 39.7/39.7 & 14.8 $\pm$ 0.1 & 14.64 & 0.15 & 1.30$\pm$0.04\\
D2 & 39.2/42.6 & 14.6 $\pm$ 0.1 & 14.21 & 0.30 & 1.25$\pm$0.03\\
D3 & 37.3/39.4& 14.5 $\pm$ 0.1 & 14.33 & 0.12 & 1.14$\pm$0.05\\
D4 & 40.1/41.8 & 14.5 $\pm$ 0.1 & 14.33 & 0.13 & 1.34$\pm$0.06\\
D5 & 43.6/43.9 & 14.5 $\pm$ 0.1 & 14.82 & 0.23 & 1.22$\pm$0.02\\
D6 & 40.4/38.9 & 12.7 $\pm$ 0.1 & 12.74 & 0.12 & 1.35$\pm$0.04\\
D7 & 41.9/42.5 & 14.8 $\pm$ 0.1 & 14.55 & 0.18 & 1.38$\pm$0.04\\
D8 & 41.6/43.4 & 16.0 $\pm$ 0.1 & 15.81 & 0.21 & 1.30$\pm$0.05\\
\hline\hline
\end{tabular}
\caption{\label{tab:lightyield} Total light collection in the ANAIS-112 modules as estimated before the data taking started~\cite{Olivan:2017akd} and the average value after the periodical monitoring during the three years of operation. The energy resolution of ANAIS-112 modules at 3.2~keV $^{40}$K line obtained from the full three years exposure and the quantum efficiency for every PMT unit used, as provided by the manufacturer, are also given.}
\end{table*}
\par
\begin{figure*}[htb]
\centering 
\includegraphics[width=.9\textwidth]{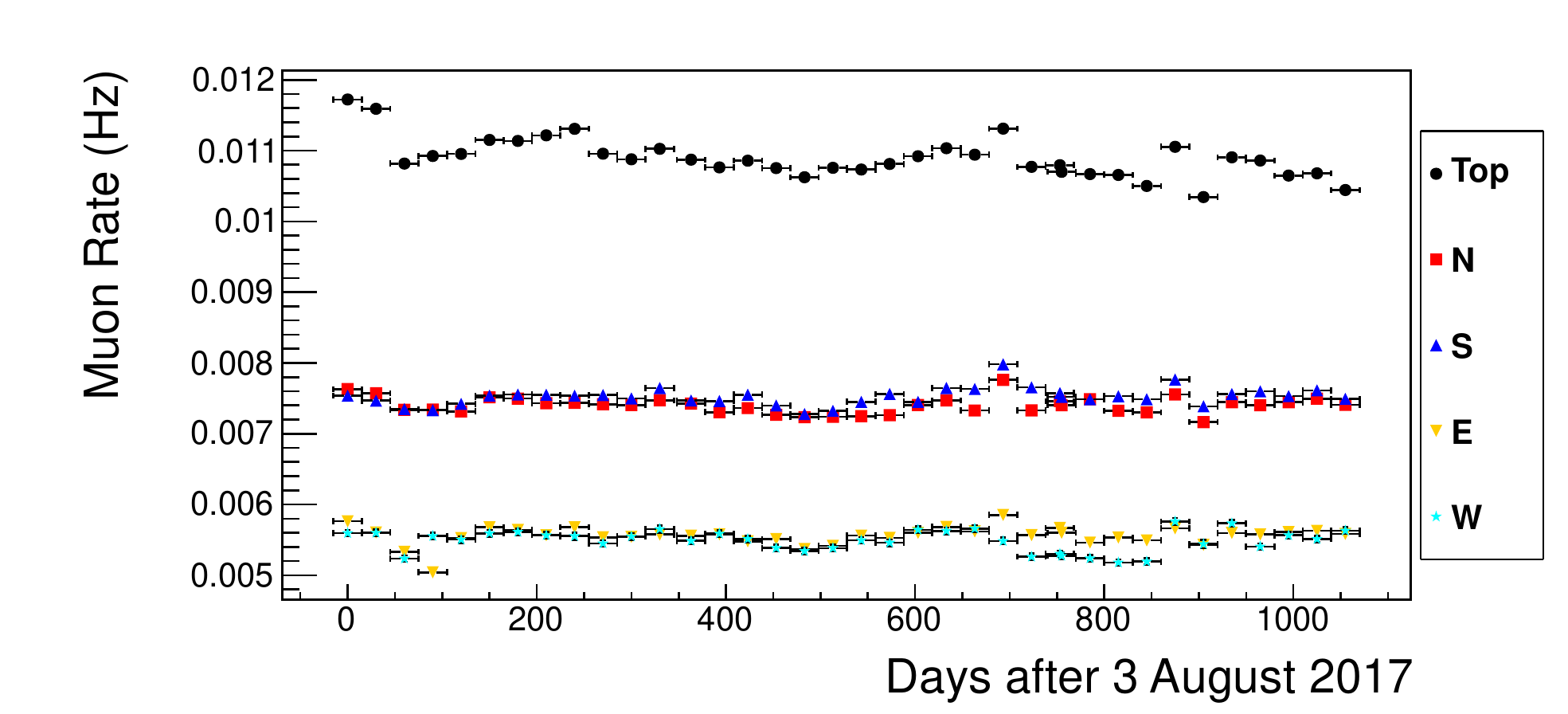}
\includegraphics[width=.9\textwidth]{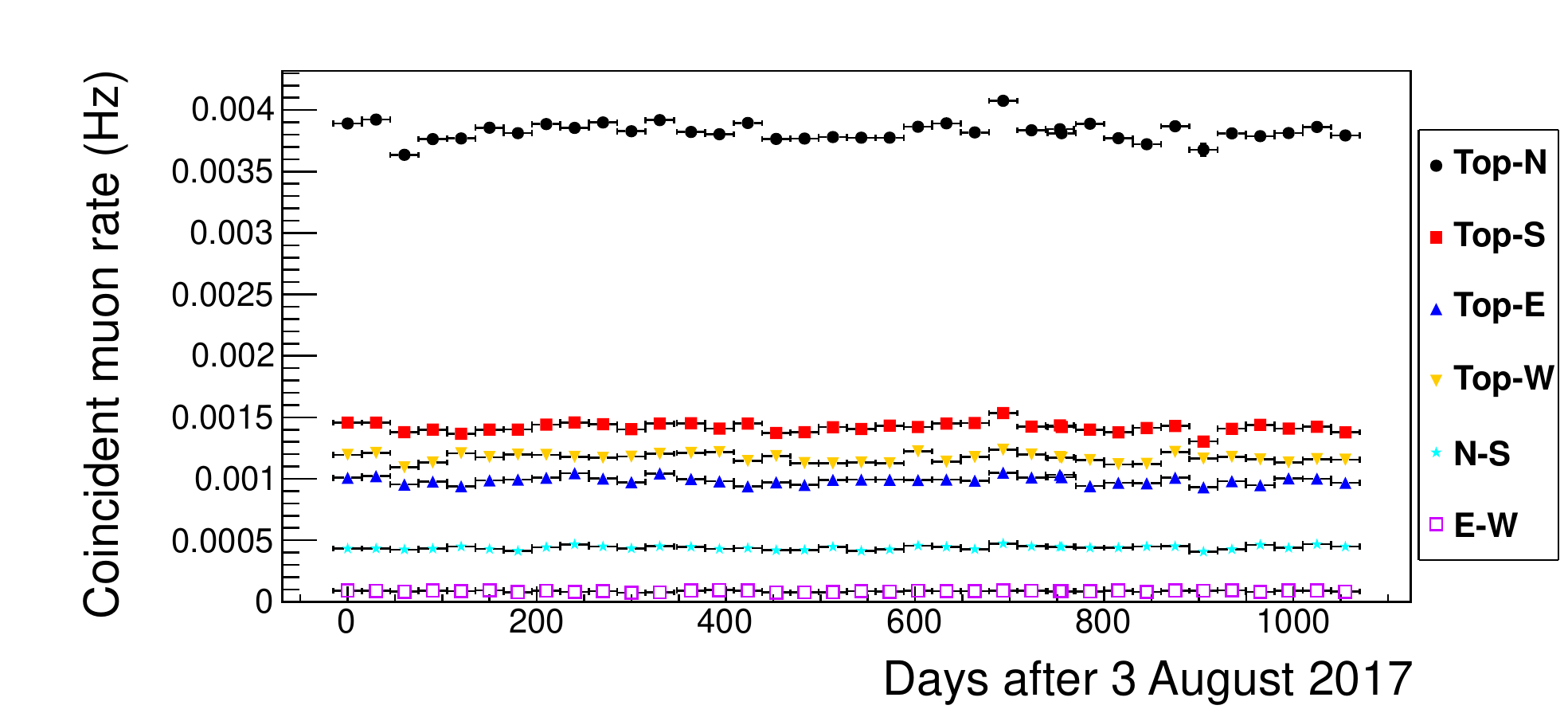}
\caption{\label{fig:murates} Evolution of the muon rate on each side of the ANAIS-112 scintillator veto system (top panel) and the rate of coincidences on two sides (lower panel) on a monthly basis since August 2017. N, S, W and E stand for north, south, west and east sides of the setup.  
} 
\end{figure*}

The ANAIS-112 setup includes an active veto system consisting of 16 plastic scintillators covering the top and the four sides of the shielding in order to tag muons and then, to enable the removal of muon related events from data. The plastic scintillators have different dimensions and properties, but all of them are 5~cm thick, and the equivalent surface of each system side is 2.1~m$^2$ (top) and 1.5~m$^2$ (north, south, east and west). 
A specific data acquisition system is used for reading the 16 signals and a pulse shape analysis is applied to select events attributable to muons~\cite{MAThesis}. This veto system allows for monitoring the muon rates \textit{onsite} along the ANAIS-112 data taking. Figure~\ref{fig:murates} shows the rates measured by each side of the veto system (top panel), and the rates of coincidences between two sides (lower panel), both expressed in muons/second, on a monthly basis. 
The rock overburden above the LSC underground facilities is strongly asymmetric in the north-south axis, while it is almost flat in the east-west direction, explaining properly the rate of coincidences between two sides of the veto system shown in the bottom panel of Figure~\ref{fig:murates}: a much higher muon rate is observed for top-north coincidences than for the other combinations. It also explains why the total rates observed in the north and south sides are higher than those corresponding to the east and west sides, as shown in the top panel of Fig.~\ref{fig:murates}. However, it has to be remarked that our veto system was designed to tag efficiently muons, at the cost of having a relatively low energy threshold, which implies a possible contamination from highly energetic gammas from the environment in our measured rates. The possible contribution of such kind of events has not been estimated. On the other hand, coincidence events can be attributed to muons undoubtedly and they will allow tagging muon-related events in our data.
Hints on an annual modulation in the coincident muon rate at the ANAIS-112 position are found, peaking around June, 7~$\pm$~10~days. A more detailed study is in progress. 
The derived residual muon flux is fully compatible with the published value, $5.26\times10^{-3}$~s$^{-1}$~m$^{-2}$ \cite{Trzaska:2019kuk}, which corresponds to LSC Hall A.
\par
The ANAIS-112 setup includes a passive neutron shielding consisting of polyethylene bricks and water tanks. In the next months (along 2021), the neutron flux at the ANAIS-112 position will be monitored in order to identify any possible seasonal fluctuation performed by a new Collaboration, HENSA-ANAIS. Previous measurement of the neutron flux at LSC can be found in \cite{Jordan:2013exa}.
\par
The environmental parameters are continuously monitored along the data taking. We monitor relative humidity and temperature in different positions of the ANAIS-112 setup, radon content in the laboratory air (inside ANAIS hut, but outside ANAIS-112 shielding), high voltage supplied to PMTs, low voltage supplied to preamplifiers, and electronic modules, etc. We have not identified any correlation between the trigger rate or the rate in the ROI with those environmental parameters. 
Only relative humidity and radon content in the laboratory air are correlated and show a seasonal modulation. 
It is worth reminding here that the ANAIS-112 inner shielding is continuously flushed with radon-free gas. An upper limit of $0.04~$Bq/m$^3$ at 95\% C.L. was determined by screening the same nitrogen gas used for the flushing into the ANAIS shielding using an HPGe detector at LSC. After January 2019, a different flushing system has been in use, and this upper limit does not apply anymore. Nevertheless, we are confident that the radon content inside the ANAIS-112 shielding is at a similar level because the nitrogen gas used since then is produced by boiling-off liquid nitrogen. According to the ANAIS-112 background model, this upper limit would contribute to the ROI rate  below $3 \times 10^{-4}$~cpd/kg/keV.

\section{Data analysis}
\label{sec:dataAna}
In a first analysis step, we calculate for each event different pulse parameters (area, amplitude, pulse shape parameters, etc.), the time since the last muon veto, and we also apply a peak-finding algorithm to identify individual peaks (associated to individual p.e.) in low energy pulses. The energy of an event is derived from the addition of the pulse areas of both PMT signals. 
\par
Concerning the energy calibration at low energy (LE), we profit from the calibrations with $^{109}$Cd external sources discussed in the previous section, but also from known lines present in the background at 3.2 and 0.9~keV (from $^{40}$K and $^{22}$Na crystal contamination, respectively). These events are tagged by coincident energy depositions at high energy in a second module. This results in a remarkable increase of the calibration accuracy in the ROI, and of the reliability of the ANAIS-112 analysis energy threshold, set at 1~keV~\cite{Amare:2018sxx}. 
Our calibration procedure is the following: first, we calibrate with the external $^{109}$Cd sources the detectors every two weeks to control the gain stability and to correct possible drifts; second, in order to increase the statistics of the $^{40}$K and $^{22}$Na peaks we add up one and a half months of data to carry out the final LE calibration. The peaks registered during calibration runs are fitted to Gaussian line shapes, while for the $^{40}$K and $^{22}$Na peaks, with a lower number of events, we take the median of their distributions. Then we perform a linear regression on the expected energies against the peak's positions for every detector and recalibrate the low energy events (below 50~keV).
\par
\begin{figure}[htb]
\centering 
\includegraphics[width=.48\textwidth]{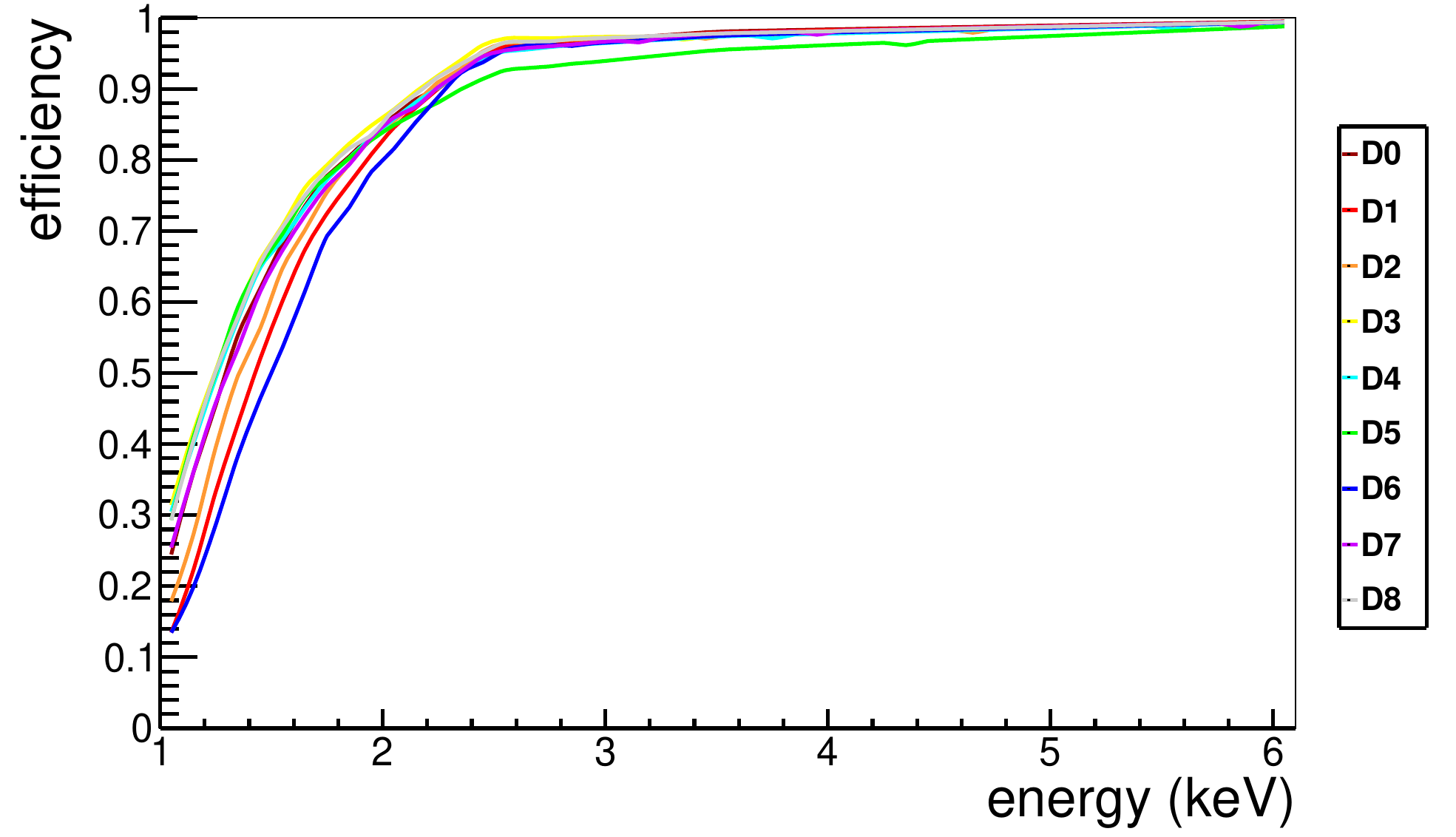}
\caption{\label{fig:efficiencies} Total efficiency, $\epsilon(E,d)$, for event detection in all the ANAIS-112 modules as a function of energy. It has been obtained by combining the trigger efficiency, the asymmetry cut efficiency, and the pulse shape cut efficiency, all of them recalculated for the full analyzed exposure corresponding to three years by following the same procedure established in~\cite{Amare:2018sxx}. 
} 
\end{figure}

The development of robust protocols for the selection of events corresponding to bulk scintillation in sodium iodide produced by particle energy depositions has been crucial to fulfill the ANAIS-112 goal because the trigger rate in the ROI is dominated by other events, some of them with origin in the PMTs, others still unexplained (see Sec.~\ref{sec:anais}). These protocols have been thoroughly explained in \cite{Amare:2018sxx}; therefore, below, we just summarize the steps of the event selection and update the corresponding efficiency. The efficiency, recalculated with the full available statistics, is shown in Fig.~\ref{fig:efficiencies}.
\par
A blind analysis strategy is applied to all ANAIS-112 data after the first analysis step described above. We calibrate the energy response of every detector at LE and HE, where the LE variable is kept hidden for events corresponding to single hits (M1 events). We use calibration events, coincident events and events outside the ROI to recalculate our efficiencies for the selection procedures before unblinding the ROI. 
The event selection in the ROI starts by removing events arriving less than one second after the last muon veto trigger, correcting the total live time by subtracting one second per muon veto
trigger. The live time used for the annual modulation analysis is 1018.6~days.
Then events in the ROI are selected by imposing the following criteria: 
\begin{itemize}
\item{single hit events (M1);} 
\item{a pulse shape cut combining the fraction of the pulse area in [100-600]~ns after the event trigger, defined following~\cite{2008NIMPA.592..297B}, and the logarithm of the mean time of the distribution of the individual p.e. arrival times in the digitized window~\cite{Kim:2018wcl}; }
\item{an asymmetry cut: events having a strongly asymmetric light sharing between the two PMT signals are removed by requiring that the number of peaks identified in each PMT is larger than 4.}
\end{itemize}
\par
The total detection efficiency, $\epsilon(E,d)$, calculated independently for every detector $d$ as a function of the energy, $E$, can be written~\cite{Amare:2018sxx} as
\begin{equation} \label{eq:eff}
\epsilon(E,d)=\epsilon_{trg}(E,d)\times\epsilon_{PSA}(E,d)\times\epsilon_{asy}(E,d).
\end{equation}
The trigger efficiency $\epsilon_{trg}(E,d)$ is calculated from Monte Carlo (MC) simulations, while the efficiencies of the pulse shape cut ($\epsilon_{PSA}(E,d)$) and the asymmetry cut ($\epsilon_{asy}(E,d)$)  are evaluated from nonblinded populations accumulated for the whole exposure time: the 3.2 and 0.9~keV events selected by the coincidence with the high energy gammas following $^{40}$K and $^{22}$Na decays for the pulse shape cut, and $^{109}$Cd calibration events for the asymmetry cut. The total detection efficiency ranges from 0.15 to 0.30 at 1~keV, depending on the detector, increases up to 0.8 at 2~keV and is close to 1 at 4~keV for all the modules. 
Statistical errors in the total efficiency vary from 2\% to 3\% at 1~keV down to 0.1\% at 6~keV. Comparing different methods for the efficiency calculation we have also estimated a systematic uncertainty that amounts up to 20\% at 1--1.2~keV and is negligible above 1.5~keV. 
\subsection{Control populations}
\label{sec:controlPopulations}
\begin{figure}[htbp]
\centering 
\includegraphics[width=.48\textwidth]{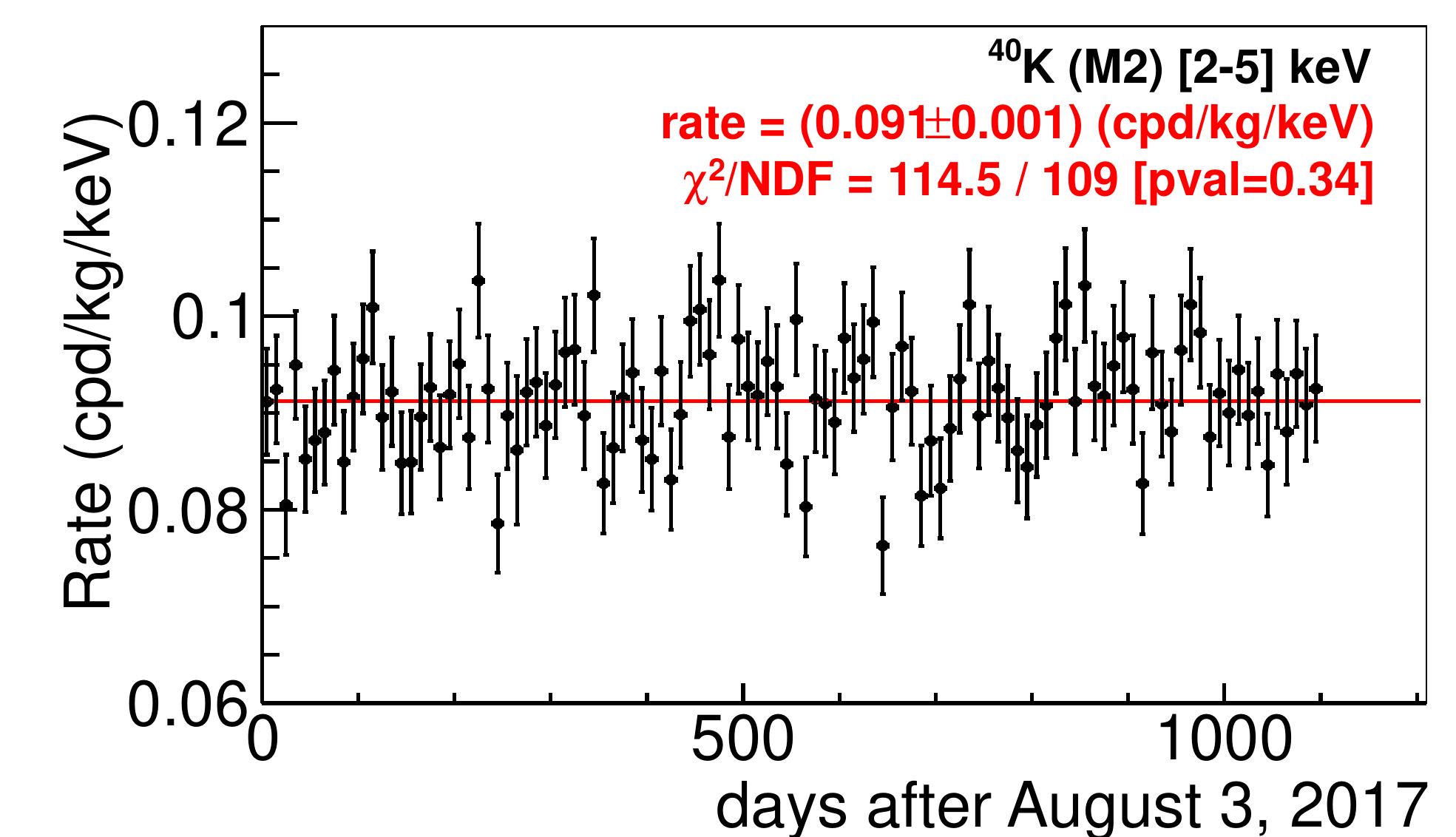}
\includegraphics[width=.48\textwidth]{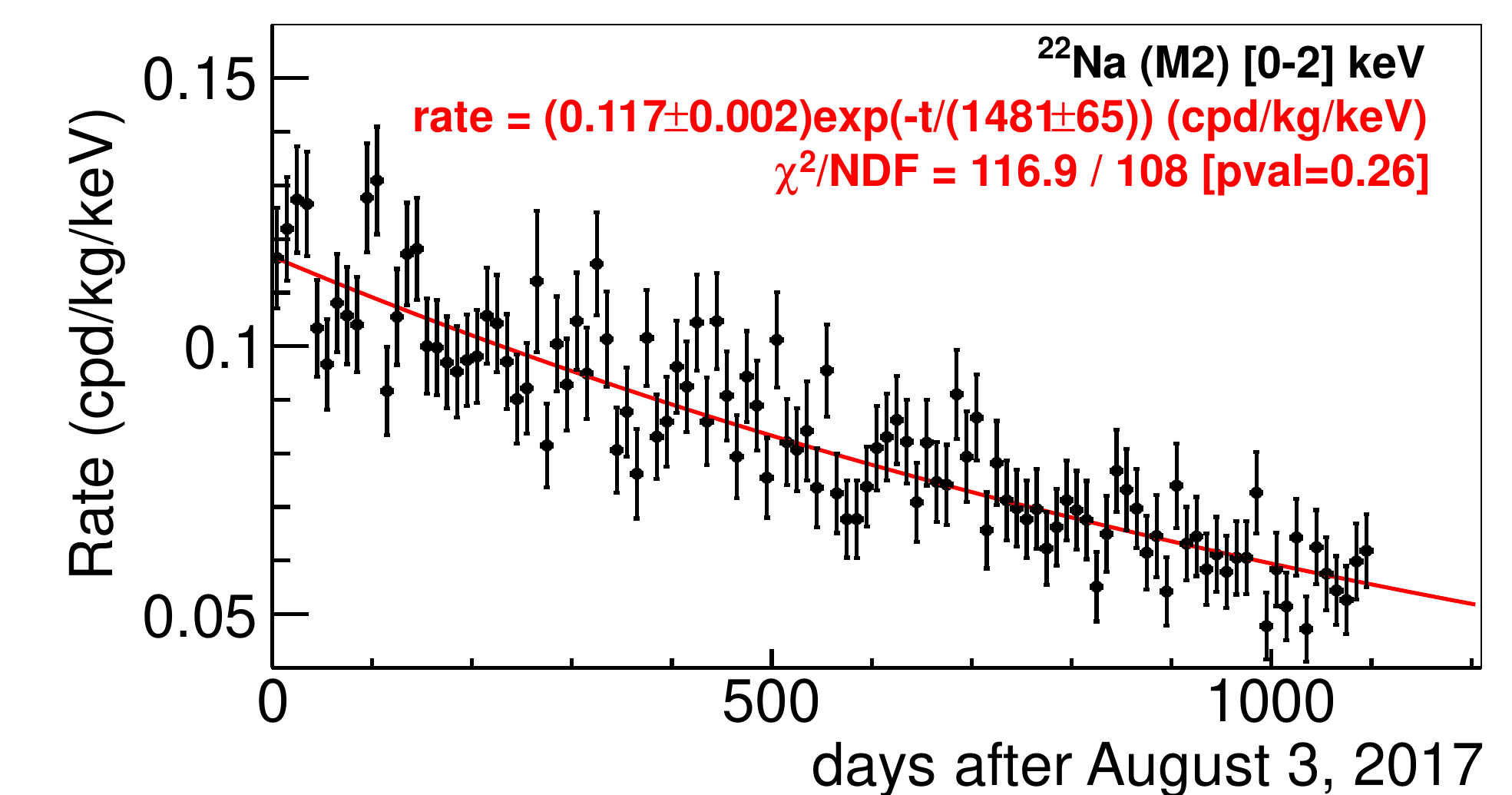}
\caption{\label{fig:stab1} Time evolution of the rate of events corresponding to $^{40}$K (upper panel) and $^{22}$Na (lower panel) at low energy identified by the coincidence with the corresponding HE gamma in a second module. 
} 
\end{figure}

We use specific control populations to monitor the stability of the event selection efficiencies and to assess continuously that the background model provides a good description of data. Figure~\ref{fig:stab1} shows coincident events in any two modules (events with multiplicity equal to 2, labeled M2) identified as compatible with $^{40}$K and $^{22}$Na decays in one of the modules. They are tagged by the detection of the high-energy gamma ray emitted in the EC decay in a high percentage of the cases, which after escaping from that module, interacts and is fully absorbed in a second module. Both isotopes are internal radioactive contaminants present in the NaI crystal, by contamination of the raw powder in the case of $^{40}$K and by activation while the powder/crystal was exposed to cosmic rays before moving the detectors underground for $^{22}$Na. These event populations show the time evolution corresponding to the parent isotopes half-lives, compatible with a constant rate in the case of $^{40}$K events and an exponential decay in the case of those following $^{22}$Na decay. For the latter, the fit provides a lifetime of 1481$\pm$65~days, at 1.7~$\sigma$ from the $^{22}$Na lifetime. The identification of these event populations requires triggering and selecting events properly down to 3.2~keV and 0.9~keV, respectively. This provides a good checking of the efficiency estimate and its stability along the full experiment's exposure. 
\par
Moreover, we use these populations to check the stability of the calibration procedure. We calculate the residuals of the energy associated to the $^{40}$K and $^{22}$Na peaks with respect to their nominal energy, gathering data every 90~days. We obtain average residuals of 0.01~keV at the 3.2~keV line of $^{40}$K and --0.04~keV at the 0.87~keV line of $^{22}$Na, the latter below the ROI. 
The standard deviation is about 0.015~keV for both distributions. The average energy resolutions (FWHM) are 1.36~keV at 3.2~keV and 0.63~keV at 0.87~keV, with 0.02~keV standard deviation in both cases. These numbers are in full agreement with our first year estimation~\cite{Amare:2018sxx}, 
so we can conclude that the resolution is constant after 3 years of data taking, a symptom of a stable calibration. This result supports that the low energy calibration procedure followed is robust and stable, and the possible systematic uncertainty associated in the annual modulation result is small. We are working on a more elaborated analysis of the possible contribution of other uncertainties in the energy calibration in the final result, since the decreasing statistics in the $^{22}$Na peak will force us to adapt the present calibration procedure in the five years analysis.

\subsection{Background understanding}
\label{sec:bkg}
A good knowledge of the background in the ROI is required to perform the annual modulation analysis in the ANAIS-112 data, as it will be stressed in Sec.~\ref{sec:annmodulation}. The ANAIS-112 background model explains in a consistent way all the energy regions and populations (M1 and M2 events).
\par 
The model presented in \cite{Amare:2018ndh} is used throughout the following analysis. It has been slightly improved by correcting the contribution of the cosmogenically induced isotopes $^{121m}$Te, $^{113}$Sn, and $^{109}$Cd in modules D3 to D8. 
According to our model~\cite{Amare:2018ndh,Amare:2014bea,Villar:2018ymt}, these isotopes were considered to have reached saturation while the modules were exposed to cosmic rays at the surface in the crystal growing and detector building steps. However, we have recently identified a lower initial activity consistent with M2 events having the specific signature of the $^{121m}$Te decay, hinting at no saturation in the isotope production. Although $^{109}$Cd and $^{113}$Sn events are not identified in M2 events in a similar way, we apply the corresponding correction according to the respective half-lives, considering for them the same activation exposure at the surface. These isotopes affect only very slightly the ROI, but are relevant to explain the time evolution of M2 events in detectors D6, D7, and D8, and, in the case of $^{109}$Cd, also that of M1 events outside the ROI.
A thorough revision of the background model is underway, profiting from all the accumulated exposure in the three years of data taking, in particular the contribution of $^{210}$Pb, which dominates the background below 70~keV. Any revision of the contributions in the ROI from long-life isotopes, as $^{40}$K and even $^{210}$Pb, could be reabsorbed as a constant term in our analysis, not affecting our conclusions in the search for a modulation in ANAIS-112 data. See Sec.~\ref{sec:annmodulation} for more details.
\par
\begin{figure}[hb]
\centering 
\includegraphics[width=.38\textwidth]{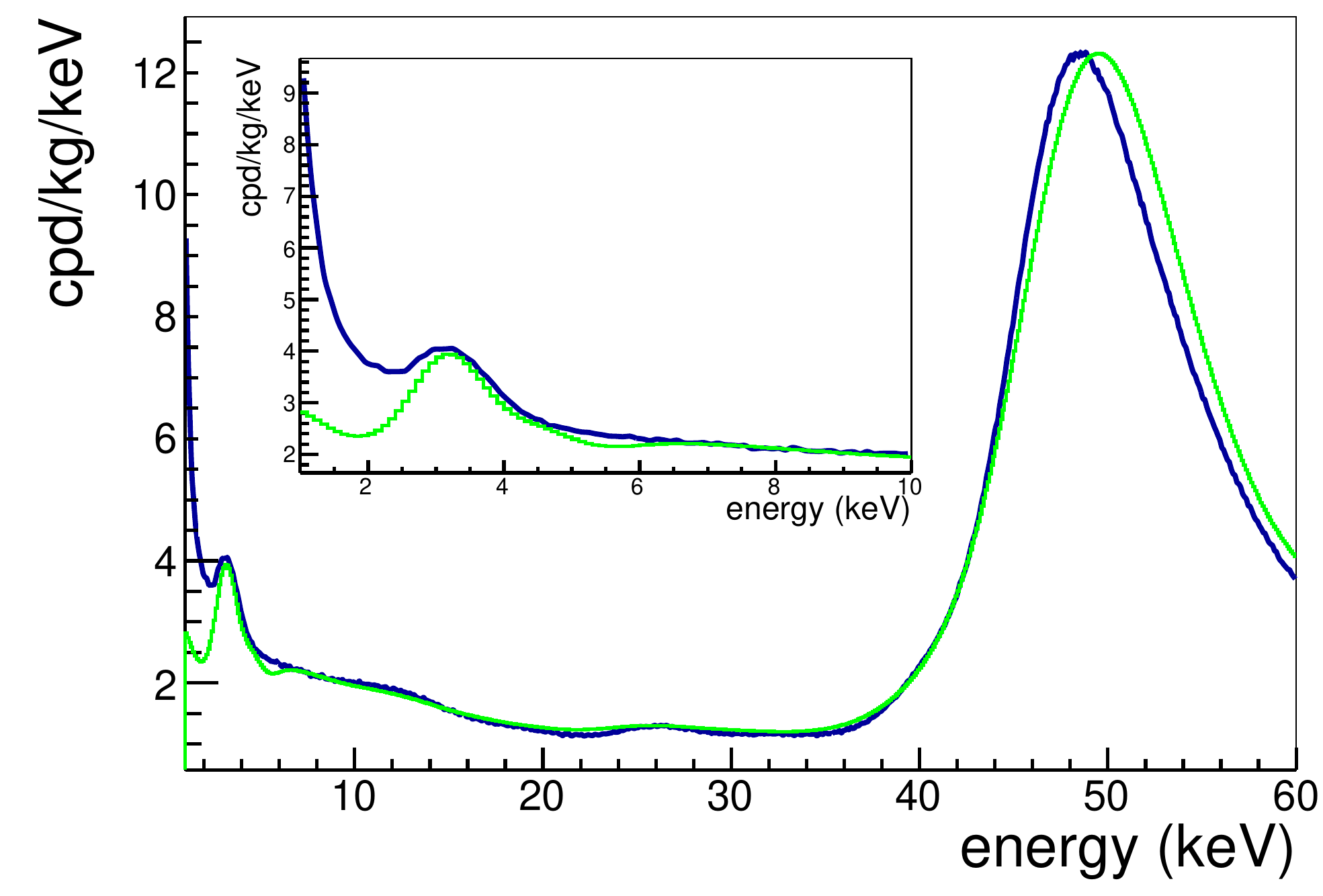}
\includegraphics[width=.4\textwidth]{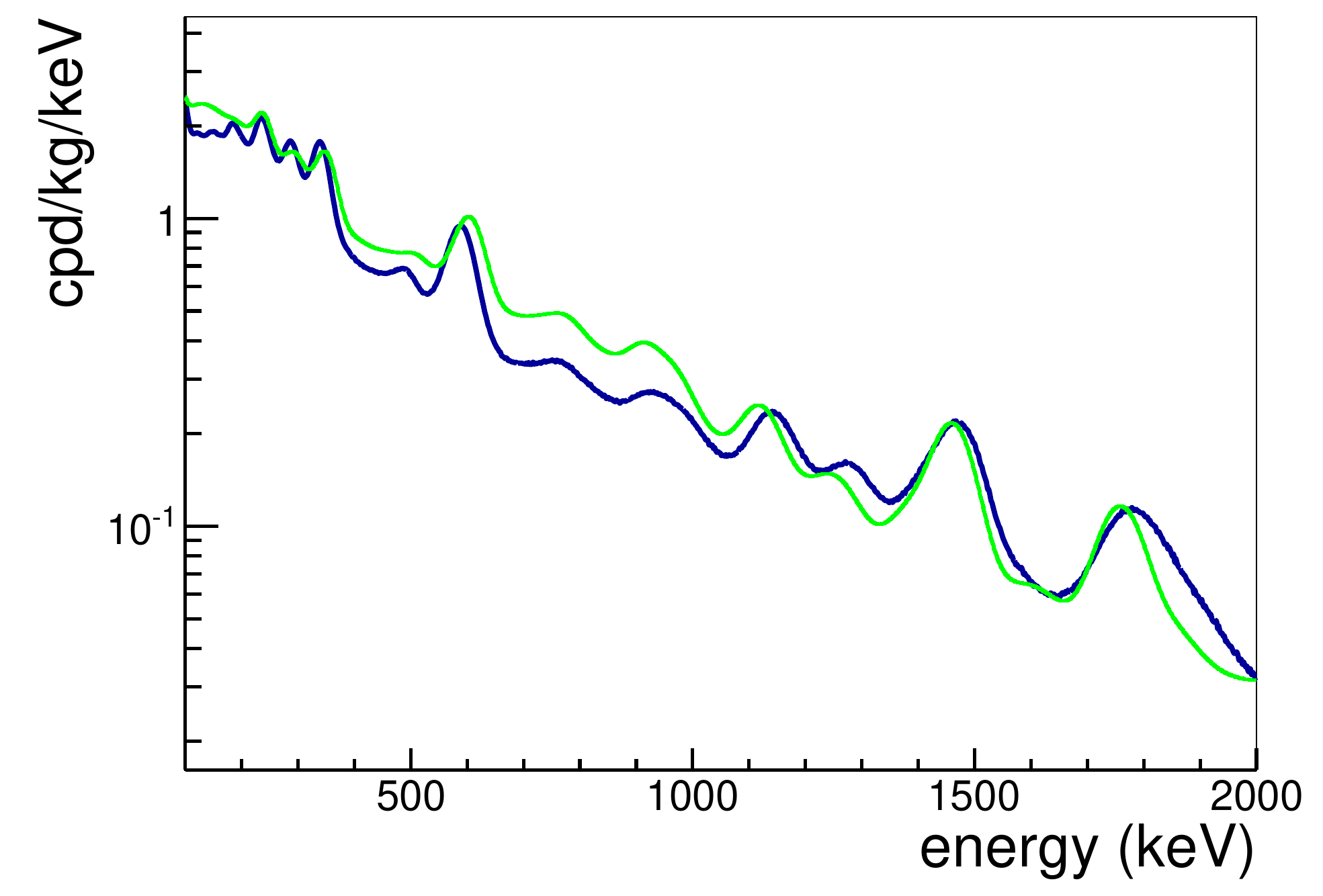}
\caption{\label{fig:bkg} ANAIS-112 full exposure background   after unblinding in the low (upper panel) and high (bottom panel) energy regions. Inset in upper panel shows the background in the ROI. Data correspond to M1 events surviving our filter protocols and corrected with the efficiency. Background model estimate is depicted in green. 
} 
\end{figure}

Figure~\ref{fig:bkg} shows the average background corresponding to the full exposure after unblinding in the low and high energy regions (upper and lower panels, respectively). The data correspond to M1 events surviving our filtering protocols and corrected with the corresponding efficiency. Also depicted in the figure are our background model estimates without any correction or normalization. 
\par
Very good agreement is observed, where it is worth remarking that fitting has not been attempted. The approach followed in \cite{Amare:2018ndh} to build the background model takes as input independent estimates of the different contaminations.
\par
\begin{figure}[htb]
\centering 
\includegraphics[width=.48\textwidth]{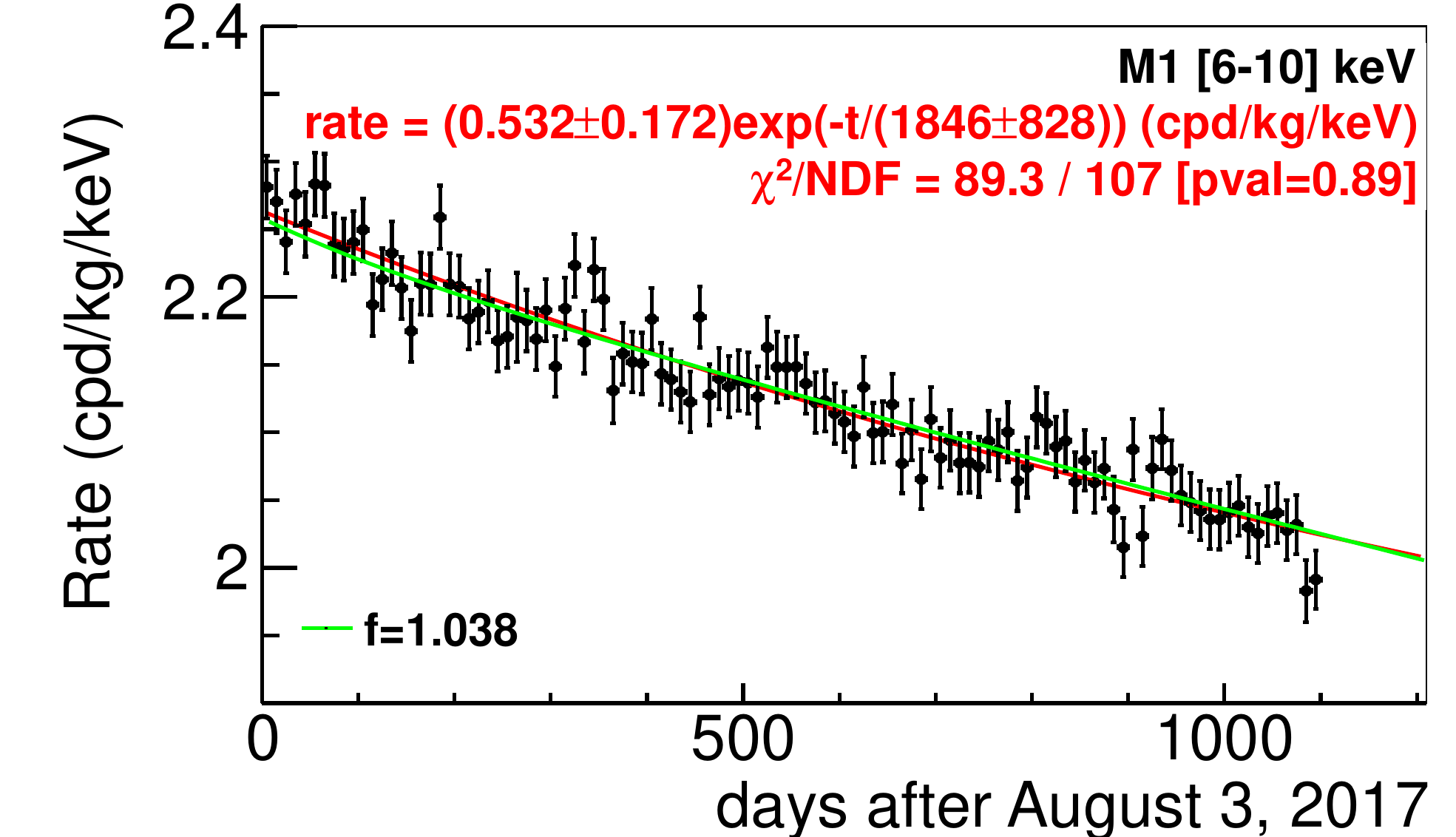}
\includegraphics[width=.48\textwidth]{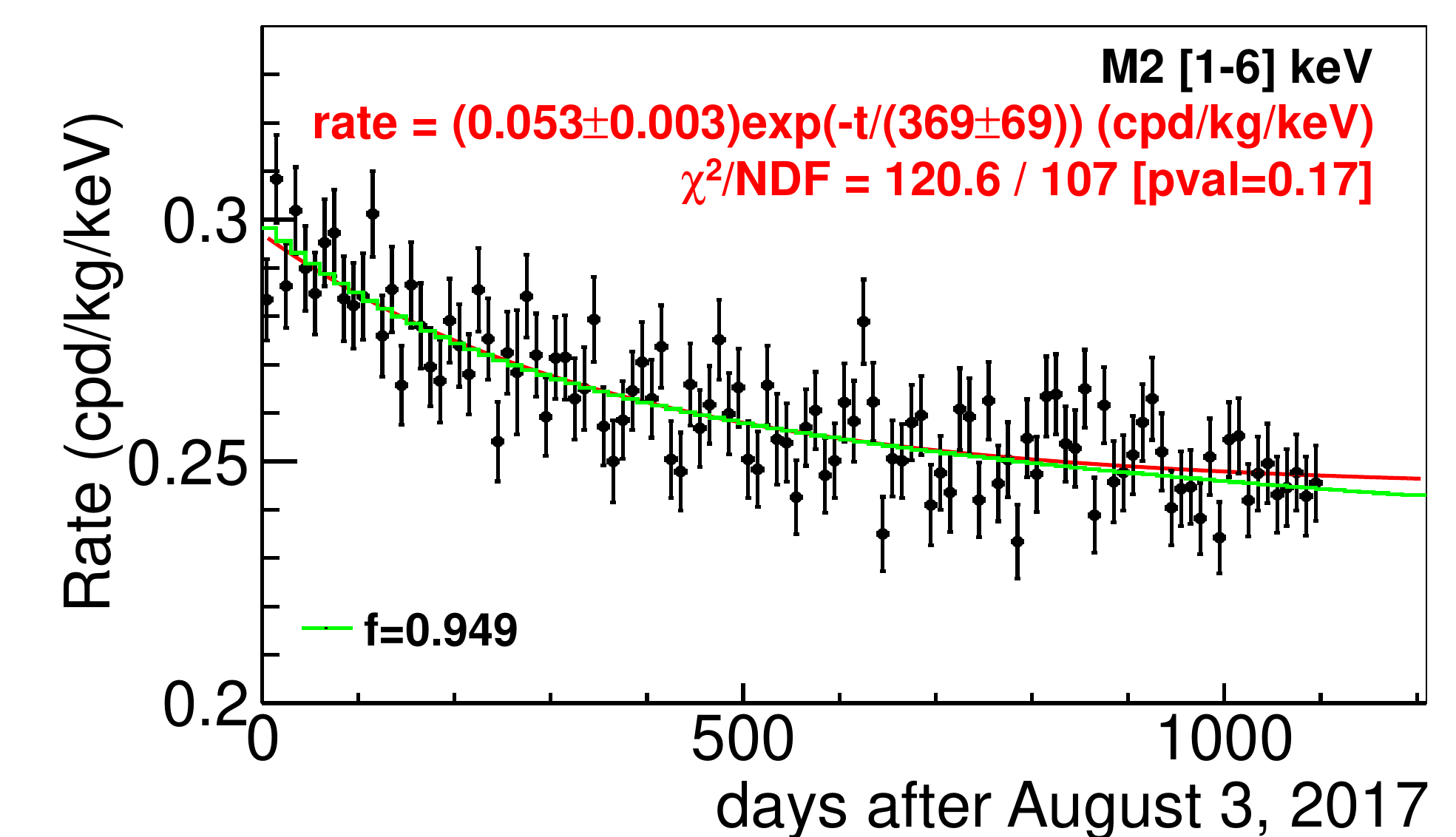}
\caption{\label{fig:stab2} Time evolution of the rate corresponding to M1 events above the ROI (upper panel) and M2 events in the ROI (lower panel). Both rates show exponential decays in time with very different "effective" lifetimes (fit is shown in red). Estimates of our background model are also shown as green solid lines, conveniently  normalized by a factor, f, displayed in both panels.
} 
\end{figure}
\begin{figure}[htb]
\centering 
\includegraphics[width=0.48\textwidth]{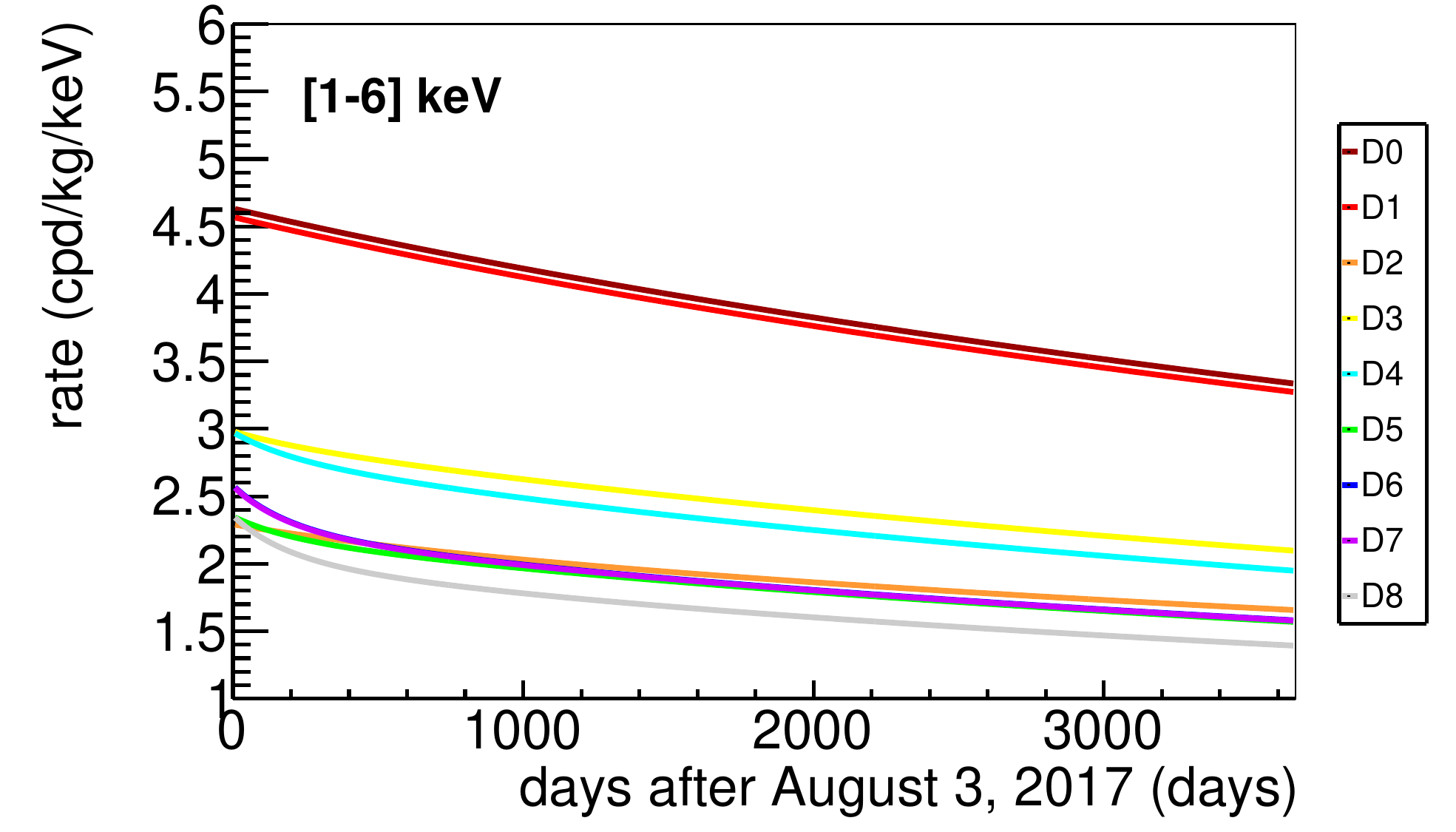}
\caption{\label{fig:MCevo} Evolution in time predicted by our background model in the ROI, [1--6]~keV energy region, for every ANAIS-112 module.
} 
\end{figure}

One important asset of our background model is that it allows for predicting the time evolution of the rate of events for the different populations. 
Figure~\ref{fig:stab2} shows the time evolution of the rate of events corresponding to M1 events in the [6--10]~keV region, just above the ROI (upper panel) and M2 events in the ROI (lower panel). Both rates show exponential decays in time with very different "effective" lifetimes: 1846$\pm$828~days and 369$\pm$69~days, respectively. Figure~\ref{fig:stab2} also compares the predictions of our background model for the evolution of the corresponding rates. The background model has been corrected by a normalization factor, f, in order to better reproduce the observations. This factor is at the level of a few percent. Coincident events are good tracers of the radioactive backgrounds. They are basically free from other populations that  could be leaking at the lowest energies in the ROI and that could explain the higher background affecting the region from 1 to 2~keV. 
As observed in Fig.~\ref{fig:stab2}, the background model reproduces quite satisfactorily M2 events from 1 to 6 keV.
Figure~\ref{fig:MCevo} shows the time evolution of the backgrounds in the ROI according to this background model for M1 events. 
\section{Blind annual modulation analysis}
\label{sec:modulation}
We have recently unblinded the third year of data and carried out the corresponding annual modulation analysis, which is presented in this section. This analysis uses 1047.4~days  
live time. After the muon cut, the effective live time is 1018.6~days, corresponding to an effective exposure of 313.95~kg$\times$y. The approach followed is similar to the previous analysis~\cite{Amare:2019jul, Amare:2019ncj}, but in this case, we include an improved background description and we 
 extend the analysis by fitting simultaneously the nine detectors with free background parameters for every module but the same modulation amplitude.
\par
Our analysis approach is quite different from DAMA/LIBRA's one, which is based on the subtraction of the average detection rate in 1~keV bins inside the ROI in data-taking cycles of roughly one year duration~\cite{Bernabei:2008yi}. As shown before, the ANAIS background is clearly time dependent, so this approach can produce a time modulation in data as an artifact caused by the interplay between a time-dependent background and the analysis method~\cite{Buttazzo:2020bto,Messina_2020}. Because of that, the development of a robust background model is mandatory. Moreover, we stress that our fitting procedure, detailed in subsequent subsections, allows us to include time-dependent backgrounds in a consistent way without subtraction of any component.
\subsection{Annual modulation search}
\label{sec:annmodulation}
We present in this section the results of the model independent analysis searching for annual modulation in the same regions explored by DAMA/LIBRA Collaboration ([1--6]~keV and [2--6]~keV). 
In order to perform a direct comparison with the results from the DAMA/LIBRA analysis with one
free parameter~\cite{Bernabei:2018yyw}, we fix the period to 1 year and the maximum of the modulation to June 2.  
\par
In a first search, the data from the nine modules are added together, grouped in bins of 10 days. We construct the $\chi^2$ function,
\begin{equation}
    \chi^2 = \sum_i \frac{(n_i - \mu_i)^2}{\sigma_i^2} ,
\end{equation}
where $n_i$ is the  number of events in the time bin $t_i$ obtained by correcting the measured number of events by the live time at that temporal bin and the detector efficiency [Eq.~\ref{eq:eff}], $\sigma_i$ is the corresponding Poisson uncertainty, accordingly corrected by live time and efficiency, and $\mu_i$ is the expected number of events at that time bin, that can be written as
\begin{equation}
    \mu_i=[R_0\phi_{bkg}(t_i)+S_m cos(\omega(t_i-t_0))]M\Delta E \Delta t .
\end{equation}
Here, $R_0$ is a free parameter that represents the unmodulated rate in the detector, $\phi_{bkg}$ is the  probability distribution function (PDF) of any unmodulated  component, $S_m$ is the modulation amplitude, $\omega$ is fixed to $2\pi/365$\,d\,$=0.01721$\,rad\,d$^{-1}$, $t_0$ to $-62.2$\,d (corresponding the cosine maximum to June 2, when taking as time origin August 3, 2017), $M$ is the total detector mass, $\Delta E$ is the energy interval width, and $\Delta t$ the time bin width (ten days, in our case). S$_m$ is fixed to 0 for the null hypothesis and left unconstrained (positive or negative) for the modulation hypothesis.
\par
In modeling the experimental background, we follow two different approaches. In the first one, following our previous analysis~\cite{Amare:2019jul, Amare:2019ncj}, we approximate the background evolution with an exponential decay:
\begin{equation}\label{eq:fit1}
    \phi_{bkg}(t_i)=1+fe^{-t_i/\tau},
\end{equation}
where $f$ and $\tau$ are free parameters.
The second approach exploits our Monte Carlo background model~\cite{Amare:2018ndh} in order to compute the background evolution in time, which is not a simple exponential, as it is a sum of different components. This evolution is converted into a probability distribution function $\phi^{MC}_{bkg}(t)$, so the background PDF can be expressed as:  
\begin{equation}\label{eq:fit2}
    \phi_{bkg}(t_i)=1+f\phi^{MC}_{bkg}(t_i).
\end{equation}
It is worth remarking that in this approach the number of nuisance parameters is reduced by one with respect to Eq.~\ref{eq:fit1}. The constant terms in both equations represent any nonvarying rate, including the unmodulated term of an hypothetical WIMP component.  
\par
\begin{figure*}[htbp]
\centering %
\includegraphics[width=.49\textwidth]{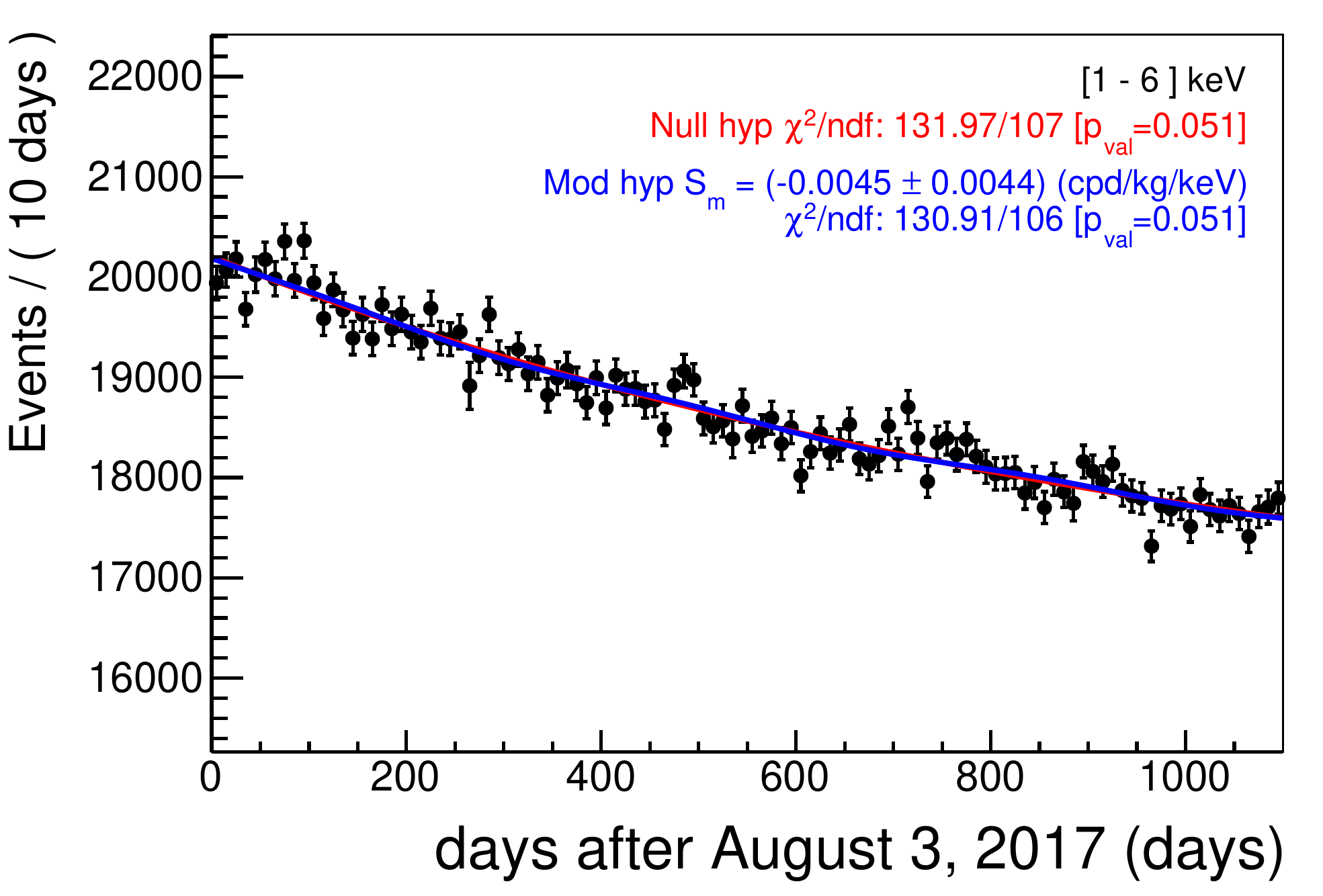}
\includegraphics[width=.49\textwidth]{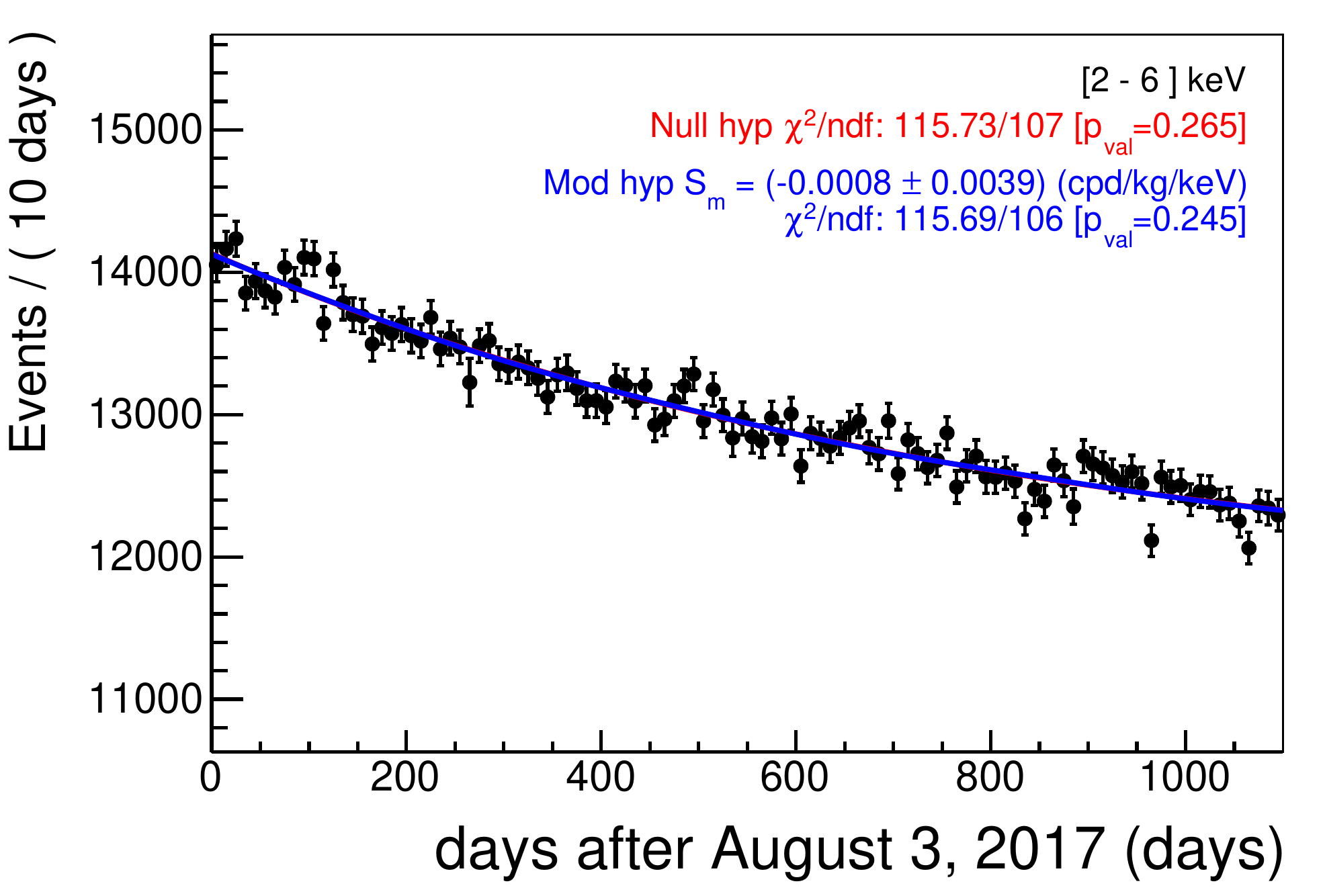}
\includegraphics[width=.49\textwidth]{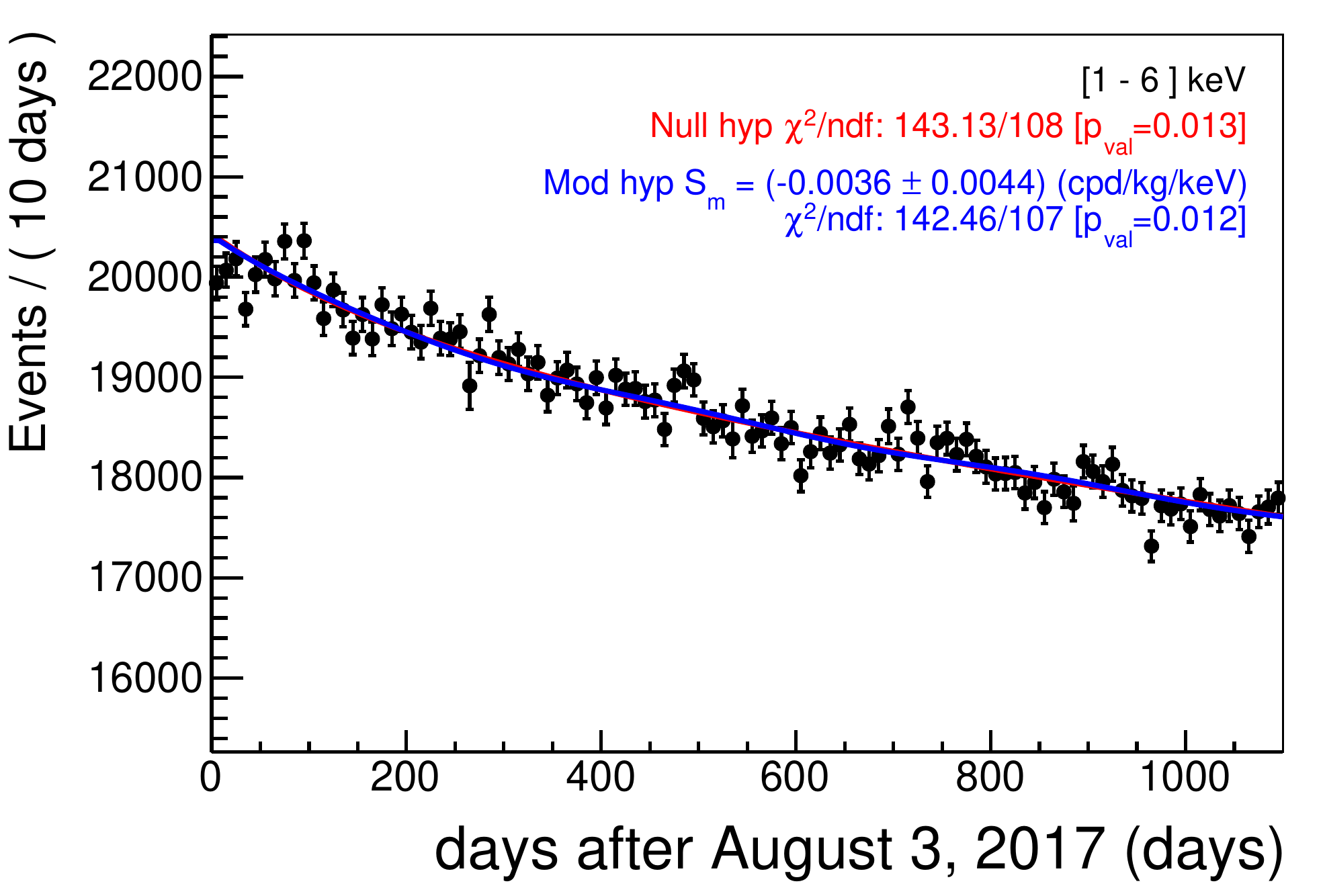}
\includegraphics[width=.49\textwidth]{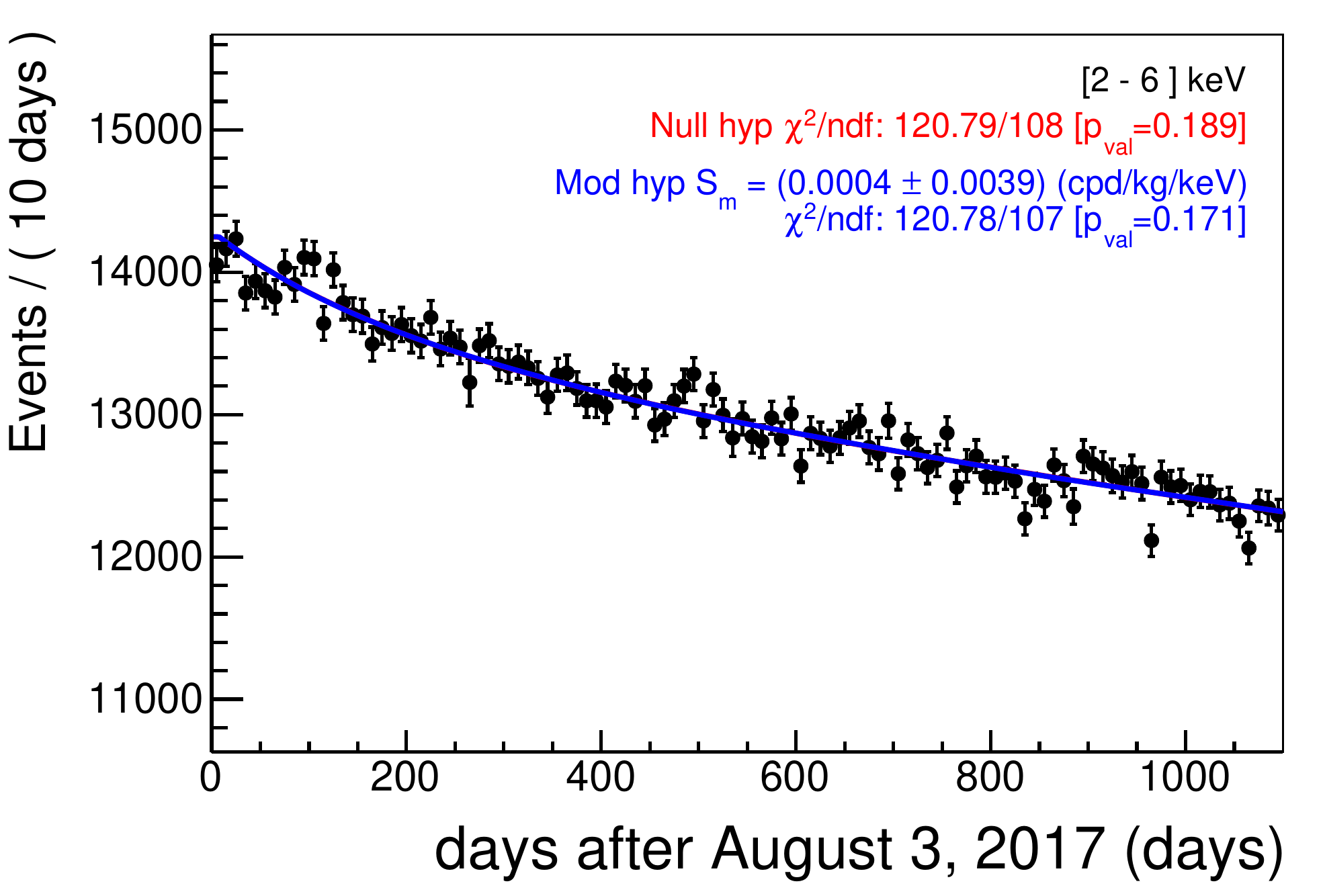}
\caption{Upper panels: ANAIS-112 fit results for three years of data in the [1--6]~keV (left) and [2--6]~keV (right) energy regions, both in the modulation (blue) and null hypothesis (red) when the background is described by Eq.~\ref{eq:fit1}. Lower panels: same, but using the background described by Eq.~\ref{eq:fit2}. Best fits for $S_m$, $\chi^2$ and p values are also shown.}
\label{fig:fit1}
\end{figure*}

The results of the $\chi^2$ minimization are shown in Fig.~\ref{fig:fit1} for [1--6]~keV (left) and [2--6]~keV (right) energy regions. The upper panels correspond to the fit with the exponential background hypothesis, while the lower panels show the results for the Monte Carlo background model. The $\chi^2$ and p values of the fit for the null (modulation) hypothesis are also shown in red (blue), together with the best fit for $S_m$. The best fit values are also collected in Tables~\ref{tab:results} and \ref{tab:resultsNuisance}. 
 \begin{table*}[htbp]
 \centering
 \begin{tabular}{ccccccc}
 \hline\hline
 \multirow{2}{*}{Energy region} & \multirow{2}{*}{Model} & $\chi^2$/NDF  & Nuisance & $S_m$ & \multirow{2}{*}{p value mod}  & \multirow{2}{*}{p value null}  \\
 & &null hyp & params.& cpd/kg/keV & &  \\
 \hline 
 \multirow{3}{1.5cm}{[1--6] keV} & eq.~\ref{eq:fit1} & 132 / 107 & 3 & -0.0045$\pm$0.0044 &0.051 & 0.051 \\
 & eq.~\ref{eq:fit2} & 143.1 / 108 & 2 & -0.0036$\pm$0.0044 &0.012 & 0.013 \\
 & eq.~\ref{eq:fit3} & 1076 / 972 & 18 & -0.0034$\pm$0.0042 &0.011 & 0.011 \\
 \hline 
 \multirow{3}{1.5cm}{[2--6] keV} & eq.~\ref{eq:fit1} & 115.7 / 107 & 3 & -0.0008$\pm$0.0039 &0.25 & 0.27 \\
 & eq.~\ref{eq:fit2} & 120.8 / 108 & 2 & 0.0004$\pm$0.0039 &0.17 & 0.19 \\
 & eq.~\ref{eq:fit3} & 1018 / 972 & 18 & 0.0003$\pm$0.0037 &0.14 & 0.15 \\
 \hline \hline
 \end{tabular} 
 \caption{\label{tab:results} Summary of the fits searching for an annual modulation with fixed phase in the three years of ANAIS-112 data for different background modeling (see text for more details).} 
 \end{table*} 
\setlength{\tabcolsep}{2em}
\begin{table*}[htbp]
 \centering
 \begin{tabular}{cccccc}
 \hline\hline
 \multirow{2}{*}{Energy region} & \multirow{2}{*}{Model} & \multirow{2}{*}{detector} & bkg index & \multirow{2}{*}{f} & $\tau$  \\
 & & & cpd/kg/keV & & days \\
 \hline 
 \multirow{11}{1.5cm}{[1--6] keV} & eq.~\ref{eq:fit1} & all &3.605$\pm$0.003 & 0.24$\pm$0.03 & 1034$\pm$200 \\
 & eq.~\ref{eq:fit2} & all &3.605$\pm$0.003 & 0.85$\pm$0.02 & - \\ 
 & \multirow{9}{*}{ eq.~\ref{eq:fit3}} & 0 & 5.10$\pm$0.01  & 0.96$\pm$0.06 & - \\ 
 & & 1 & 5.09$\pm$0.01  & 1.00$\pm$0.08 & - \\ 
 & & 2 & 2.880$\pm$0.009  & 0.96$\pm$0.08 & - \\ 
 & & 3 & 3.885$\pm$0.009  & 0.71$\pm$0.06 & - \\ 
 & & 4 & 3.548$\pm$0.009  & 0.76$\pm$0.05 & - \\ 
 & & 5 & 3.490$\pm$0.009  & 0.62$\pm$0.05 & - \\ 
 & & 6 & 3.184$\pm$0.009  & 0.82$\pm$0.04 & - \\ 
 & & 7 & 2.803$\pm$0.008  & 0.85$\pm$0.04 & - \\ 
 & & 8 & 2.426$\pm$0.007  & 0.88$\pm$0.04 & - \\ 
 \hline 
 \multirow{11}{1.5cm}{[2--6] keV} & eq.~\ref{eq:fit1} & all &3.145$\pm$0.003 & 0.21$\pm$0.02 & 821$\pm$100 \\
 & eq.~\ref{eq:fit2} & all &3.145$\pm$0.003  & 0.91$\pm$0.02 & - \\ 
 & \multirow{9}{*}{ eq.~\ref{eq:fit3}} & 0 & 4.584$\pm$0.010  & 1.00$\pm$0.09 & - \\ 
 & & 1 & 4.667$\pm$0.010  & 1.00$\pm$0.08 & - \\ 
 & & 2 & 2.434$\pm$0.007  & 0.87$\pm$0.09 & - \\ 
 & & 3 & 3.149$\pm$0.008  & 0.80$\pm$0.08 & - \\ 
 & & 4 & 3.067$\pm$0.008  & 0.76$\pm$0.05 & - \\ 
 & & 5 & 2.929$\pm$0.008  & 0.73$\pm$0.06 & - \\ 
 & & 6 & 2.786$\pm$0.008  & 0.87$\pm$0.04 & - \\ 
 & & 7 & 2.495$\pm$0.007  & 0.98$\pm$0.04 & - \\ 
 & & 8 & 2.172$\pm$0.007  & 0.92$\pm$0.04 & - \\ 
 \hline \hline
 \end{tabular} 
 \caption{\label{tab:resultsNuisance} Summary of the nuisance parameters obtained in the fits searching for an annual modulation with fixed phase in the three years of ANAIS-112 data taking for different background modeling (see text for more details).}
 \end{table*} 
In the [2\nobreakdash--6]~keV region data are well described by the null hypothesis in both models (p values of 0.265 and 0.189). Smaller p values (0.051 and 0.013) are obtained in [1--6]~keV region. We will comment on this later. For the modulation hypothesis, we obtain in all cases best fit modulation amplitudes compatible with zero at 1$\sigma$. The standard deviation of the modulation amplitude $\sigma(S_m)$ is the same for the two background modeling approaches in the
[1--6]~keV (0.0044~cpd/kg/keV) and [2--6]~keV (0.0039~cpd/kg/keV) energy regions.
\par
\begin{figure*}[htbp]
\centering %
\includegraphics[width=.70\textwidth]{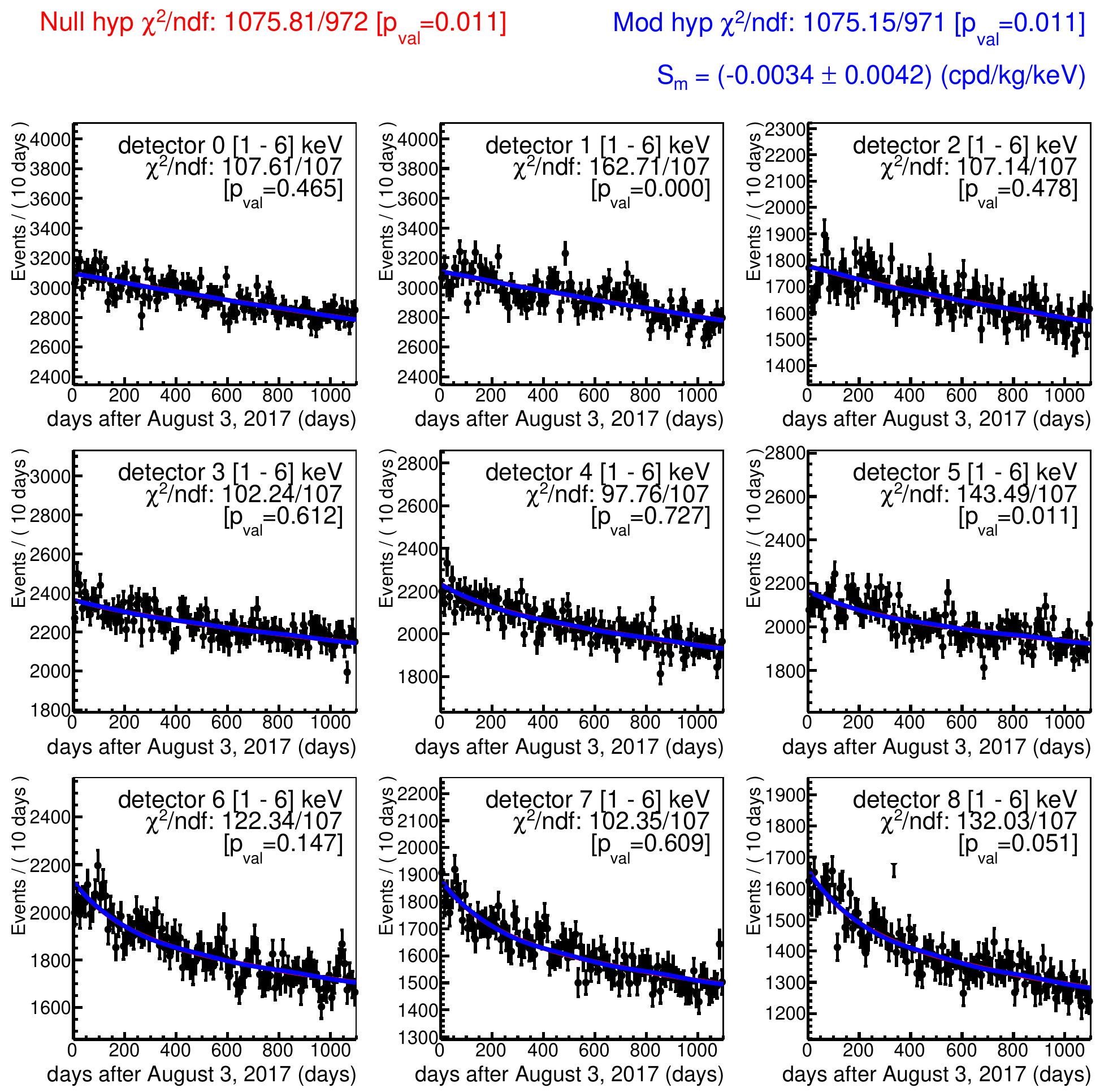}\\
\caption{Results of the fit of the nine modules data using Eq.~\ref{eq:fit3} in the [1--6]~keV energy region, in the modulation (blue) and null hypothesis (red). Best fits for $S_m$, $\chi^2$ and p values are also shown.
}
\label{fig:simulfit16}
\end{figure*}
\begin{figure*}[htbp]
\centering %
\includegraphics[width=.70\textwidth]{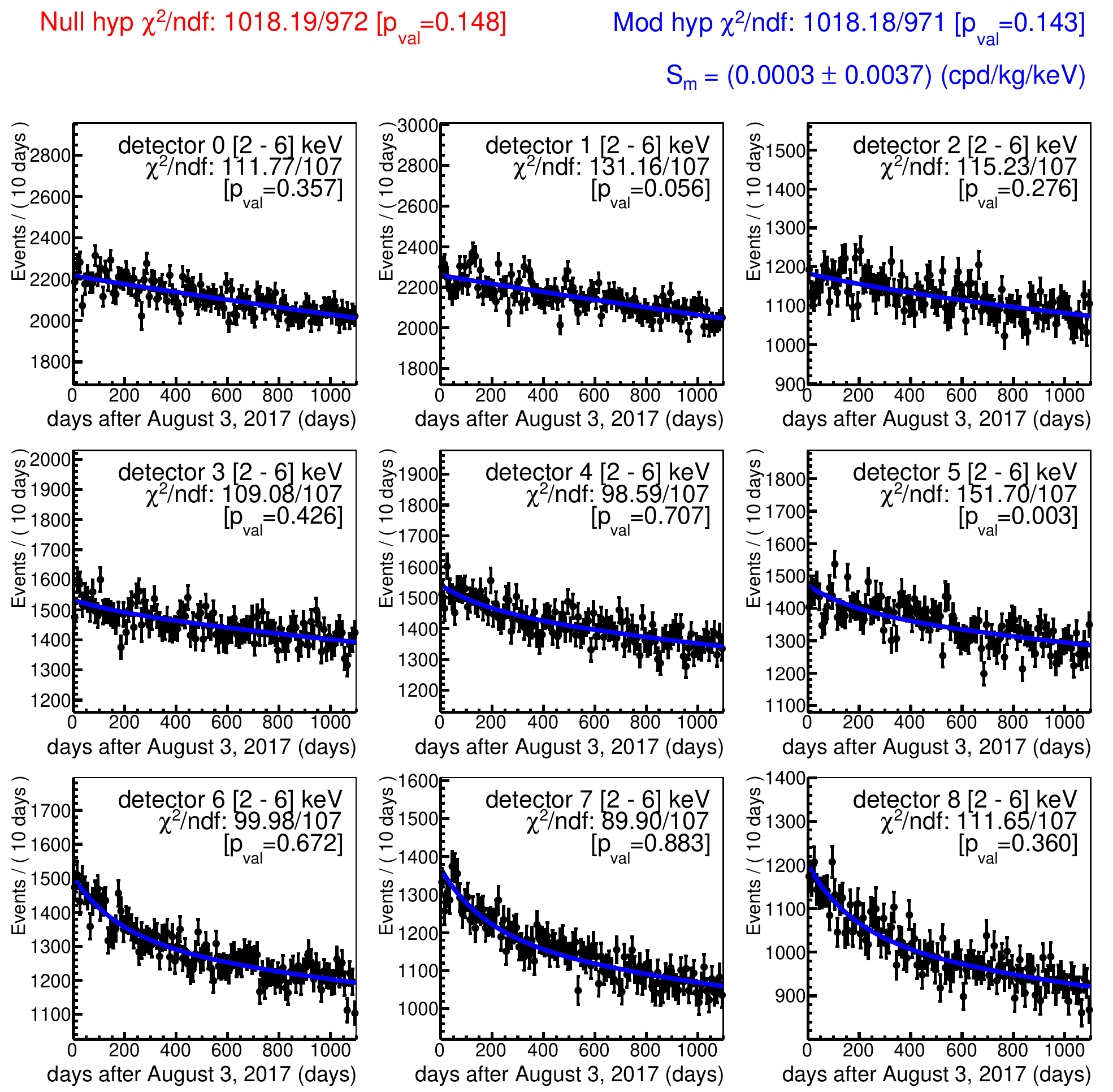}
\caption{Same as Fig.~\ref{fig:simulfit16} for  [2--6]~keV energy region.
}
\label{fig:simulfit26}
\end{figure*}

In order to account for systematic effects related to the differences in backgrounds and efficiencies among detectors, we apply a third approach in which the number of measured events of every module, $n_{i,d}$, is considered independently. The summation in the $\chi^2$ expression is therefore performed also over detectors. The expected number of events for every time bin $t_i$ and detector $d$ is written as
\begin{equation}\label{eq:fit3}
    \mu_{i,d}=[R_{0,d}(1+f_d\phi_{bkg,d}^{MC}(t_i))+S_m cos(\omega(t_i-t_0))]M_d\Delta E \Delta t,
\end{equation}
where $M_d$ is the mass of every module, $\phi_{bkg,d}^{MC}$ is the PDF sampled from the MC background evolution in time calculated independently for every module, and $R_{0,d}$ and $f_d$ are free parameters. In this case, the number of nuisance parameters is 18. The results of the fit are displayed in Figs.~\ref{fig:simulfit16} and \ref{fig:simulfit26} and the best fit parameters and p values summarized in Tables~\ref{tab:results} and \ref{tab:resultsNuisance}.
The $\chi^2$, number of degrees of freedom (NDF) and p values are also calculated separately for the data of every module and displayed in the legend of each panel.
It is worth noting that in general the fits obtained for the individual detectors are
good, with p values larger than 0.05 in all cases except for D1 and D5. The results
are compatible with those of the previous methods, with slightly lower values of $\sigma(S_m)$, both in the [1--6]~keV  and [2--6]~keV energy regions, as expected from our sensitivity analysis \cite{Coarasa:2018qzs}. Therefore, in the following we select this method to quote our final result.

\par
The results for the best fit of the amplitude modulation collected in Table~\ref{tab:results} for the three background models show a variation below 0.0011 (0.0012) cpd/kg/keV for the [1--6] keV ([2--6] keV) energy region, which is much lower than the statistical uncertainty. In the following we will not use this estimate as a systematical error but as a limit to the possible systematic contribution of the background modeling.
\par
\begin{figure}[htpb]
\centering 
\includegraphics[width=.48\textwidth]{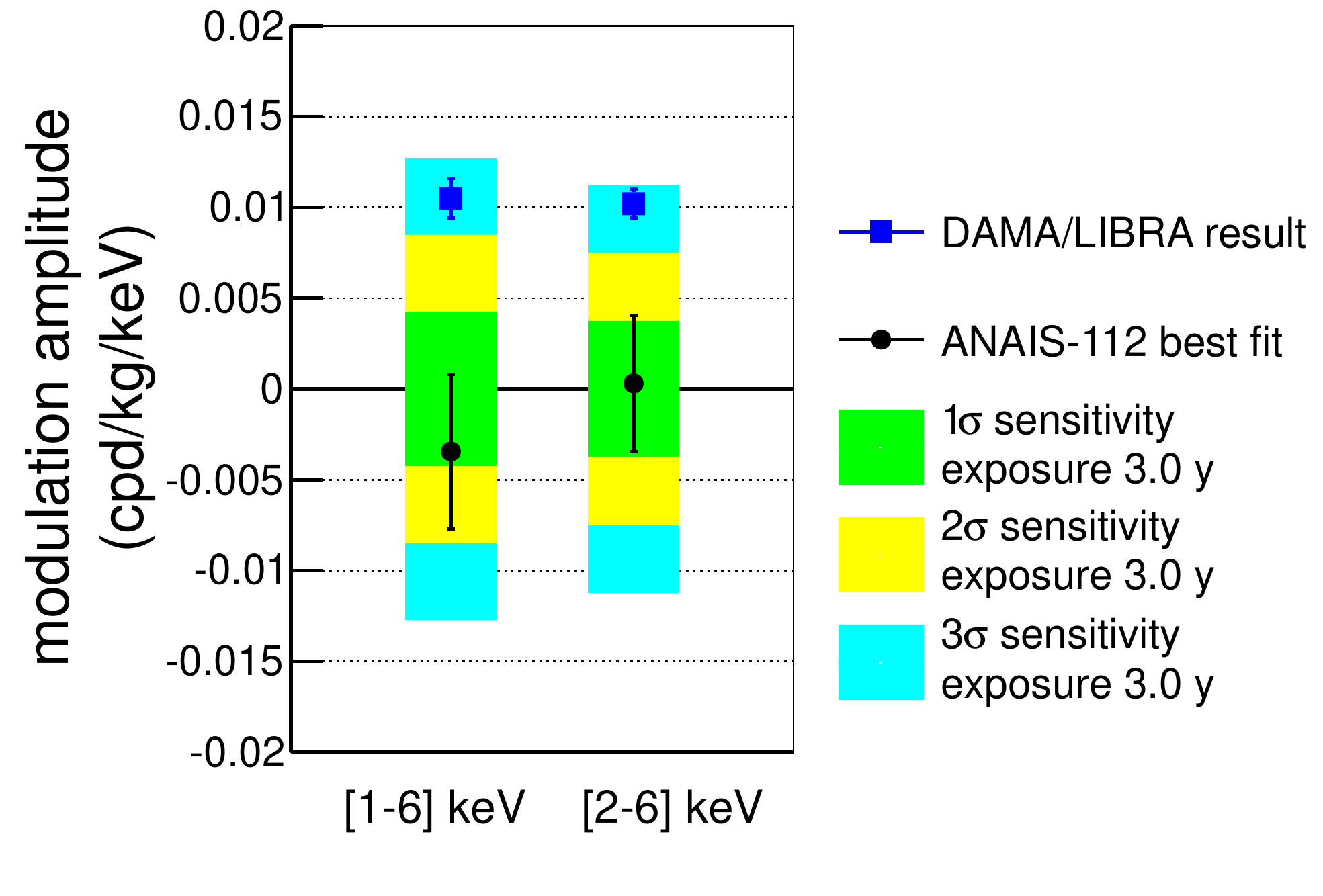}
\caption{\label{fig:summary} 
Comparison between ANAIS-112 results on annual modulation using three years of data and DAMA/LIBRA modulation best fit. Estimated sensitivity is shown
at different confidence levels as colored bands: green at 1$\sigma$, yellow
at 2$\sigma$, and cyan at 3$\sigma$. 
}
\end{figure}

Finally, Fig.~\ref{fig:summary} summarizes the results of the ANAIS-112 annual modulation analysis with three years of exposure in comparison with the DAMA/LIBRA best fit~\cite{Bernabei:2018yyw}. Our best fits 
are incompatible with the DAMA/LIBRA result at 3.3 (2.6)~$\sigma$, for a sensitivity of 2.5 (2.7)~$\sigma$ at [1--6]~keV ([2--6]~keV).
\subsection{Goodness of fit and consistency checks}\label{sec:consistency}
In the three analyzed scenarios the null hypothesis is well supported by the $\chi^2$ test in the [2--6]~keV energy region, but it provides a poorer description of the data in the [1--6]~keV energy range. The results of the fit to Eq.~\ref{eq:fit3} suggest that this could be ascribed to instabilities in two of the modules, D1 and D5. In particular, 
the PMTs of D5 were operated at a higher voltage during the first year, producing instabilities in the gain during that period, see Fig.~\ref{fig:calstab}.
If we remove from the analysis the D1 and D5 data, 
the goodness of the fit in the [1--6]~keV region improves for the null hypothesis in the three cases, with ($\chi^2$/NDF, p value) equal to  (108.1/107, 0.45), (113.1/108, 0.35) and (769.2/756, 0.36), respectively, while the modulation hypothesis yields again amplitudes compatible with zero ($S_m$=(0$\pm$0.0048, 0.0014$\pm$0.0048, 0.00001$\pm$0.0047)~cpd/kg/keV).
The anomalous behavior of the rate evolution in these two modules could be a symptom of noise leaking in the very low energy bin [1-2]~keV which is not removed by our filtering protocols. This hypothesis is  also supported by the poor agreement between our background model and the measured events rate in the [1-2]~keV energy region~\cite{Amare:2018ndh}. We are working on the application of machine learning techniques in order to improve the rejection of nonbulk scintillation events below 2~keV.
\par
\begin{figure}[htbp]
\centering 
\includegraphics[width=.48\textwidth]{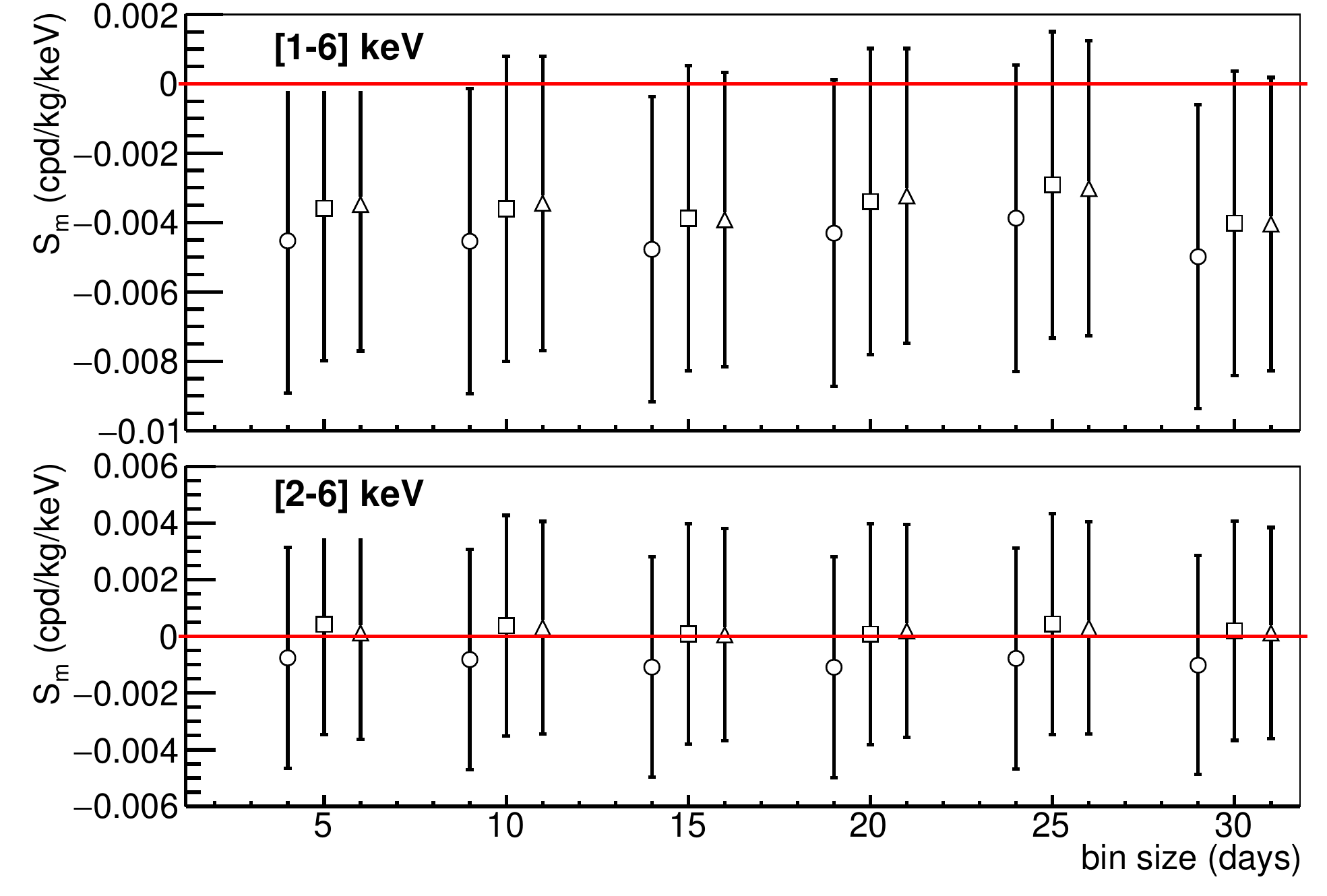}
\caption{Best fit values for the three methods in the [1--6]~keV (upper panel) and [2--6]~keV (lower panel) energy regions for different choices of the time binning. Circles, squares and triangles correspond to fits to eqs.~\ref{eq:fit1}, \ref{eq:fit2} and \ref{eq:fit3}, respectively.}
\label{fig:timeBin}
\end{figure}

Next, we repeat the previous analysis, but considering only the last two years of data taking. In this way, we study any systematic effect related to the decreasing event rate, which is more pronounced in the first year. The results are collected in Table~\ref{tab:results2y}. For the first method, they are compatible with the two years results reported in \cite{Amare:2019ncj}. We obtain larger p values in the [1--6]~keV energy range, which could be partially explained by the more stable behavior of D5 in this time period.  
 \setlength{\tabcolsep}{1.5em}
  \begin{table*}[p]
 \centering
 \begin{tabular}{ccccccc}
 \hline\hline
 \multirow{2}{*}{Energy region} & \multirow{2}{*}{Model} &  $\chi^2$/NDF & nuisance & $S_m$ & \multirow{2}{*}{p value mod}  & \multirow{2}{*}{p value null}  \\
 & &null hyp & params& cpd/kg/keV & &  \\
 \hline 
 \multirow{3}{1.5cm}{[1--6] keV} & eq.~\ref{eq:fit1} & 81.23 / 70 & 3 & -0.0056$\pm$0.0055 &0.17 & 0.17 \\
 & eq.~\ref{eq:fit2} & 81.37 / 71 & 2 & -0.0057$\pm$0.0055 &0.19 & 0.19 \\
 & eq.~\ref{eq:fit3} & 621.7 / 639 & 18 & -0.0100$\pm$0.0051 &0.71 & 0.68 \\
 \hline 
 \multirow{3}{1.5cm}{[2--6] keV} & eq.~\ref{eq:fit1} & 81.65 / 70 & 3 & 0.0032$\pm$0.0049 &0.15 & 0.16 \\
 & eq.~\ref{eq:fit2} & 81.82 / 71 & 2 & 0.0034$\pm$0.0049 &0.17 & 0.18 \\
 & eq.~\ref{eq:fit3} & 604.1 / 639 & 18 & 0.0013$\pm$0.0046 &0.83 & 0.84 \\
 \hline \hline
 \end{tabular} 
 \caption{\label{tab:results2y} Summary of the fits searching for an annual modulation with fixed phase in the last two years of ANAIS-112 data for different background modeling, as in Table~\ref{tab:results}. }
 \end{table*} 
Finally, we take into account the effect of the choice of the time binning. We repeat all the fits for time bins ranging from 5 to 30 days. The best fits for the modulation amplitude are summarized in 
Fig.~\ref{fig:timeBin}. The conclusion is that the binning choice has a minor effect on the results presented in the previous section, being negligible compared to the differences related to the fitting method used.
\par
In order to check whether the fits are unbiased, we perform a large set of Monte Carlo pseudoexperiments, sampled from the background models
with the best fit background parameters (see Table~\ref{tab:resultsNuisance}) and $S_m=0$. Then we fit the MC data with our model leaving $S_m$ unconstrained. The resulting best fit values, $\hat{S}_m$,
follow a Gaussian distribution that we use to calculate the bias of the fit for the null hypothesis. In addition, we generate a new set of MC data, but in this case, we introduce a modulation amplitude equal to that observed by DAMA/LIBRA in the corresponding energy interval (0.0105$\pm$0.0011~cpd/kg/keV at [1--6]~keV and 0.0102$\pm$0.0008~cpd/kg/keV at [2--6]~keV). We follow the same procedure as before to compute the bias at the DAMA/LIBRA observed modulation. The results are collected in Table~\ref{tab:bias}. In all cases, the bias is compatible with zero or negligible. We also show in the last column the standard deviation of $\hat{S}_m$ obtained from the distributions, that
agrees with our estimates presented in Sec.~\ref{sec:sensitivity}. 
\setlength{\tabcolsep}{2em}
  \begin{table*}[htbp]
 \centering
 \begin{tabular}{ccccc}
 \hline\hline
 \multirow{2}{*}{Energy region} & \multirow{2}{*}{Model} & Bias[null hypothess] & Bias[DAMA Sm] & $\sigma(S_m)$  \\
 & & cpd/kg/keV & cpd/kg/keV& cpd/kg/keV \\
 \hline 
 \multirow{3}{1.5cm}{[1--6] keV} & eq.~\ref{eq:fit1} &(-3$\pm$6)$\times$10$^{-5}$ & (-1$\pm$6)$\times$10$^{-5}$ & (430$\pm$4)$\times$10$^{-5}$ \\
 & eq.~\ref{eq:fit2} &(-7$\pm$6)$\times$10$^{-5}$ & (3$\pm$6)$\times$10$^{-5}$ & (439$\pm$4)$\times$10$^{-5}$ \\
 & eq.~\ref{eq:fit3} &(-26$\pm$6)$\times$10$^{-5}$ & (31$\pm$6)$\times$10$^{-5}$ & (425$\pm$4)$\times$10$^{-5}$ \\
 \hline 
 \multirow{3}{1.5cm}{[2--6] keV} & eq.~\ref{eq:fit1} & (3$\pm$5)$\times$10$^{-5}$ & (-10$\pm$5)$\times$10$^{-5}$ & (386$\pm$4)$\times$10$^{-5}$ \\
 & eq.~\ref{eq:fit2} &(8$\pm$6)$\times$10$^{-5}$ & (-10$\pm$6)$\times$10$^{-5}$ & (388$\pm$4)$\times$10$^{-5}$ \\
 & eq.~\ref{eq:fit3} &(-28$\pm$5)$\times$10$^{-5}$ & (29$\pm$5)$\times$10$^{-5}$ & (371$\pm$4)$\times$10$^{-5}$ \\
 \hline \hline
 \end{tabular} 
 \caption{\label{tab:bias} Bias (true value - fitted value) of the fitting procedures derived from MC simulations assuming no modulation present (third column) and DAMA/LIBRA observed modulation (fourth column), in the six analyzed scenarios. The last column is the standard deviation of the best fit modulation amplitudes obtained from the MC.}  
 \end{table*} 
\subsection{Phase-free annual modulation analysis}\label{sec:phaseFree}
\begin{figure*}[htbp]
\centering 
\includegraphics[width=.32\textwidth]{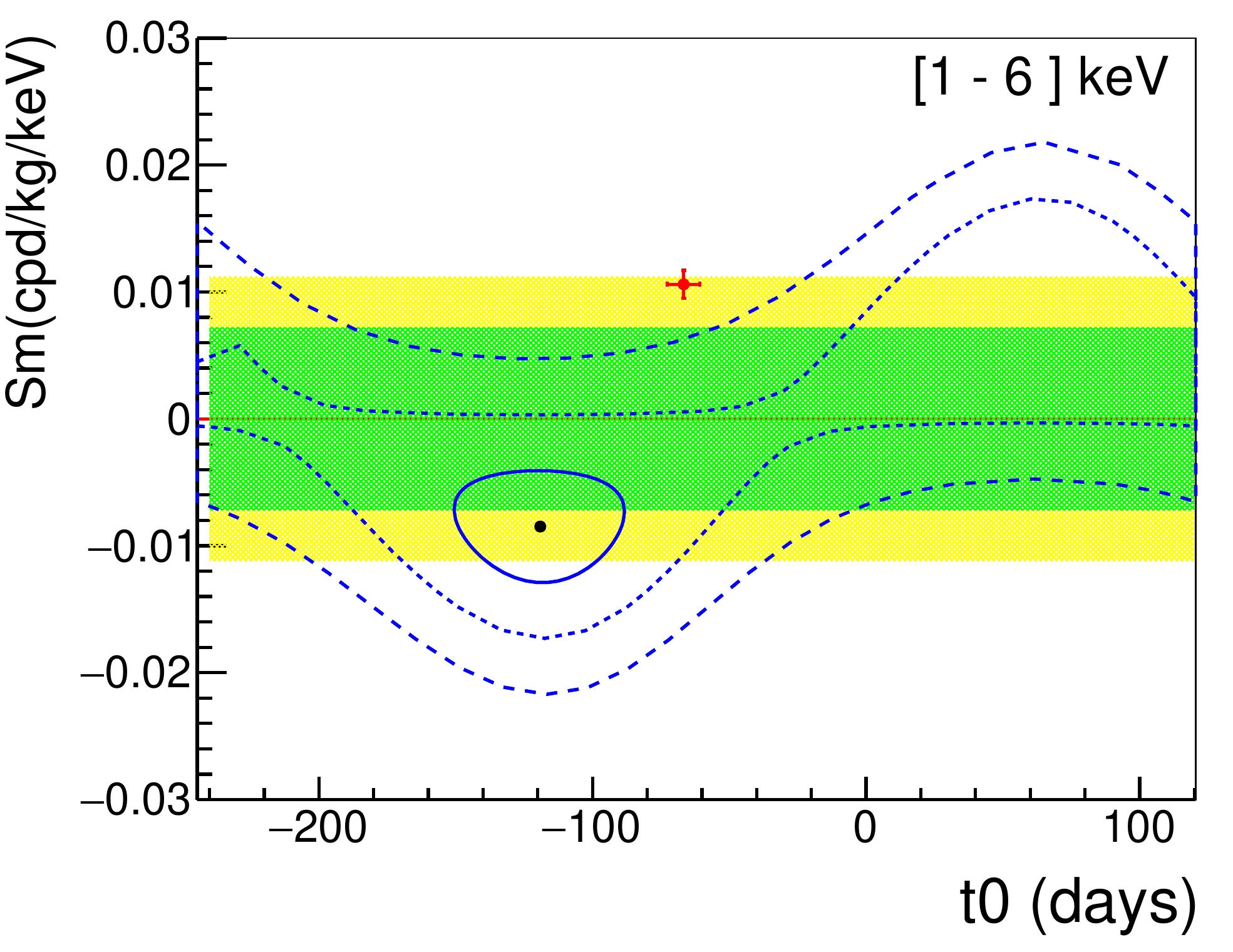}
\includegraphics[width=.32\textwidth]{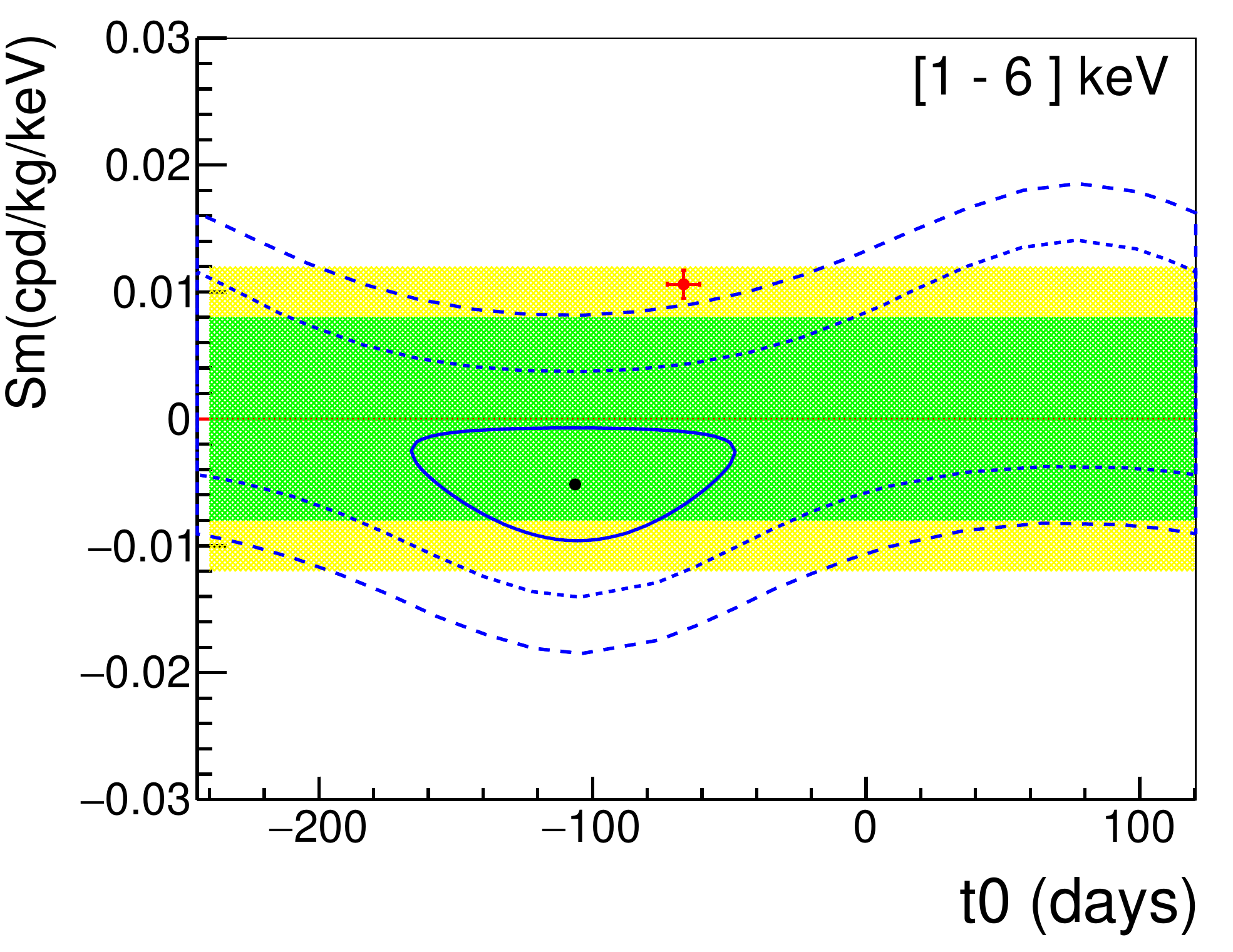}
\includegraphics[width=.32\textwidth]{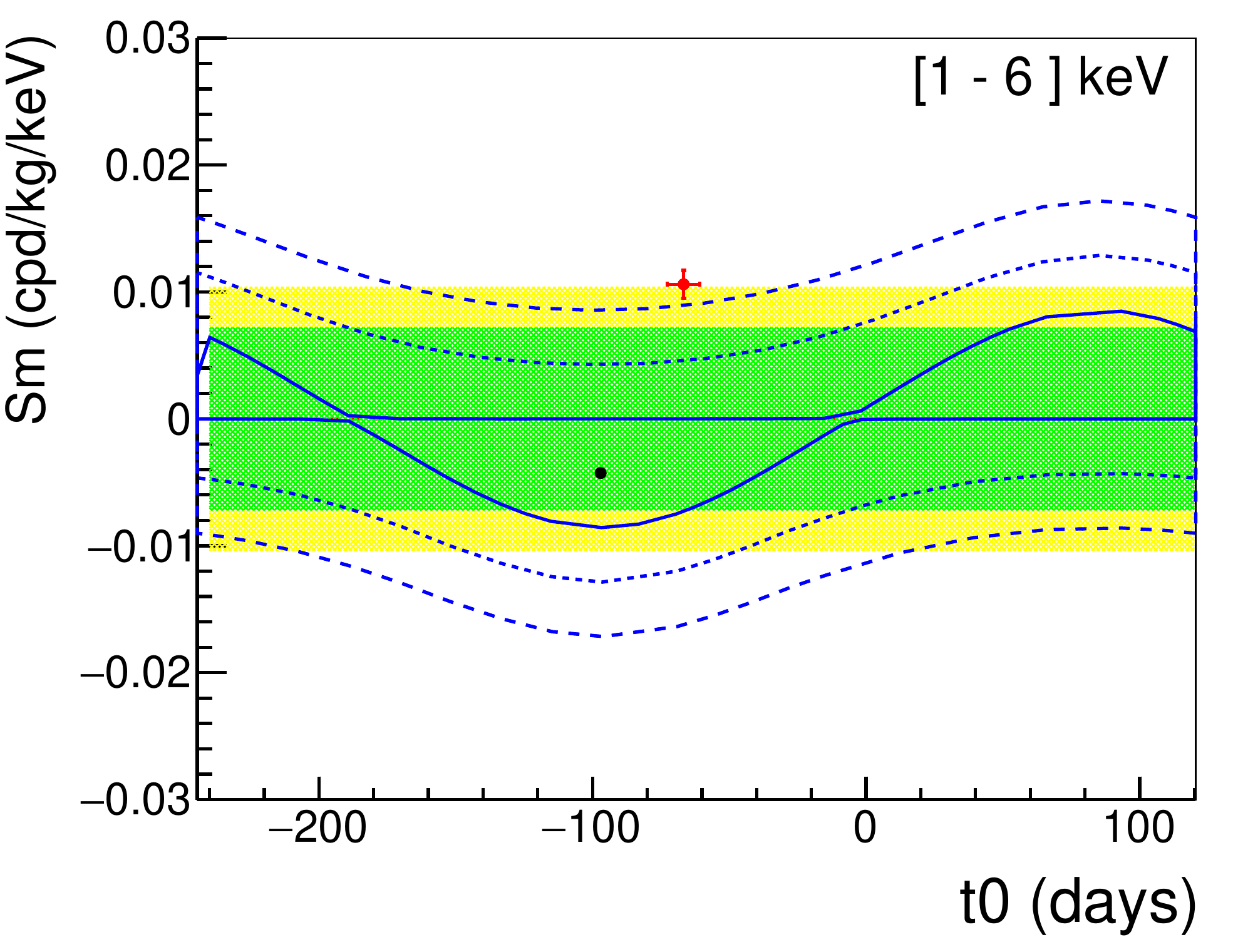}
\includegraphics[width=.32\textwidth]{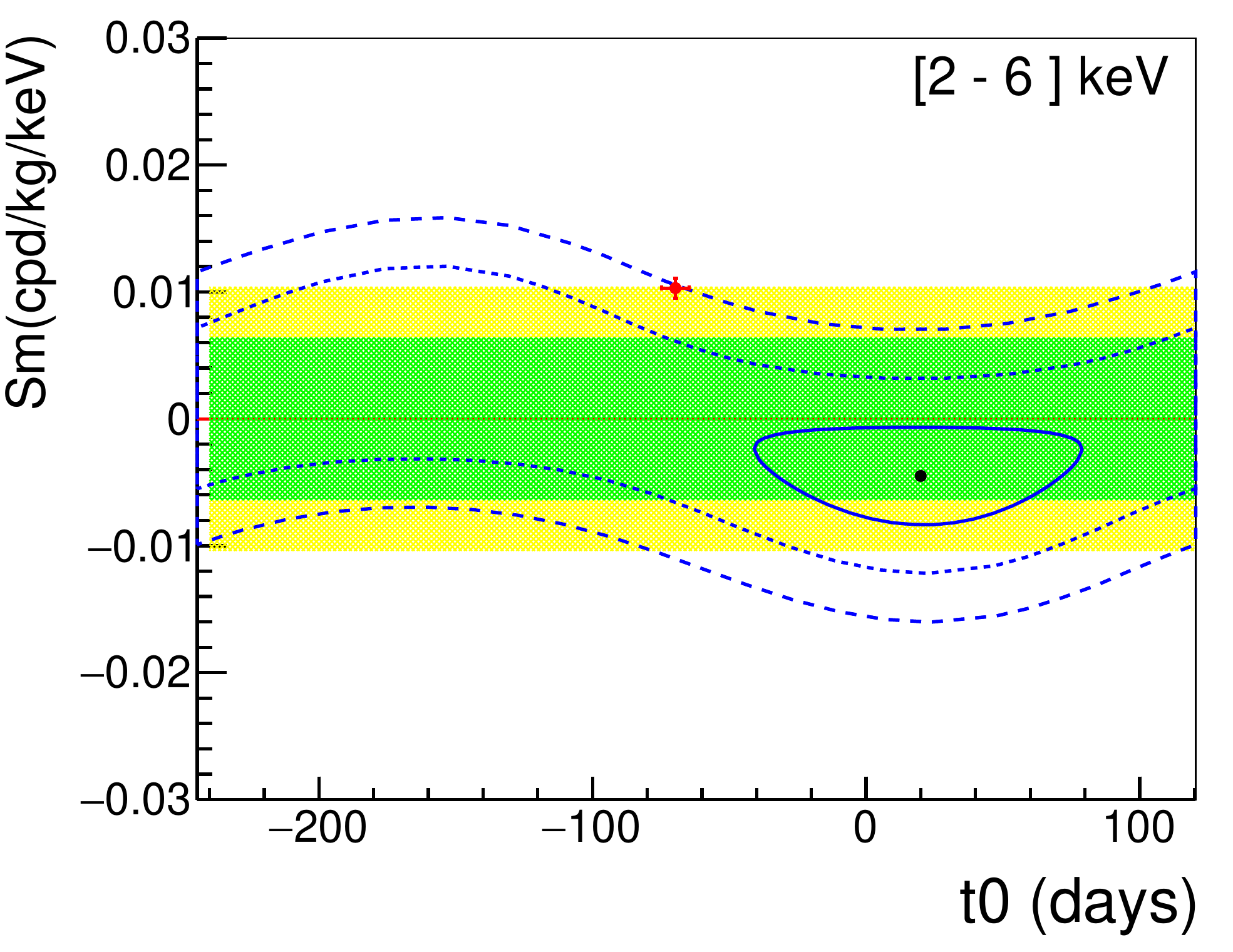}
\includegraphics[width=.32\textwidth]{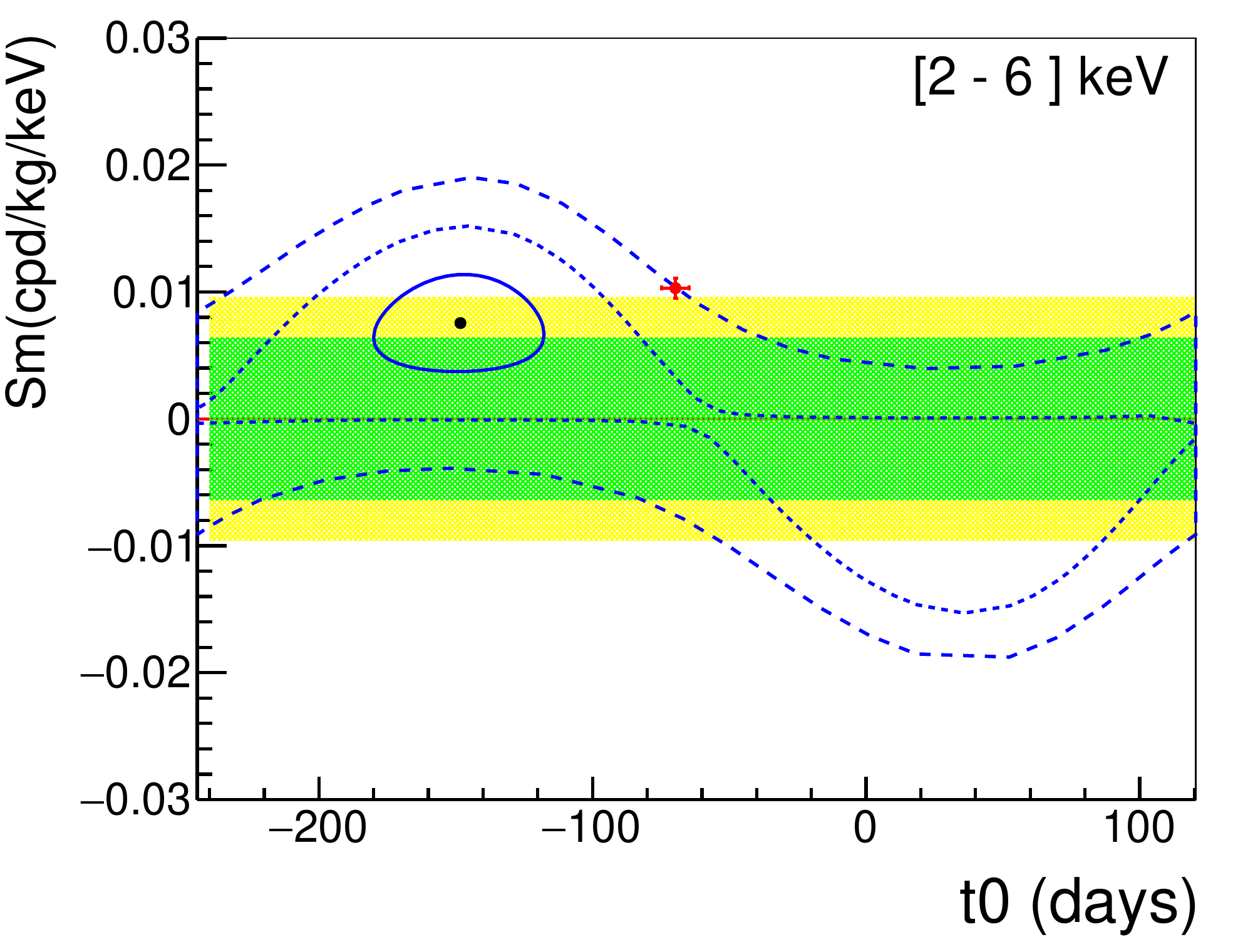}
\includegraphics[width=.32\textwidth]{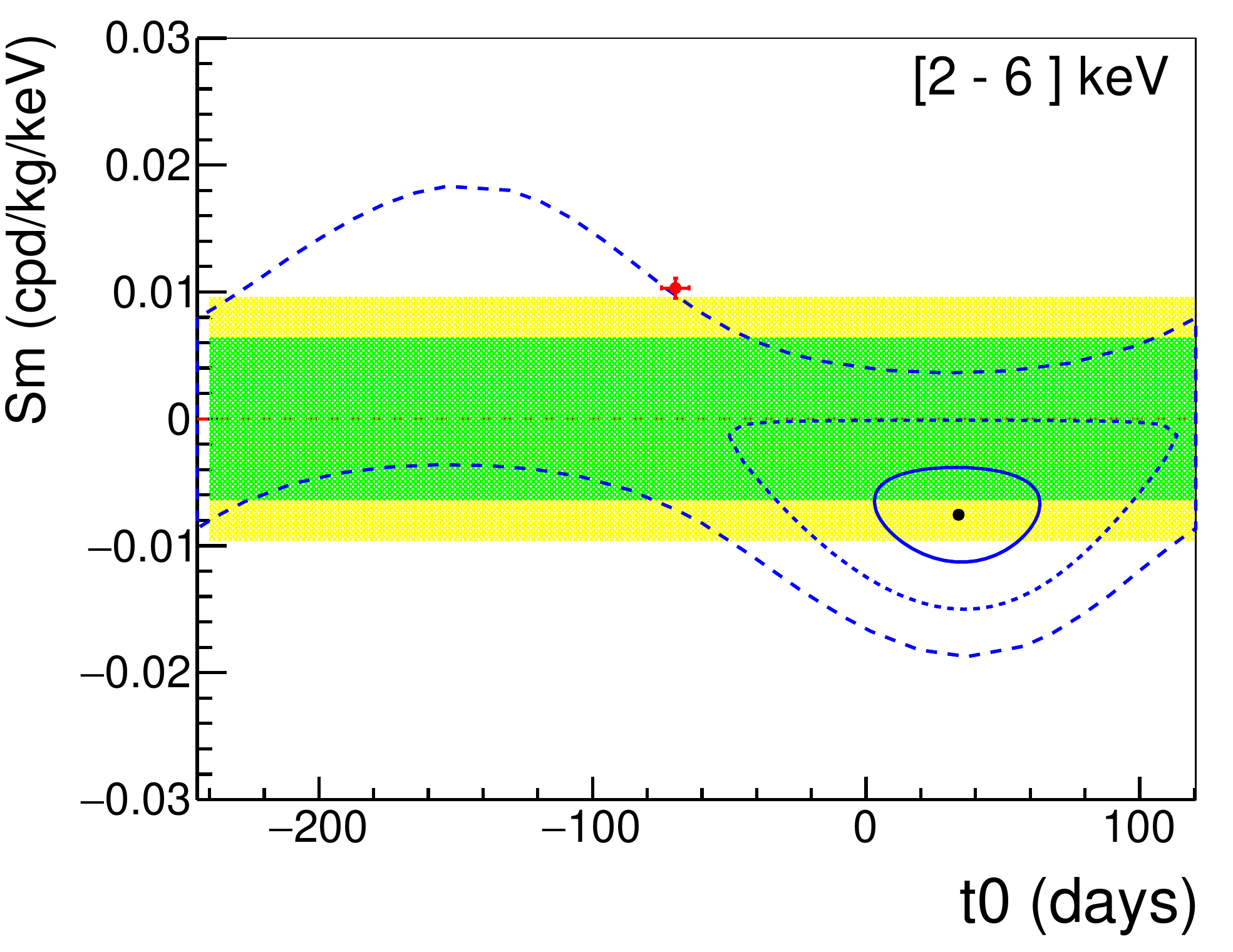}
\caption{\label{fig:fit2d} Black points: best fit results for three years of data in [1--6]~keV (upper panels) and [2--6]~keV (lower panels) energy regions in the ($S_m$,$t_0$) plane for the three fitting procedures (left, middle and right panels correspond to best fits to functions~\ref{eq:fit1},~\ref{eq:fit2}, and \ref{eq:fit3}, respectively). Exclusion contours at 1, 2 and 3$\sigma$ are depicted as blue solid, dotted and dashed lines. DAMA/LIBRA results are shown for comparison (red cross). 1 and 2~$\sigma$ biased contours extracted from the MC simulation for each fit are shown as solid green and yellow regions.}
\end{figure*}

\begin{figure}[htbp]
\centering 
\includegraphics[width=.48\textwidth]{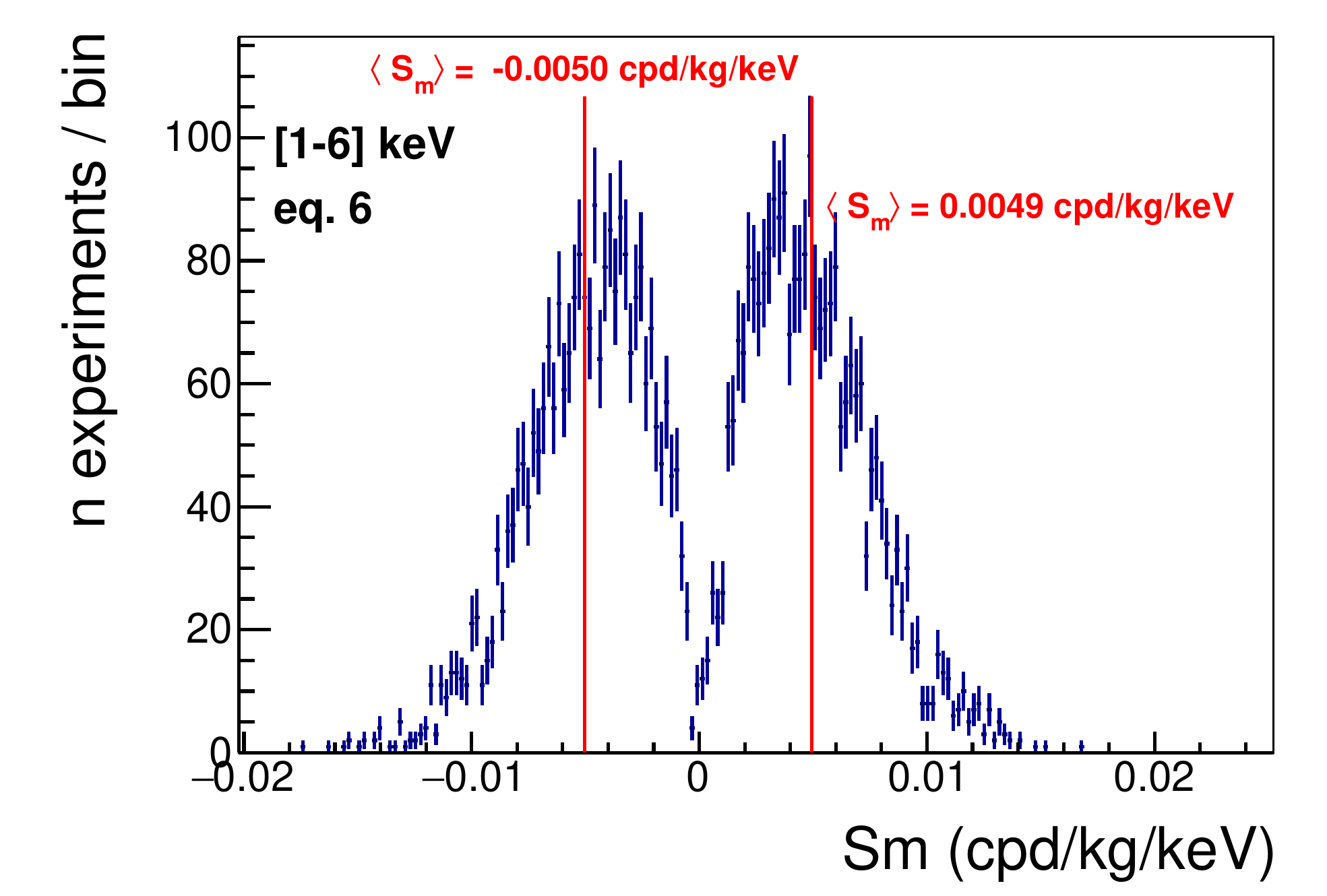}
\caption{ Distribution of the best fit modulation amplitudes for a large set of MC experiments generated for [1--6]~keV with $S_m$=0 and fitted to eq.~\ref{eq:fit3} leaving free both, $S_m$ and $t_0$. In red, mean value of every lobe of the bimodal distribution.}
\label{fig:bias}
\end{figure}

We extend the analysis presented in previous sections by taking $t_0$ as a free parameter. The best fits are presented in Fig.~\ref{fig:fit2d} for the three fitting procedures (left, middle and right panels correspond to  Eqs.~\ref{eq:fit1},~\ref{eq:fit2}, and \ref{eq:fit3}, respectively) in the [1\nobreakdash--6] keV (upper panels) and [2\nobreakdash--6]~keV (lower panels) regions. The exclusion contours at 1, 2 and 3$\sigma$ are depicted as blue solid, dotted and dashed lines.
In all cases the best fits are $\sim$3$\sigma$ away from the DAMA/LIBRA result, but it has to be highlighted that in this case the fit is biased, as it would be expected due to the nonlinearity of the model~\cite{2009Meas...42..748A}. In the Appendix we calculate the expected bias in the absence of modulation to be $\sqrt{\frac{\pi}{2}}\sigma(S_m)$, where 
$\sigma(S_m)$ is the standard deviation of the modulation amplitude for a fixed phase. We check this result by simulating a large set of Monte Carlo pseudoexperiments without modulation and fitting them with unconstrained $S_m$ and $t_0$. We present in Fig.~\ref{fig:bias}, as an example, the result for the [1--6]~keV energy region and the model given by eq.~\ref{eq:fit3} (the results for the other cases are very similar). When positive and negative values are allowed for $S_m$, a bimodal distribution is obtained, where the mean value of each lobe is $S_m=\pm$0.0050~cpd/kg/keV, which is in agreement with $\sqrt{\frac{\pi}{2}}\sigma(S_m)$ for $\sigma(S_m)$=0.0042~cpd/kg/keV (see the Appendix).   
Depending on the available statistics, the identified bias can be large and it requires, in our opinion, the revision of the
results from different dark matter experiments looking for dark matter modulation with an unconstrained phase~\cite{Akerib:2018zoq,Ahmed:2012vq,Fox:2011px, Yang:2019lao,Aprile:2015ibr,PhysRevLett.123.031302}.
We depict in Fig.~\ref{fig:fit2d} the 1 and 2$\sigma$ bands extracted from the MC simulation for each fit as solid green and yellow contours. They represent the biased contours expected from the corresponding fit. Correcting the best fits with the calculated bias, the results are compatible with no modulation in all cases.
\subsection{Frequency analysis}
\label{sec:freq}
\begin{figure*}[htbp]
\centering %
\includegraphics[width=.32\textwidth]{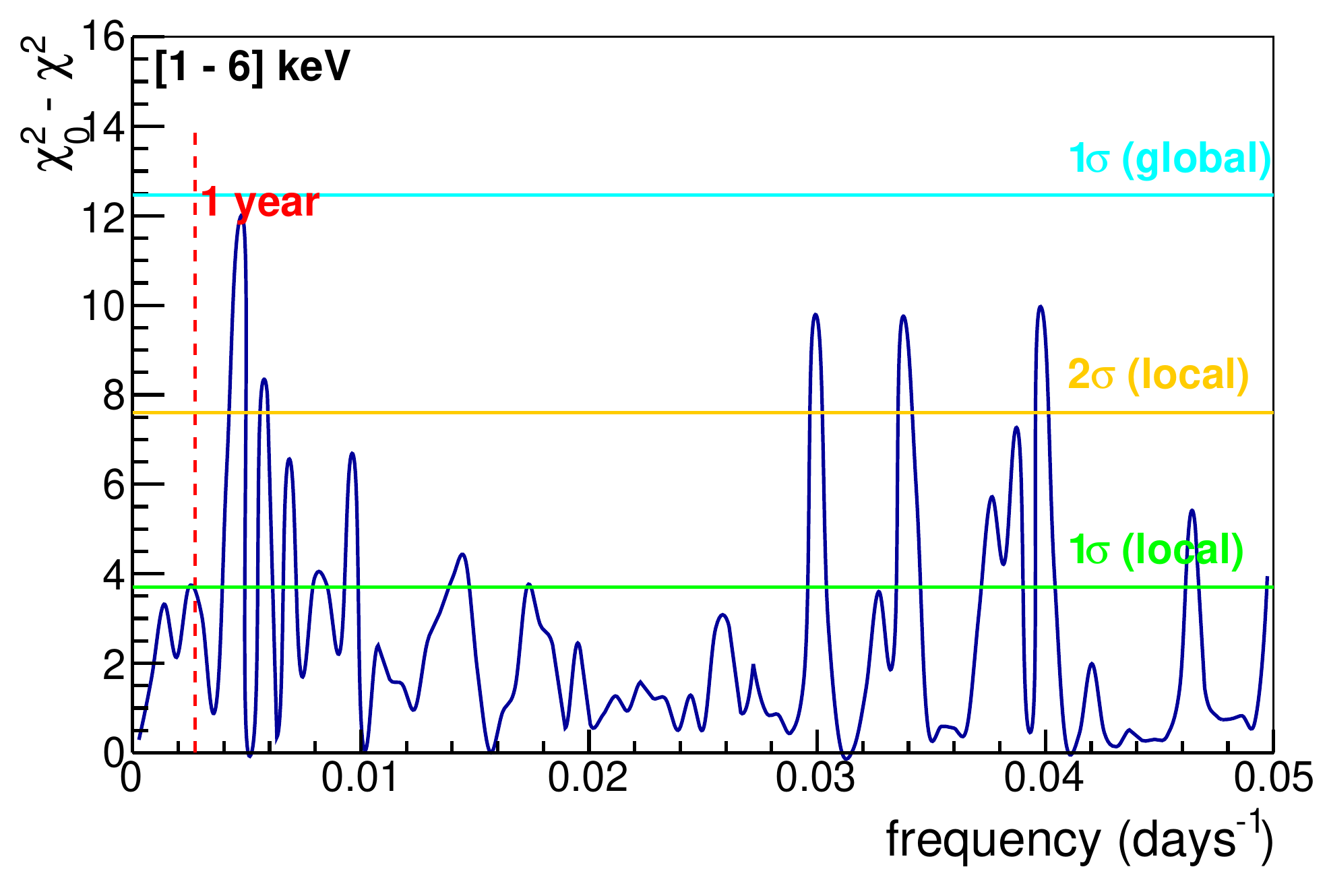}
\includegraphics[width=.32\textwidth]{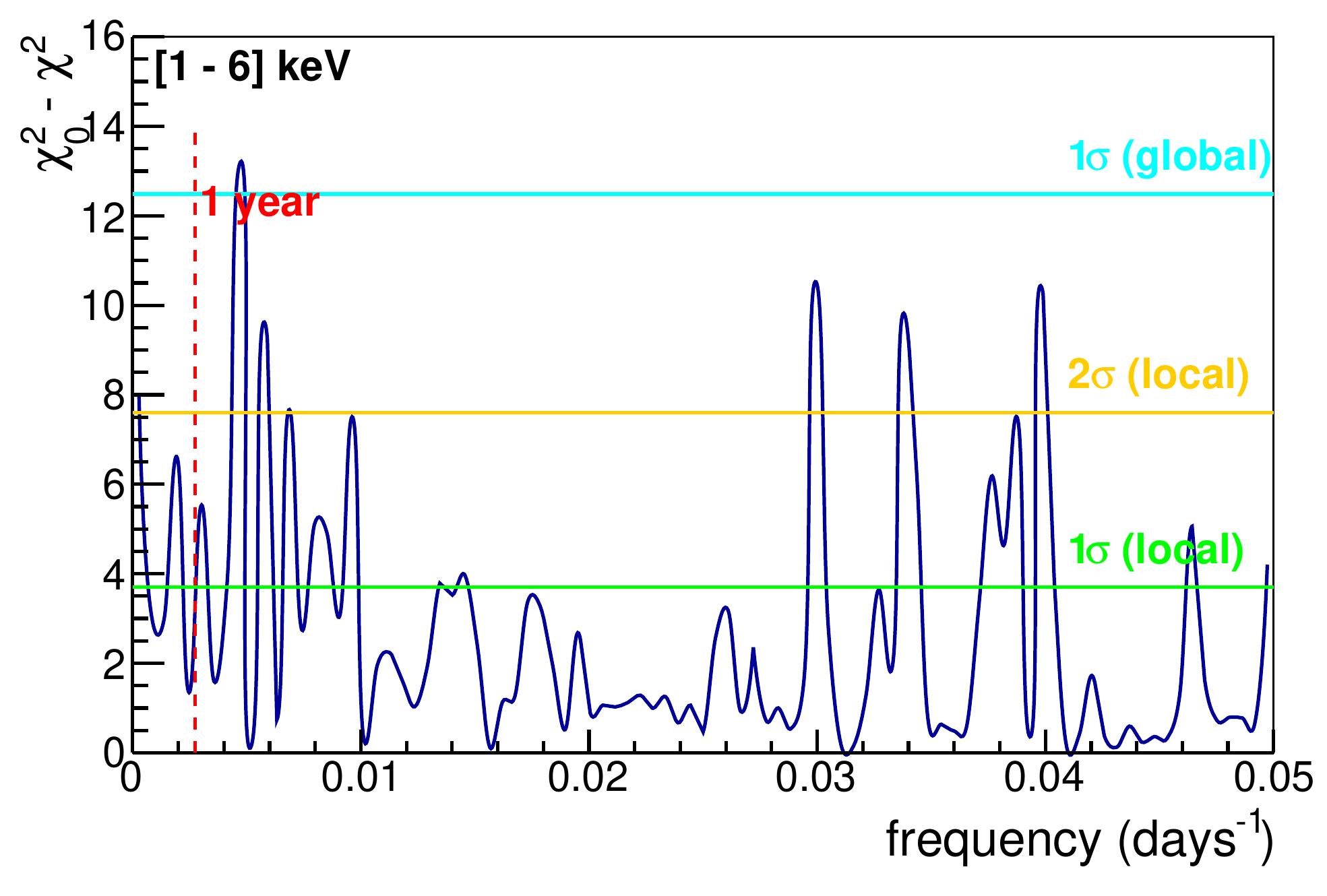}
\includegraphics[width=.32\textwidth]{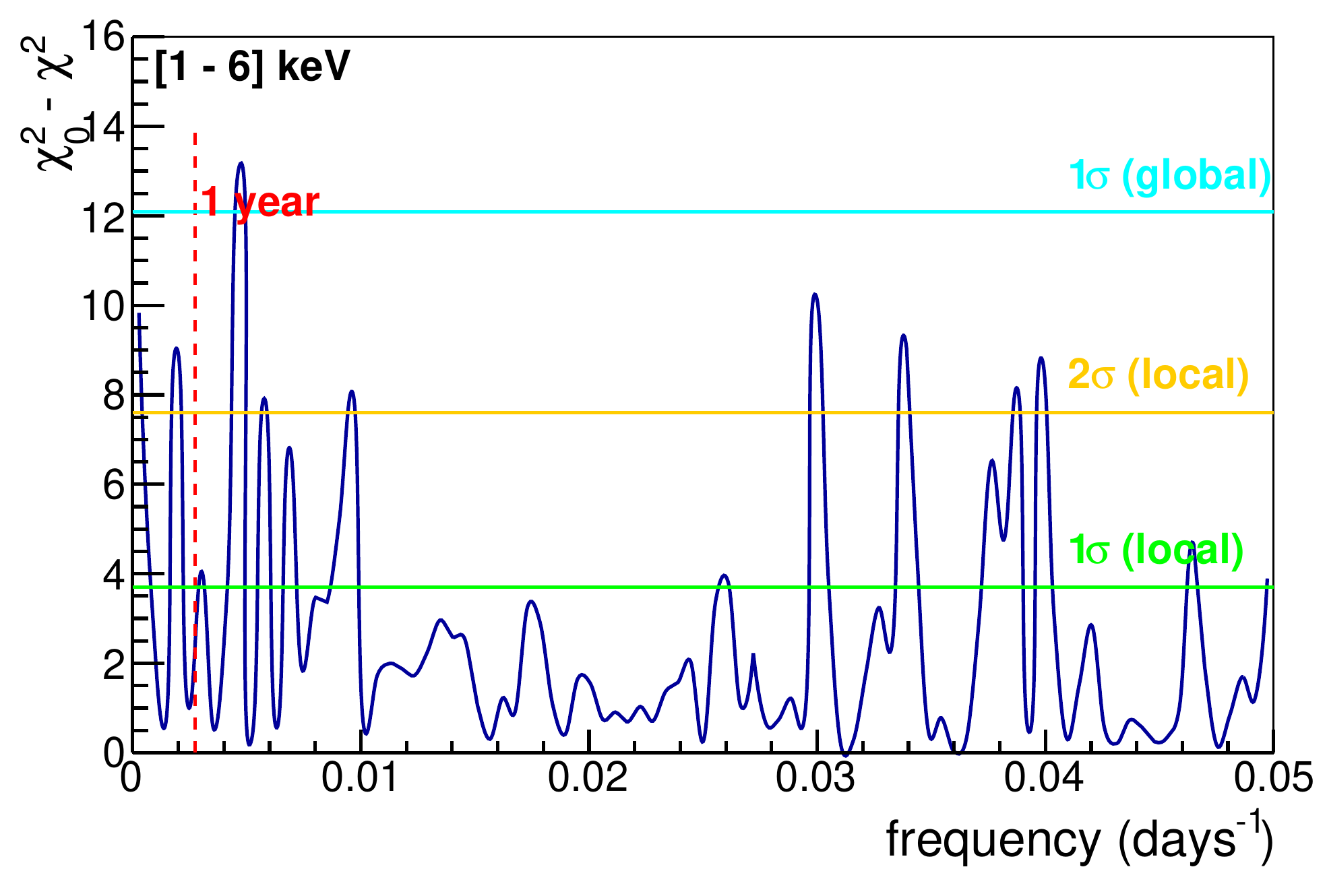}
\includegraphics[width=.32\textwidth]{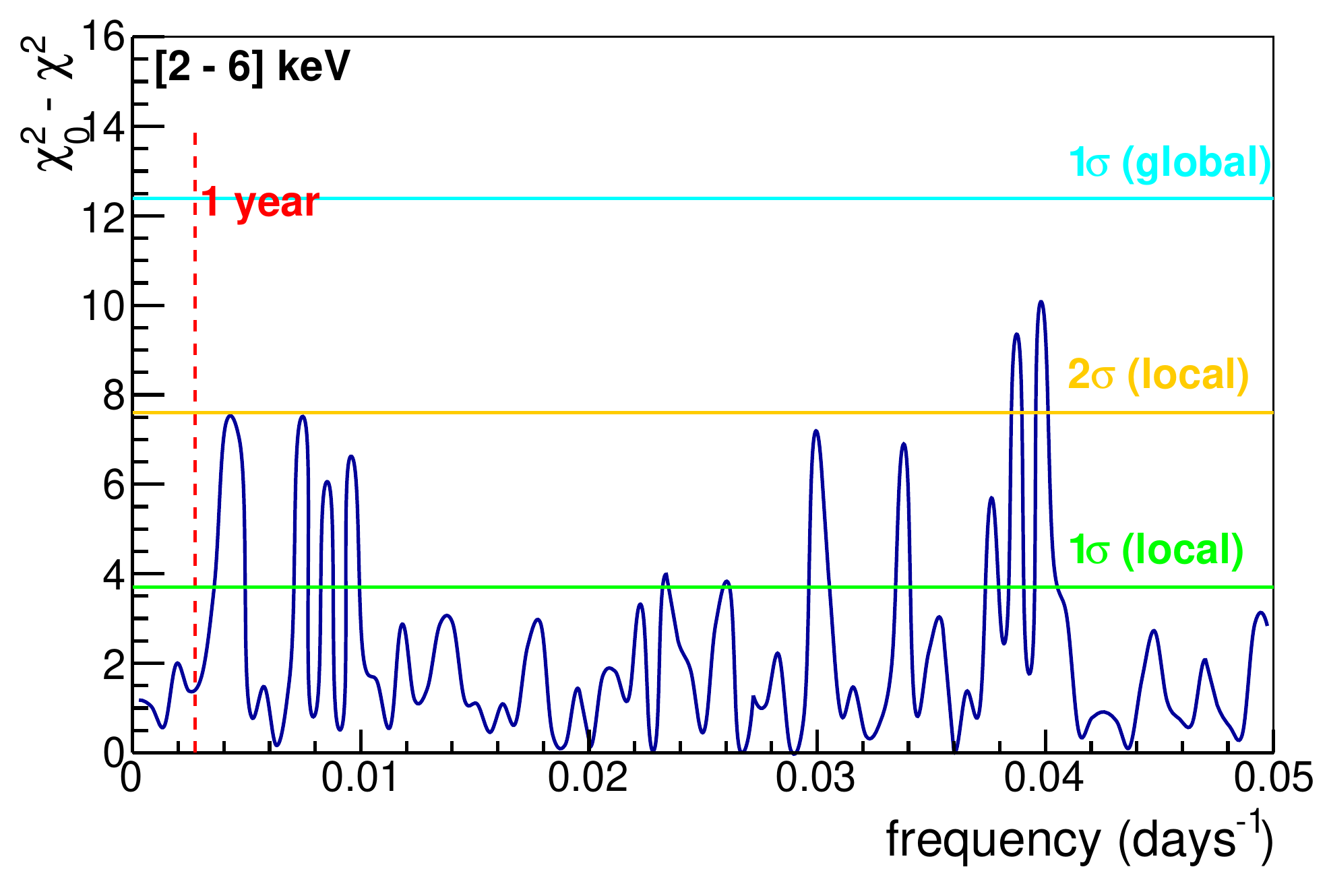}
\includegraphics[width=.32\textwidth]{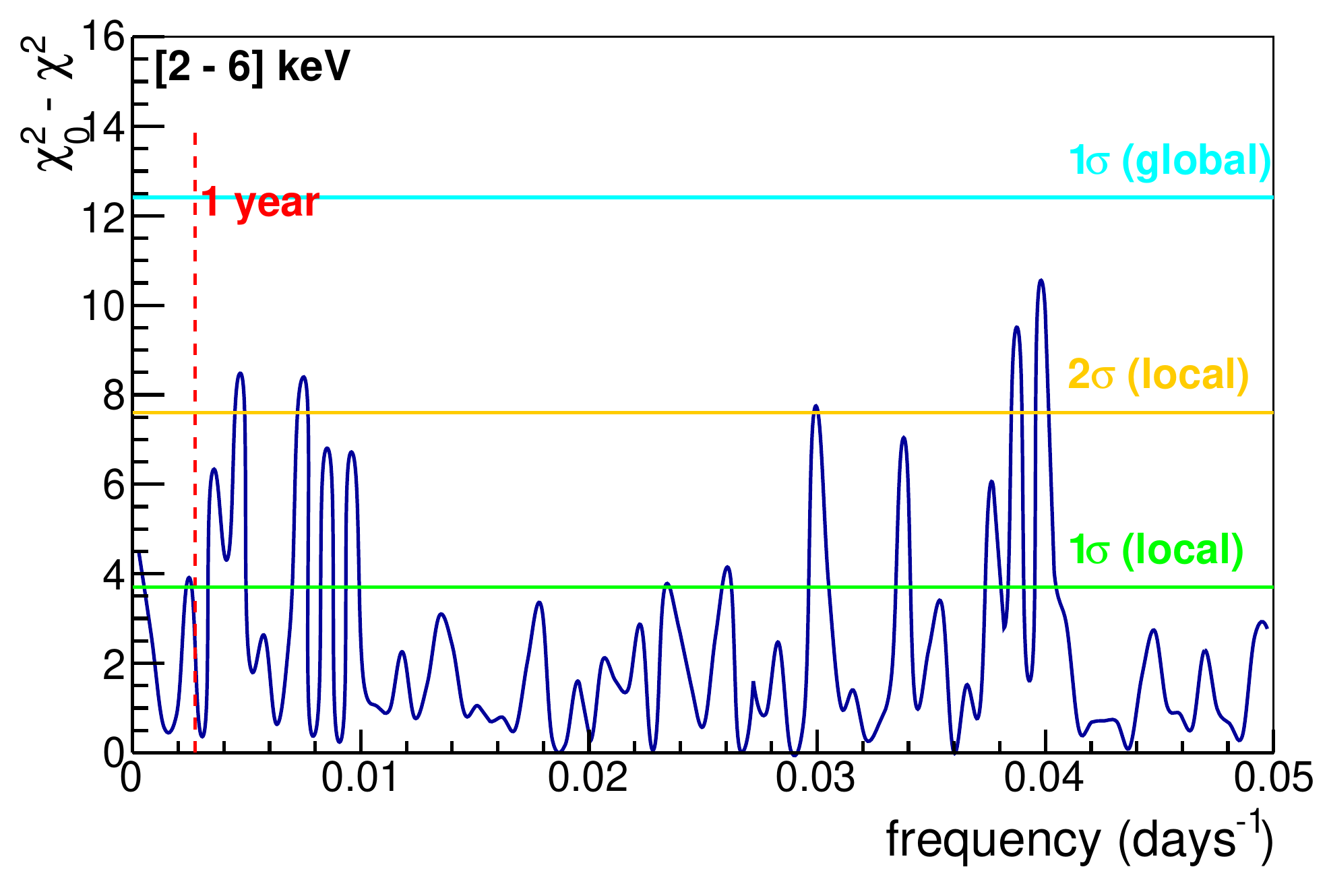}
\includegraphics[width=.32\textwidth]{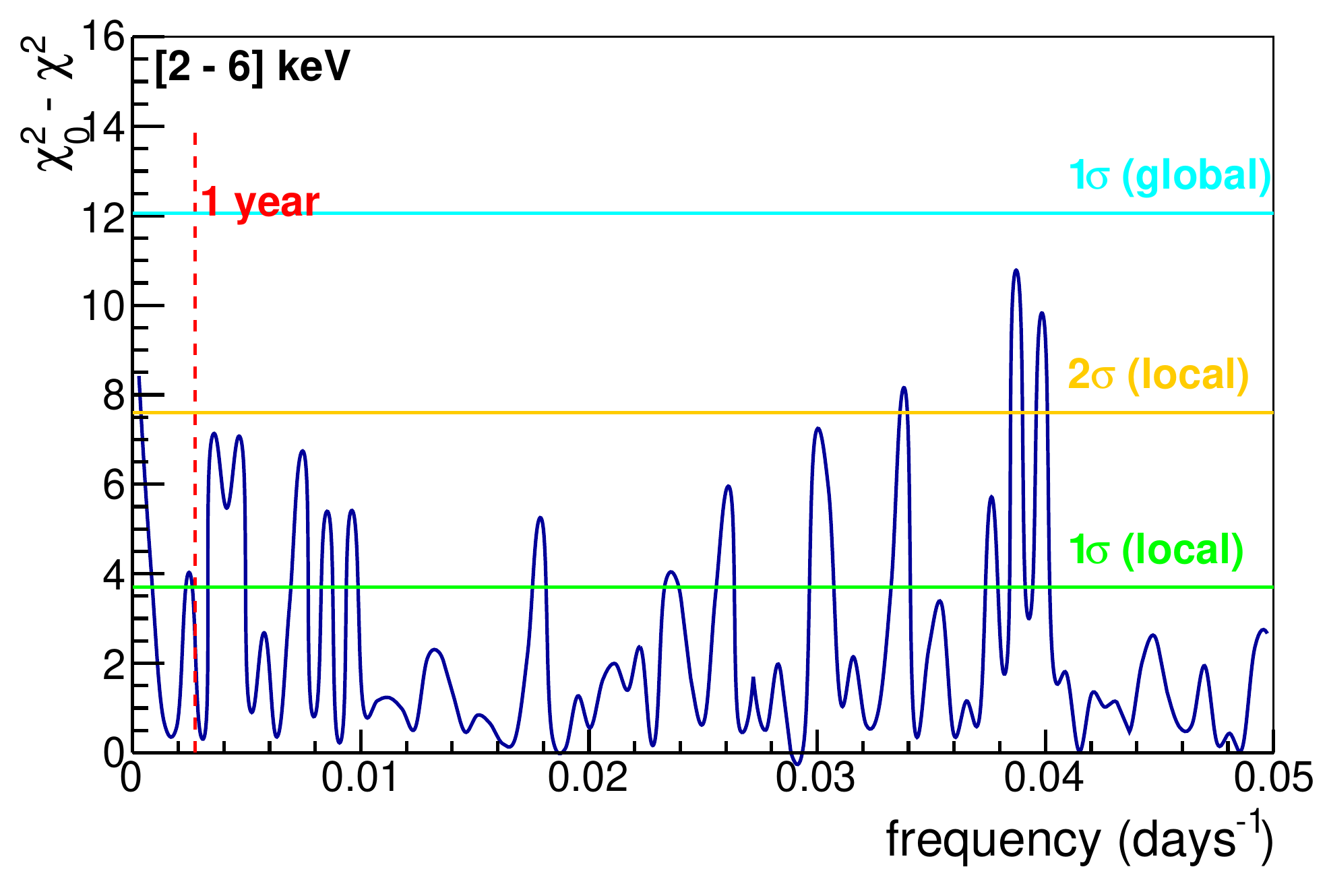}
\caption{$\chi^2_0-\chi^2$ periodograms for three years of data in [1--6]~keV (upper panels) and [2--6]~keV (lower panels) energy regions. Left, middle and right panels correspond to best fits to functions~\ref{eq:fit1},~\ref{eq:fit2}, and \ref{eq:fit3}, respectively. Local and global significances are calculated as described in the text.}
\label{fig:periodogram}
\end{figure*}

We search for the presence of a periodic signal 
in our data using the least-square periodogram, which is equivalent to the
Lomb-Scargle technique~\cite{2018ApJS..236...16V}. 
We perform a scan in frequency from 0 to the Nyquist frequency, which corresponds to 0.05~days$^{-1}$ for 10-days bins. For every frequency we fit the data to the null and the modulation hypothesis, where $S_m$ and $t_0$ are free in the latter. Our test statistics is the difference in  $\chi^2$ between the null and the modulation hypothesis, $\chi_0^2 - \chi^2$. 
Figure~\ref{fig:periodogram} shows the periodograms obtained for the six studied scenarios.  
With MC simulations we have checked that our test statistic is distributed as a $\chi^2$ with 2 degrees of freedom, so the local significance at 1$\sigma$ and 2$\sigma$ can be easily calculated using the probability distribution $P(\chi^2, 2)$ where we find 3.7 and 7.6, respectively. These values are  displayed in the figure as green and yellow lines. Nevertheless, as the search is performed across a continuous range in frequency, we need to take into account the probability of the signal to occur anywhere within the search range ("look elsewhere effect"). Following~\cite{Lista:2016chp}, we calculate the global significance 
from the distribution of the maximum of the local test statistics in the whole search range for  a large number of MC experiments. The 1$\sigma$ level is represented in the figure as a cyan line.
Several peaks are present in the periodograms but none of them is statistically significant. We can conclude there are no statistically significant modulation in the frequency range analyzed in the ANAIS-112 data. 
\section{Sensitivity projection}
\label{sec:sensitivity}
The statistical significance of our result is determined by the standard deviation of the modulation amplitude distribution, $\sigma(S_m)$, which would be obtained in a large number of experiments like ANAIS-112 for a given exposure. Then, we quote our sensitivity to DAMA/LIBRA result as the ratio $S_m^{DAMA}/\sigma(S_m)$, which directly gives in $\sigma$ units the C.L. at which we can test the DAMA/LIBRA signal.
At present, our result  $\sigma(S_m)=0.0042~(0.0037)$~cpd/kg/keV for [1--6]~keV
([2--6]~keV) corresponds to a sensitivity of 2.5~$\sigma$~(2.7~$\sigma$) with respect to the DAMA/LIBRA signal. 
\par
\begin{figure}[htbp]
\centering 
\includegraphics[width=.48\textwidth]{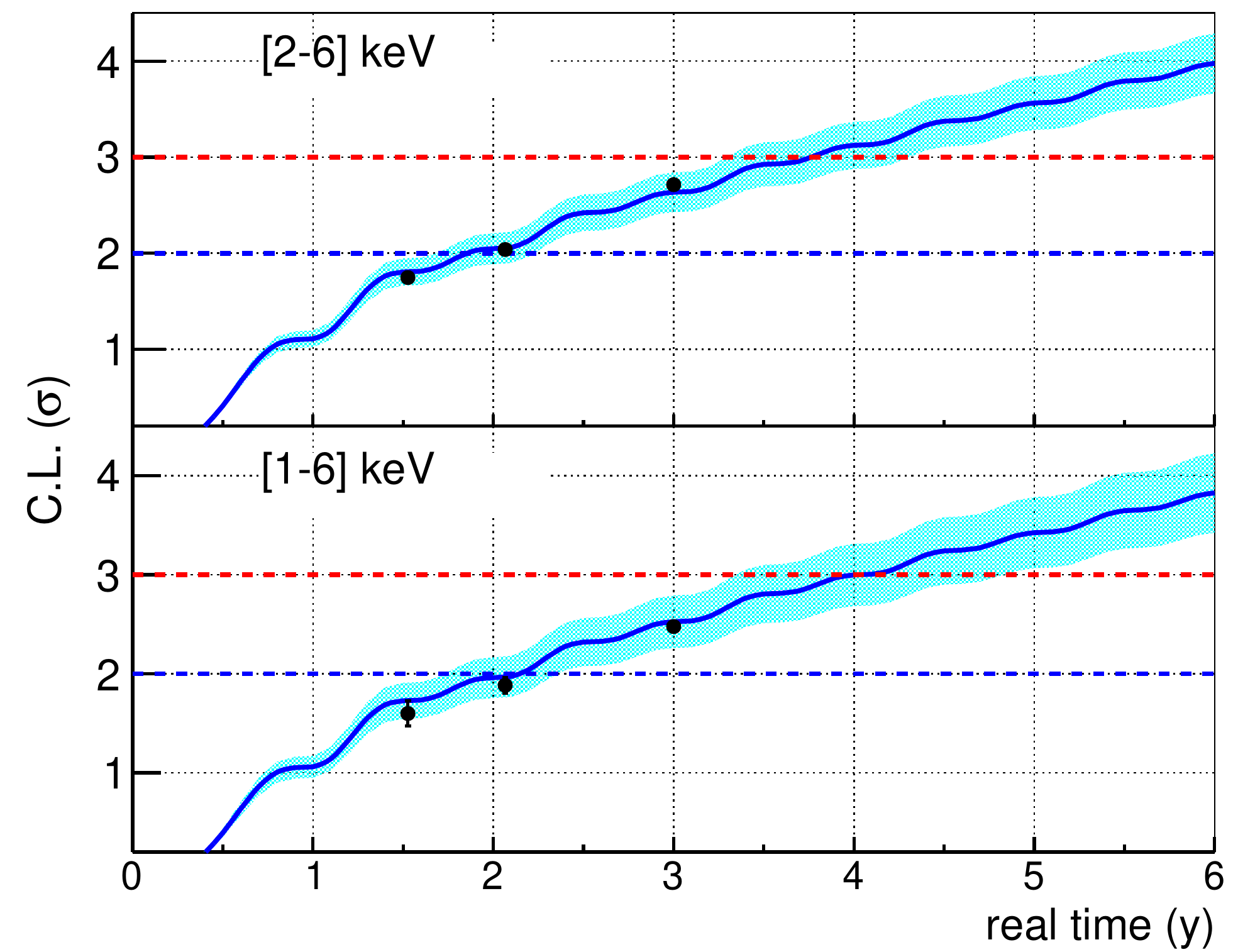}
\caption{\label{fig:sensitivity} 
ANAIS-112 sensitivity to the DAMA/LIBRA signal in $\sigma$ C.L. units (see text) as a function of real time in the [1--6]~keV (lower panel) and [2--6]~keV (upper panel) energy regions. The black dots are the sensitivities measured experimentally. The blue bands represent the 68\% C.L. DAMA/LIBRA uncertainty.
}
\end{figure}

Figure~\ref{fig:sensitivity} (dark blue lines) displays our sensitivity projection calculated following Ref.~\cite{Coarasa:2018qzs} for the two studied energy ranges, whereas the blue bands represent the 68\% uncertainty in $S_m^{DAMA}$ as reported in Ref.~\cite{Bernabei:2018yyw}. In the calculation, we take into account the ANAIS-112 live time distribution, the background reduction expected due to decaying isotopes (according to our background model) and the statistical error in the detection efficiency. 
The black dots are the sensitivities derived in this and the previous analysis~\cite{Amare:2019jul,Amare:2019ncj}. The results perfectly agree with our estimates, where the 3-years point is slightly above the expected sensitivity because the curve is calculated for the model that combines the data of the nine detectors before the fit, as we did for the 1.5 and 2 years results. 
In conclusion, our data confirm the ANAIS-112 projected sensitivity to the DAMA/LIBRA result. A 3$\sigma$ sensitivity should be in reach before completing the scheduled 5~years of data taking. In order to reach 5$\sigma$ sensitivity a much longer measurement time is required, 10~years in total. 
\section{Conclusions}
\label{sec:conclusions}
We have unblinded three years of ANAIS-112 data, updating and completing the annual modulation analysis to test the DAMA/LIBRA result. Our results confirm previous analyses, and the statistical significance increases according to our sensitivity estimates, supporting our prospects of reaching 3$\sigma$ within the scheduled 5-years operation (see Fig.~\ref{fig:sensitivity}). 
\par
We obtain for the best fit a modulation amplitude of --0.0034$\pm$0.0042 (0.0003$\pm$0.0037) cpd/keV/kg in the [1\nobreakdash--6]~keV ([2\nobreakdash--6]~keV) energy region, supporting the absence of modulation in our data, and being incompatible with the DAMA/LIBRA result at 3.3 (2.6)~$\sigma$, with a sensitivity of 2.5 (2.7)~$\sigma$. We have developed consistency checks for all of our analysis procedures and, using MC pseudoexperiments, shown that they are not biased. 
The possible systematic effect on the standard deviation of the modulation amplitude, and therefore on the experimental sensitivity, due to background modeling, time binning, and considered modules or periods of data taking has been studied, supporting that the corresponding uncertainty is much lower than the evaluated statistical uncertainty.
\par
We have also carried out two analyses with free phase or free frequency, which are both compatible with the absence of modulation in our data. Although the former has shown to be biased, the best fits corrected with the calculated bias are compatible with no modulation.
\par
The ANAIS-112 experiment will be able to provide 
3$\sigma$ C.L. test on DAMA/LIBRA annual modulation signal 
by autumn 2022, according to our sensitivity estimates, which are confirmed with the results presented here. 
\acknowledgments
This work has been financially supported by the Spanish Ministerio de Econom\'{\i}a y Competitividad and the European Regional Development Fund (MINECO-FEDER) under Grant No. FPA2017-83133-P, the Ministerio de Ciencia e
Innovaci\'on -- Agencia Estatal de Investigaci\'on under Grant No. PID2019-104374GB-I00, the Consolider-Ingenio 2010 Programme under Grants No. MultiDark CSD2009-
00064 and No. CPAN CSD2007-00042, the LSC Consortium, and the Gobierno de Arag\'on and the European Social Fund (Group in Nuclear and Astroparticle Physics and I. Coarasa predoctoral grant). We thank the support of the Spanish Red Consolider MultiDark FPA2017-90566-REDC.
The authors would like to acknowledge the use of Servicio General de Apoyo a la Investigaci\'on-SAI, Universidad de Zaragoza, and technical support from LSC and GIFNA staff. The authors are very thankful to Sven Heinemeyer for proof reading.
\appendix
\section{}
\label{sec:app}
Under the hypothesis of an annual modulation component with amplitude $S_m$ and phase $\phi$, the rate in the ROI can be written as
\begin{equation}
R(t)=R_0+S_m cos(\omega(t-t_0))=R_0 + A\cos\omega t + B\sin\omega t
\label{eq:app1}
\end{equation}
where $\phi=\omega t_0$, $A=S_m \cos\phi$, and $B=S_m \sin\phi$. Note that $(A,B)$ can be viewed as Cartesian coordinates and $(S_m,\phi)$ as their corresponding polar coordinates.
The least squares estimator (LSE) for the parameters A and B is unbiased because, as shown in eq.~\ref{eq:app1}, the rate is linear in both parameters, while this is not the case for the LSE for $S_m$ and $\phi$~\cite{Eadie_1971}.
This can be proven in a similar way as described in our previous article \cite{Coarasa:2018qzs}. When the data are evenly distributed along an integer number of years, the covariance matrix is diagonal, and the estimators of A and B have the same variance, $\sigma^2$. Furthermore, when no modulation is present, the estimators of A and B are independent Gaussian distributions with a null mean and variance $\sigma^2$. Under these assumptions, the bias of the LSE of $S_m$ can be calculated because the joint probability of $(A,B)$
\begin{multline}
    P(A,B) dA dB= \\
    \frac{1}{\sqrt{2\pi}\sigma}exp(-A^2/2\sigma^2)\frac{1}{\sqrt{2\pi}\sigma}exp(-B^2/2\sigma^2) dA dB
\end{multline}
can be written in polar coordinates $(S_m,\phi)$
\begin{equation}
    P(S_m,\phi) dS_m d\phi=\frac{1}{2\pi\sigma^2}exp(-S_m^2/2\sigma^2) S_m dS_m d\phi
\end{equation}
The expectation value of $S_m=+\sqrt{A^2+B^2}$ is therefore,
\begin{multline}
    E(S_m)=\int_{-\infty}^\infty\int_{-\infty}^\infty S_m P(A,B) dA dB = \\ \int_0^\infty\int_0^{2\pi} S_m P(S_m,\phi) S_m dS_m d\phi = \sqrt{\frac{\pi}{2}}\sigma.
\end{multline}

\begin{thebibliography}{63}%
\makeatletter
\providecommand \@ifxundefined [1]{%
 \@ifx{#1\undefined}
}%
\providecommand \@ifnum [1]{%
 \ifnum #1\expandafter \@firstoftwo
 \else \expandafter \@secondoftwo
 \fi
}%
\providecommand \@ifx [1]{%
 \ifx #1\expandafter \@firstoftwo
 \else \expandafter \@secondoftwo
 \fi
}%
\providecommand \natexlab [1]{#1}%
\providecommand \enquote  [1]{``#1''}%
\providecommand \bibnamefont  [1]{#1}%
\providecommand \bibfnamefont [1]{#1}%
\providecommand \citenamefont [1]{#1}%
\providecommand \href@noop [0]{\@secondoftwo}%
\providecommand \href [0]{\begingroup \@sanitize@url \@href}%
\providecommand \@href[1]{\@@startlink{#1}\@@href}%
\providecommand \@@href[1]{\endgroup#1\@@endlink}%
\providecommand \@sanitize@url [0]{\catcode `\\12\catcode `\$12\catcode
  `\&12\catcode `\#12\catcode `\^12\catcode `\_12\catcode `\%12\relax}%
\providecommand \@@startlink[1]{}%
\providecommand \@@endlink[0]{}%
\providecommand \url  [0]{\begingroup\@sanitize@url \@url }%
\providecommand \@url [1]{\endgroup\@href {#1}{\urlprefix }}%
\providecommand \urlprefix  [0]{URL }%
\providecommand \Eprint [0]{\href }%
\providecommand \doibase [0]{http://dx.doi.org/}%
\providecommand \selectlanguage [0]{\@gobble}%
\providecommand \bibinfo  [0]{\@secondoftwo}%
\providecommand \bibfield  [0]{\@secondoftwo}%
\providecommand \translation [1]{[#1]}%
\providecommand \BibitemOpen [0]{}%
\providecommand \bibitemStop [0]{}%
\providecommand \bibitemNoStop [0]{.\EOS\space}%
\providecommand \EOS [0]{\spacefactor3000\relax}%
\providecommand \BibitemShut  [1]{\csname bibitem#1\endcsname}%
\let\auto@bib@innerbib\@empty
\bibitem [{\citenamefont {Tanabashi}\ \emph {et~al.}(2018)\citenamefont
  {Tanabashi} \emph {et~al.}}]{Tanabashi:2018oca}%
  \BibitemOpen
  \bibfield  {author} {\bibinfo {author} {\bibfnamefont {M.}~\bibnamefont
  {Tanabashi}} \emph {et~al.} (\bibinfo {collaboration} {Particle Data
  Group}),\ }\bibfield  {title} {\enquote {\bibinfo {title} {{Review of
  Particle Physics}},}\ }\href {\doibase 10.1103/PhysRevD.98.030001} {\bibfield
   {journal} {\bibinfo  {journal} {Phys. Rev. D}\ }\textbf {\bibinfo {volume}
  {98}},\ \bibinfo {pages} {030001} (\bibinfo {year} {2018})}\BibitemShut
  {NoStop}%
\bibitem [{\citenamefont {Bertone}\ and\ \citenamefont
  {Hooper}(2018)}]{Bertone:2016nfn}%
  \BibitemOpen
  \bibfield  {author} {\bibinfo {author} {\bibfnamefont {Gianfranco}\
  \bibnamefont {Bertone}}\ and\ \bibinfo {author} {\bibfnamefont {Dan}\
  \bibnamefont {Hooper}},\ }\bibfield  {title} {\enquote {\bibinfo {title}
  {{History of dark matter}},}\ }\href {\doibase 10.1103/RevModPhys.90.045002}
  {\bibfield  {journal} {\bibinfo  {journal} {Rev. Mod. Phys.}\ }\textbf
  {\bibinfo {volume} {90}},\ \bibinfo {pages} {045002} (\bibinfo {year}
  {2018})},\ \Eprint {http://arxiv.org/abs/1605.04909} {arXiv:1605.04909
  [astro-ph.CO]} \BibitemShut {NoStop}%
\bibitem [{\citenamefont {Battaglieri}\ \emph {et~al.}(2017)\citenamefont
  {Battaglieri} \emph {et~al.}}]{Battaglieri:2017aum}%
  \BibitemOpen
  \bibfield  {author} {\bibinfo {author} {\bibfnamefont {Marco}\ \bibnamefont
  {Battaglieri}} \emph {et~al.},\ }\bibfield  {title} {\enquote {\bibinfo
  {title} {{US Cosmic Visions: New Ideas in Dark Matter 2017: Community
  Report}},}\ }in\ \href@noop {} {\emph {\bibinfo {booktitle} {{U.S. Cosmic
  Visions: New Ideas in Dark Matter}}}}\ (\bibinfo {year} {2017})\ \Eprint
  {http://arxiv.org/abs/1707.04591} {arXiv:1707.04591 [hep-ph]} \BibitemShut
  {NoStop}%
\bibitem [{\citenamefont {Zyla}\ \emph {et~al.}(2020)\citenamefont {Zyla} \emph
  {et~al.}}]{Zyla:2020zbs}%
  \BibitemOpen
  \bibfield  {author} {\bibinfo {author} {\bibfnamefont {P.A.}\ \bibnamefont
  {Zyla}} \emph {et~al.} (\bibinfo {collaboration} {Particle Data Group}),\
  }\bibfield  {title} {\enquote {\bibinfo {title} {{Review of Particle
  Physics}},}\ }\href {\doibase 10.1093/ptep/ptaa104} {\bibfield  {journal}
  {\bibinfo  {journal} {PTEP}\ }\textbf {\bibinfo {volume} {2020}},\ \bibinfo
  {pages} {083C01} (\bibinfo {year} {2020})}\BibitemShut {NoStop}%
\bibitem [{\citenamefont {Marrodán~Undagoitia}\ and\ \citenamefont
  {Rauch}(2016)}]{Undagoitia:2015gya}%
  \BibitemOpen
  \bibfield  {author} {\bibinfo {author} {\bibfnamefont {Teresa}\ \bibnamefont
  {Marrodán~Undagoitia}}\ and\ \bibinfo {author} {\bibfnamefont {Ludwig}\
  \bibnamefont {Rauch}},\ }\bibfield  {title} {\enquote {\bibinfo {title}
  {{Dark matter direct-detection experiments}},}\ }\href {\doibase
  10.1088/0954-3899/43/1/013001} {\bibfield  {journal} {\bibinfo  {journal} {J.
  Phys.}\ }\textbf {\bibinfo {volume} {G43}},\ \bibinfo {pages} {013001}
  (\bibinfo {year} {2016})},\ \Eprint {http://arxiv.org/abs/1509.08767}
  {arXiv:1509.08767 [physics.ins-det]} \BibitemShut {NoStop}%
\bibitem [{\citenamefont {Schumann}(2019)}]{Schumann:2019eaa}%
  \BibitemOpen
  \bibfield  {author} {\bibinfo {author} {\bibfnamefont {M.}~\bibnamefont
  {Schumann}},\ }\bibfield  {title} {\enquote {\bibinfo {title} {{Direct
  Detection of WIMP Dark Matter: Concepts and Status}},}\ }\href {\doibase
  10.1088/1361-6471/ab2ea5} {\bibfield  {journal} {\bibinfo  {journal} {J.
  Phys. G: Nucl. Part. Phys.}\ }\textbf {\bibinfo {volume} {46}},\ \bibinfo
  {pages} {103003} (\bibinfo {year} {2019})},\ \Eprint
  {http://arxiv.org/abs/1903.03026} {arXiv:1903.03026 [astro-ph.CO]}
  \BibitemShut {NoStop}%
\bibitem [{\citenamefont {Drukier}\ \emph {et~al.}(1986)\citenamefont
  {Drukier}, \citenamefont {Freese},\ and\ \citenamefont
  {Spergel}}]{Drukier:1986tm}%
  \BibitemOpen
  \bibfield  {author} {\bibinfo {author} {\bibfnamefont {A.~K.}\ \bibnamefont
  {Drukier}}, \bibinfo {author} {\bibfnamefont {Katherine}\ \bibnamefont
  {Freese}}, \ and\ \bibinfo {author} {\bibfnamefont {D.~N.}\ \bibnamefont
  {Spergel}},\ }\bibfield  {title} {\enquote {\bibinfo {title} {{Detecting Cold
  Dark Matter Candidates}},}\ }\href {\doibase 10.1103/PhysRevD.33.3495}
  {\bibfield  {journal} {\bibinfo  {journal} {Phys. Rev.}\ }\textbf {\bibinfo
  {volume} {D33}},\ \bibinfo {pages} {3495--3508} (\bibinfo {year}
  {1986})}\BibitemShut {NoStop}%
\bibitem [{\citenamefont {Freese}\ \emph {et~al.}(1988)\citenamefont {Freese},
  \citenamefont {Frieman},\ and\ \citenamefont {Gould}}]{Freese:1987wu}%
  \BibitemOpen
  \bibfield  {author} {\bibinfo {author} {\bibfnamefont {Katherine}\
  \bibnamefont {Freese}}, \bibinfo {author} {\bibfnamefont {Joshua~A.}\
  \bibnamefont {Frieman}}, \ and\ \bibinfo {author} {\bibfnamefont {Andrew}\
  \bibnamefont {Gould}},\ }\bibfield  {title} {\enquote {\bibinfo {title}
  {{Signal Modulation in Cold Dark Matter Detection}},}\ }\href {\doibase
  10.1103/PhysRevD.37.3388} {\bibfield  {journal} {\bibinfo  {journal} {Phys.
  Rev.}\ }\textbf {\bibinfo {volume} {D37}},\ \bibinfo {pages} {3388--3405}
  (\bibinfo {year} {1988})}\BibitemShut {NoStop}%
\bibitem [{\citenamefont {Bernabei}\ \emph {et~al.}(2020)\citenamefont
  {Bernabei} \emph {et~al.}}]{Bernabei:2020mon}%
  \BibitemOpen
  \bibfield  {author} {\bibinfo {author} {\bibfnamefont {R.}~\bibnamefont
  {Bernabei}} \emph {et~al.},\ }\bibfield  {title} {\enquote {\bibinfo {title}
  {{The DAMA project: Achievements, implications and perspectives}},}\ }\href
  {\doibase 10.1016/j.ppnp.2020.103810} {\bibfield  {journal} {\bibinfo
  {journal} {Prog. Part. Nucl. Phys.}\ }\textbf {\bibinfo {volume} {114}},\
  \bibinfo {pages} {103810} (\bibinfo {year} {2020})}\BibitemShut {NoStop}%
\bibitem [{\citenamefont {Adhikari}\ \emph {et~al.}(2018)\citenamefont
  {Adhikari} \emph {et~al.}}]{Adhikari:2018ljm}%
  \BibitemOpen
  \bibfield  {author} {\bibinfo {author} {\bibfnamefont {Govinda}\ \bibnamefont
  {Adhikari}} \emph {et~al.},\ }\bibfield  {title} {\enquote {\bibinfo {title}
  {{An experiment to search for dark-matter interactions using sodium iodide
  detectors}},}\ }\href {\doibase 10.1038/s41586-018-0739-1} {\bibfield
  {journal} {\bibinfo  {journal} {Nature}\ }\textbf {\bibinfo {volume} {564}},\
  \bibinfo {pages} {83--86} (\bibinfo {year} {2018})},\ \bibinfo {note}
  {[Erratum: Nature 566, E2 (2019)]},\ \Eprint
  {http://arxiv.org/abs/1906.01791} {arXiv:1906.01791 [astro-ph.IM]}
  \BibitemShut {NoStop}%
\bibitem [{\citenamefont {Kobayashi}\ \emph {et~al.}(2019)\citenamefont
  {Kobayashi} \emph {et~al.}}]{Kobayashi:2018jky}%
  \BibitemOpen
  \bibfield  {author} {\bibinfo {author} {\bibfnamefont {M.}~\bibnamefont
  {Kobayashi}} \emph {et~al.} (\bibinfo {collaboration} {XMASS}),\ }\bibfield
  {title} {\enquote {\bibinfo {title} {{Search for sub-GeV dark matter by
  annual modulation using XMASS-I detector}},}\ }\href {\doibase
  10.1016/j.physletb.2019.06.022} {\bibfield  {journal} {\bibinfo  {journal}
  {Phys. Lett. B}\ }\textbf {\bibinfo {volume} {795}},\ \bibinfo {pages}
  {308--313} (\bibinfo {year} {2019})},\ \Eprint
  {http://arxiv.org/abs/1808.06177} {arXiv:1808.06177 [astro-ph.CO]}
  \BibitemShut {NoStop}%
\bibitem [{\citenamefont {Akerib}\ \emph {et~al.}(2018)\citenamefont {Akerib}
  \emph {et~al.}}]{Akerib:2018zoq}%
  \BibitemOpen
  \bibfield  {author} {\bibinfo {author} {\bibfnamefont {D.~S.}\ \bibnamefont
  {Akerib}} \emph {et~al.} (\bibinfo {collaboration} {LUX}),\ }\bibfield
  {title} {\enquote {\bibinfo {title} {{Search for annual and diurnal rate
  modulations in the LUX experiment}},}\ }\href {\doibase
  10.1103/PhysRevD.98.062005} {\bibfield  {journal} {\bibinfo  {journal} {Phys.
  Rev. D}\ }\textbf {\bibinfo {volume} {98}},\ \bibinfo {pages} {062005}
  (\bibinfo {year} {2018})},\ \Eprint {http://arxiv.org/abs/1807.07113}
  {arXiv:1807.07113 [astro-ph.CO]} \BibitemShut {NoStop}%
\bibitem [{\citenamefont {Abe}\ \emph {et~al.}(2018)\citenamefont {Abe} \emph
  {et~al.}}]{Abe:2018mxq}%
  \BibitemOpen
  \bibfield  {author} {\bibinfo {author} {\bibfnamefont {K.}~\bibnamefont
  {Abe}} \emph {et~al.} (\bibinfo {collaboration} {XMASS}),\ }\bibfield
  {title} {\enquote {\bibinfo {title} {{Direct dark matter search by annual
  modulation with 2.7 years of XMASS-I data}},}\ }\href {\doibase
  10.1103/PhysRevD.97.102006} {\bibfield  {journal} {\bibinfo  {journal} {Phys.
  Rev.}\ }\textbf {\bibinfo {volume} {D97}},\ \bibinfo {pages} {102006}
  (\bibinfo {year} {2018})},\ \Eprint {http://arxiv.org/abs/1801.10096}
  {arXiv:1801.10096 [astro-ph.CO]} \BibitemShut {NoStop}%
\bibitem [{\citenamefont {Aprile}\ \emph {et~al.}(2017)\citenamefont {Aprile}
  \emph {et~al.}}]{Aprile:2017yea}%
  \BibitemOpen
  \bibfield  {author} {\bibinfo {author} {\bibfnamefont {E.}~\bibnamefont
  {Aprile}} \emph {et~al.} (\bibinfo {collaboration} {XENON}),\ }\bibfield
  {title} {\enquote {\bibinfo {title} {{Search for Electronic Recoil Event Rate
  Modulation with 4 Years of XENON100 Data}},}\ }\href {\doibase
  10.1103/PhysRevLett.118.101101} {\bibfield  {journal} {\bibinfo  {journal}
  {Phys. Rev. Lett.}\ }\textbf {\bibinfo {volume} {118}},\ \bibinfo {pages}
  {101101} (\bibinfo {year} {2017})},\ \Eprint
  {http://arxiv.org/abs/1701.00769} {arXiv:1701.00769 [astro-ph.CO]}
  \BibitemShut {NoStop}%
\bibitem [{\citenamefont {Savage}\ \emph {et~al.}(2009)\citenamefont {Savage},
  \citenamefont {Freese}, \citenamefont {Gondolo},\ and\ \citenamefont
  {Spolyar}}]{Savage:2009mk}%
  \BibitemOpen
  \bibfield  {author} {\bibinfo {author} {\bibfnamefont {C.}~\bibnamefont
  {Savage}}, \bibinfo {author} {\bibfnamefont {K.}~\bibnamefont {Freese}},
  \bibinfo {author} {\bibfnamefont {P.}~\bibnamefont {Gondolo}}, \ and\
  \bibinfo {author} {\bibfnamefont {D.}~\bibnamefont {Spolyar}},\ }\bibfield
  {title} {\enquote {\bibinfo {title} {{Compatibility of DAMA/LIBRA dark matter
  detection with other searches in light of new Galactic rotation velocity
  measurements}},}\ }\href {\doibase 10.1088/1475-7516/2009/09/036} {\bibfield
  {journal} {\bibinfo  {journal} {JCAP}\ }\textbf {\bibinfo {volume} {0909}},\
  \bibinfo {pages} {036} (\bibinfo {year} {2009})},\ \Eprint
  {http://arxiv.org/abs/0901.2713} {arXiv:0901.2713 [astro-ph.CO]} \BibitemShut
  {NoStop}%
\bibitem [{\citenamefont {Aprile}\ \emph
  {et~al.}(2015{\natexlab{a}})\citenamefont {Aprile} \emph
  {et~al.}}]{Aprile:2015ade}%
  \BibitemOpen
  \bibfield  {author} {\bibinfo {author} {\bibfnamefont {E.}~\bibnamefont
  {Aprile}} \emph {et~al.} (\bibinfo {collaboration} {XENON100}),\ }\bibfield
  {title} {\enquote {\bibinfo {title} {{Exclusion of Leptophilic Dark Matter
  Models using XENON100 Electronic Recoil Data}},}\ }\href {\doibase
  10.1126/science.aab2069} {\bibfield  {journal} {\bibinfo  {journal}
  {Science}\ }\textbf {\bibinfo {volume} {349}},\ \bibinfo {pages} {851--854}
  (\bibinfo {year} {2015}{\natexlab{a}})},\ \Eprint
  {http://arxiv.org/abs/1507.07747} {arXiv:1507.07747 [astro-ph.CO]}
  \BibitemShut {NoStop}%
\bibitem [{\citenamefont {Herrero-Garcia}(2015)}]{Herrero-Garcia:2015kga}%
  \BibitemOpen
  \bibfield  {author} {\bibinfo {author} {\bibfnamefont {J.}~\bibnamefont
  {Herrero-Garcia}},\ }\bibfield  {title} {\enquote {\bibinfo {title}
  {{Halo-independent tests of dark matter annual modulation signals}},}\ }\href
  {\doibase 10.1088/1475-7516/2015/09/012} {\bibfield  {journal} {\bibinfo
  {journal} {JCAP}\ }\textbf {\bibinfo {volume} {1509}},\ \bibinfo {pages}
  {012} (\bibinfo {year} {2015})},\ \Eprint {http://arxiv.org/abs/1506.03503}
  {arXiv:1506.03503 [hep-ph]} \BibitemShut {NoStop}%
\bibitem [{\citenamefont {Baum}\ \emph {et~al.}(2019)\citenamefont {Baum},
  \citenamefont {Freese},\ and\ \citenamefont {Kelso}}]{Baum:2018ekm}%
  \BibitemOpen
  \bibfield  {author} {\bibinfo {author} {\bibfnamefont {Sebastian}\
  \bibnamefont {Baum}}, \bibinfo {author} {\bibfnamefont {Katherine}\
  \bibnamefont {Freese}}, \ and\ \bibinfo {author} {\bibfnamefont {Chris}\
  \bibnamefont {Kelso}},\ }\bibfield  {title} {\enquote {\bibinfo {title}
  {{Dark Matter implications of DAMA/LIBRA-phase2 results}},}\ }\href {\doibase
  10.1016/j.physletb.2018.12.036} {\bibfield  {journal} {\bibinfo  {journal}
  {Phys. Lett.}\ }\textbf {\bibinfo {volume} {B789}},\ \bibinfo {pages}
  {262--269} (\bibinfo {year} {2019})},\ \Eprint
  {http://arxiv.org/abs/1804.01231} {arXiv:1804.01231 [astro-ph.CO]}
  \BibitemShut {NoStop}%
\bibitem [{\citenamefont {Kang}\ \emph {et~al.}(2018)\citenamefont {Kang},
  \citenamefont {Scopel}, \citenamefont {Tomar},\ and\ \citenamefont
  {Yoon}}]{Kang:2018qvz}%
  \BibitemOpen
  \bibfield  {author} {\bibinfo {author} {\bibfnamefont {Sunghyun}\
  \bibnamefont {Kang}}, \bibinfo {author} {\bibfnamefont {Stefano}\
  \bibnamefont {Scopel}}, \bibinfo {author} {\bibfnamefont {Gaurav}\
  \bibnamefont {Tomar}}, \ and\ \bibinfo {author} {\bibfnamefont {Jong-Hyun}\
  \bibnamefont {Yoon}},\ }\bibfield  {title} {\enquote {\bibinfo {title}
  {{DAMA/LIBRA-phase2 in WIMP effective models}},}\ }\href {\doibase
  10.1088/1475-7516/2018/07/016} {\bibfield  {journal} {\bibinfo  {journal}
  {JCAP}\ }\textbf {\bibinfo {volume} {1807}},\ \bibinfo {pages} {016}
  (\bibinfo {year} {2018})},\ \Eprint {http://arxiv.org/abs/1804.07528}
  {arXiv:1804.07528 [hep-ph]} \BibitemShut {NoStop}%
\bibitem [{\citenamefont {Herrero-Garcia}\ \emph {et~al.}(2018)\citenamefont
  {Herrero-Garcia}, \citenamefont {Scaffidi}, \citenamefont {White},\ and\
  \citenamefont {Williams}}]{Herrero-Garcia:2018mky}%
  \BibitemOpen
  \bibfield  {author} {\bibinfo {author} {\bibfnamefont {Juan}\ \bibnamefont
  {Herrero-Garcia}}, \bibinfo {author} {\bibfnamefont {Andre}\ \bibnamefont
  {Scaffidi}}, \bibinfo {author} {\bibfnamefont {Martin}\ \bibnamefont
  {White}}, \ and\ \bibinfo {author} {\bibfnamefont {Anthony~G.}\ \bibnamefont
  {Williams}},\ }\bibfield  {title} {\enquote {\bibinfo {title}
  {{Time-dependent rate of multicomponent dark matter: Reproducing the
  DAMA/LIBRA phase-2 results}},}\ }\href {\doibase 10.1103/PhysRevD.98.123007}
  {\bibfield  {journal} {\bibinfo  {journal} {Phys. Rev.}\ }\textbf {\bibinfo
  {volume} {D98}},\ \bibinfo {pages} {123007} (\bibinfo {year} {2018})},\
  \Eprint {http://arxiv.org/abs/1804.08437} {arXiv:1804.08437 [hep-ph]}
  \BibitemShut {NoStop}%
\bibitem [{\citenamefont {Amar\'e}\ \emph {et~al.}(2019)\citenamefont {Amar\'e}
  \emph {et~al.}}]{Amare:2019jul}%
  \BibitemOpen
  \bibfield  {author} {\bibinfo {author} {\bibfnamefont {J.}~\bibnamefont
  {Amar\'e}} \emph {et~al.},\ }\bibfield  {title} {\enquote {\bibinfo {title}
  {{First Results on Dark Matter Annual Modulation from the ANAIS-112
  Experiment}},}\ }\href {\doibase 10.1103/PhysRevLett.123.031301} {\bibfield
  {journal} {\bibinfo  {journal} {Phys. Rev. Lett.}\ }\textbf {\bibinfo
  {volume} {123}},\ \bibinfo {pages} {031301} (\bibinfo {year} {2019})},\
  \Eprint {http://arxiv.org/abs/1903.03973} {arXiv:1903.03973 [astro-ph.IM]}
  \BibitemShut {NoStop}%
\bibitem [{\citenamefont {Amar{\'e}}\ \emph
  {et~al.}(2019{\natexlab{a}})\citenamefont {Amar{\'e}} \emph
  {et~al.}}]{Amare:2018sxx}%
  \BibitemOpen
  \bibfield  {author} {\bibinfo {author} {\bibfnamefont {J.}~\bibnamefont
  {Amar{\'e}}} \emph {et~al.},\ }\bibfield  {title} {\enquote {\bibinfo {title}
  {{Performance of ANAIS-112 experiment after the first year of data
  taking}},}\ }\href {\doibase 10.1140/epjc/s10052-019-6697-4} {\bibfield
  {journal} {\bibinfo  {journal} {Eur. Phys. J. C}\ }\textbf {\bibinfo {volume}
  {79}},\ \bibinfo {pages} {228} (\bibinfo {year} {2019}{\natexlab{a}})},\
  \Eprint {http://arxiv.org/abs/1812.01472} {arXiv:1812.01472 [astro-ph.IM]}
  \BibitemShut {NoStop}%
\bibitem [{\citenamefont {Adhikari}\ \emph {et~al.}(2019)\citenamefont
  {Adhikari} \emph {et~al.}}]{PhysRevLett.123.031302}%
  \BibitemOpen
  \bibfield  {author} {\bibinfo {author} {\bibfnamefont {G.}~\bibnamefont
  {Adhikari}} \emph {et~al.} (\bibinfo {collaboration} {COSINE-100
  Collaboration}),\ }\bibfield  {title} {\enquote {\bibinfo {title} {Search for
  a dark matter-induced annual modulation signal in NaI(Tl) with the cosine-100
  experiment},}\ }\href {\doibase 10.1103/PhysRevLett.123.031302} {\bibfield
  {journal} {\bibinfo  {journal} {Phys. Rev. Lett.}\ }\textbf {\bibinfo
  {volume} {123}},\ \bibinfo {pages} {031302} (\bibinfo {year}
  {2019})}\BibitemShut {NoStop}%
\bibitem [{\citenamefont {Antonello}\ \emph {et~al.}(2019)\citenamefont
  {Antonello} \emph {et~al.}}]{Antonello:2018fvx}%
  \BibitemOpen
  \bibfield  {author} {\bibinfo {author} {\bibfnamefont {M.}~\bibnamefont
  {Antonello}} \emph {et~al.} (\bibinfo {collaboration} {SABRE}),\ }\bibfield
  {title} {\enquote {\bibinfo {title} {{The SABRE project and the SABRE
  Proof-of-Principle}},}\ }\href {\doibase 10.1140/epjc/s10052-019-6860-y}
  {\bibfield  {journal} {\bibinfo  {journal} {Eur. Phys. J. C}\ }\textbf
  {\bibinfo {volume} {79}},\ \bibinfo {pages} {363} (\bibinfo {year} {2019})},\
  \Eprint {http://arxiv.org/abs/1806.09340} {arXiv:1806.09340
  [physics.ins-det]} \BibitemShut {NoStop}%
\bibitem [{\citenamefont {Antonello}\ \emph {et~al.}(2021)\citenamefont
  {Antonello} \emph {et~al.}}]{Antonello:2020xhj}%
  \BibitemOpen
  \bibfield  {author} {\bibinfo {author} {\bibfnamefont {M.}~\bibnamefont
  {Antonello}} \emph {et~al.},\ }\bibfield  {title} {\enquote {\bibinfo {title}
  {{Characterization of SABRE crystal NaI-33 with direct underground
  counting}},}\ }\href {\doibase 10.1140/epjc/s10052-021-09098-5} {\bibfield
  {journal} {\bibinfo  {journal} {Eur. Phys. J. C}\ }\textbf {\bibinfo {volume}
  {81}},\ \bibinfo {pages} {299} (\bibinfo {year} {2021})},\ \Eprint
  {http://arxiv.org/abs/2012.02610} {arXiv:2012.02610 [physics.ins-det]}
  \BibitemShut {NoStop}%
\bibitem [{\citenamefont {Angloher}\ \emph {et~al.}(2016)\citenamefont
  {Angloher} \emph {et~al.}}]{Angloher:2016ooq}%
  \BibitemOpen
  \bibfield  {author} {\bibinfo {author} {\bibfnamefont {G.}~\bibnamefont
  {Angloher}} \emph {et~al.},\ }\bibfield  {title} {\enquote {\bibinfo {title}
  {{The COSINUS project - perspectives of a NaI scintillating calorimeter for
  dark matter search}},}\ }\href {\doibase 10.1140/epjc/s10052-016-4278-3}
  {\bibfield  {journal} {\bibinfo  {journal} {Eur. Phys. J. C}\ }\textbf
  {\bibinfo {volume} {76}},\ \bibinfo {pages} {441} (\bibinfo {year} {2016})},\
  \Eprint {http://arxiv.org/abs/1603.02214} {arXiv:1603.02214
  [physics.ins-det]} \BibitemShut {NoStop}%
\bibitem [{\citenamefont {Angloher}\ \emph {et~al.}(2017)\citenamefont
  {Angloher} \emph {et~al.}}]{Angloher:2017sft}%
  \BibitemOpen
  \bibfield  {author} {\bibinfo {author} {\bibfnamefont {G.}~\bibnamefont
  {Angloher}} \emph {et~al.} (\bibinfo {collaboration} {COSINUS}),\ }\bibfield
  {title} {\enquote {\bibinfo {title} {{Results from the first cryogenic NaI
  detector for the COSINUS project}},}\ }\href {\doibase
  10.1088/1748-0221/12/11/P11007} {\bibfield  {journal} {\bibinfo  {journal}
  {JINST}\ }\textbf {\bibinfo {volume} {12}},\ \bibinfo {pages} {P11007}
  (\bibinfo {year} {2017})},\ \Eprint {http://arxiv.org/abs/1705.11028}
  {arXiv:1705.11028 [physics.ins-det]} \BibitemShut {NoStop}%
\bibitem [{\citenamefont {Coarasa}\ \emph {et~al.}(2019)\citenamefont {Coarasa}
  \emph {et~al.}}]{Coarasa:2018qzs}%
  \BibitemOpen
  \bibfield  {author} {\bibinfo {author} {\bibfnamefont {I.}~\bibnamefont
  {Coarasa}} \emph {et~al.},\ }\bibfield  {title} {\enquote {\bibinfo {title}
  {{ANAIS-112 sensitivity in the search for dark matter annual modulation}},}\
  }\href {\doibase 10.1140/epjc/s10052-019-6733-4} {\bibfield  {journal}
  {\bibinfo  {journal} {Eur. Phys. J. C}\ }\textbf {\bibinfo {volume} {79}},\
  \bibinfo {pages} {233} (\bibinfo {year} {2019})},\ \Eprint
  {http://arxiv.org/abs/1812.02000} {arXiv:1812.02000 [astro-ph.IM]}
  \BibitemShut {NoStop}%
\bibitem [{\citenamefont {Gerbier}\ \emph {et~al.}(1999)\citenamefont {Gerbier}
  \emph {et~al.}}]{Gerbier:1998dm}%
  \BibitemOpen
  \bibfield  {author} {\bibinfo {author} {\bibfnamefont {G.}~\bibnamefont
  {Gerbier}} \emph {et~al.},\ }\bibfield  {title} {\enquote {\bibinfo {title}
  {{Pulse shape discrimination with NaI(Tl) and results from a WIMP search at
  the Laboratoire Souterrain de Modane}},}\ }\href {\doibase
  10.1016/S0927-6505(99)00004-3} {\bibfield  {journal} {\bibinfo  {journal}
  {Astropart. Phys.}\ }\textbf {\bibinfo {volume} {11}},\ \bibinfo {pages}
  {287--302} (\bibinfo {year} {1999})}\BibitemShut {NoStop}%
\bibitem [{\citenamefont {Bernabei}\ \emph {et~al.}(2013)\citenamefont
  {Bernabei} \emph {et~al.}}]{dama-quench-hypo}%
  \BibitemOpen
  \bibfield  {author} {\bibinfo {author} {\bibfnamefont {R.}~\bibnamefont
  {Bernabei}} \emph {et~al.} (\bibinfo {collaboration} {DAMA/LIBRA
  Collaboration}),\ }\bibfield  {title} {\enquote {\bibinfo {title} {{Dark
  matter investigation by DAMA at Gran Sasso}},}\ }\href {\doibase
  10.1142/S0217751X13300226} {\bibfield  {journal} {\bibinfo  {journal} {Int.
  J. Mod. Phys. B}\ }\textbf {\bibinfo {volume} {28}},\ \bibinfo {pages}
  {1330022} (\bibinfo {year} {2013})}\BibitemShut {NoStop}%
\bibitem [{\citenamefont {Spooner}\ \emph {et~al.}(1994)\citenamefont {Spooner}
  \emph {et~al.}}]{spooner-quench}%
  \BibitemOpen
  \bibfield  {author} {\bibinfo {author} {\bibfnamefont {N.J.C.}\ \bibnamefont
  {Spooner}} \emph {et~al.},\ }\bibfield  {title} {\enquote {\bibinfo {title}
  {The scintillation efficiency of sodium and iodine recoils in a NaI(Tl)
  detector for dark matter searches},}\ }\href {\doibase
  https://doi.org/10.1016/0370-2693(94)90343-3} {\bibfield  {journal} {\bibinfo
   {journal} {Physics Letters B}\ }\textbf {\bibinfo {volume} {321}},\ \bibinfo
  {pages} {156--160} (\bibinfo {year} {1994})}\BibitemShut {NoStop}%
\bibitem [{\citenamefont {Tovey}\ \emph {et~al.}(1998)\citenamefont {Tovey}
  \emph {et~al.}}]{tovey-quench}%
  \BibitemOpen
  \bibfield  {author} {\bibinfo {author} {\bibfnamefont {D.R.}\ \bibnamefont
  {Tovey}} \emph {et~al.},\ }\bibfield  {title} {\enquote {\bibinfo {title}
  {Measurement of scintillation efficiencies and pulse-shapes for nuclear
  recoils in {NaI(Tl)} and {CaF2(Eu)} at low energies for dark matter
  experiments},}\ }\href {\doibase
  https://doi.org/10.1016/S0370-2693(98)00643-1} {\bibfield  {journal}
  {\bibinfo  {journal} {Physics Letters B}\ }\textbf {\bibinfo {volume}
  {433}},\ \bibinfo {pages} {150 -- 155} (\bibinfo {year} {1998})}\BibitemShut
  {NoStop}%
\bibitem [{\citenamefont {Chagani}\ \emph {et~al.}(2008)\citenamefont
  {Chagani}, \citenamefont {Majewski}, \citenamefont {Daw}, \citenamefont
  {Kudryavtsev},\ and\ \citenamefont {Spooner}}]{Chagani:2008in}%
  \BibitemOpen
  \bibfield  {author} {\bibinfo {author} {\bibfnamefont {H.}~\bibnamefont
  {Chagani}}, \bibinfo {author} {\bibfnamefont {P.}~\bibnamefont {Majewski}},
  \bibinfo {author} {\bibfnamefont {E.~J.}\ \bibnamefont {Daw}}, \bibinfo
  {author} {\bibfnamefont {V.~A.}\ \bibnamefont {Kudryavtsev}}, \ and\ \bibinfo
  {author} {\bibfnamefont {N.~J.~C.}\ \bibnamefont {Spooner}},\ }\bibfield
  {title} {\enquote {\bibinfo {title} {{Measurement of the quenching factor of
  Na recoils in NaI(Tl)}},}\ }\href {\doibase 10.1088/1748-0221/3/06/P06003}
  {\bibfield  {journal} {\bibinfo  {journal} {JINST}\ }\textbf {\bibinfo
  {volume} {3}},\ \bibinfo {pages} {P06003} (\bibinfo {year} {2008})},\ \Eprint
  {http://arxiv.org/abs/0806.1916} {arXiv:0806.1916 [physics.ins-det]}
  \BibitemShut {NoStop}%
\bibitem [{\citenamefont {Collar}(2013)}]{collar-quench}%
  \BibitemOpen
  \bibfield  {author} {\bibinfo {author} {\bibfnamefont {J.~I.}\ \bibnamefont
  {Collar}},\ }\bibfield  {title} {\enquote {\bibinfo {title} {{Quenching and
  channeling of nuclear recoils in NaI(Tl): Implications for dark-matter
  searches}},}\ }\href {\doibase 10.1103/physrevc.88.035806} {\bibfield
  {journal} {\bibinfo  {journal} {Phys. Rev. C}\ }\textbf {\bibinfo {volume}
  {88}},\ \bibinfo {pages} {035806} (\bibinfo {year} {2013})}\BibitemShut
  {NoStop}%
\bibitem [{\citenamefont {Xu}\ \emph {et~al.}(2015)\citenamefont {Xu} \emph
  {et~al.}}]{Xu:2015wha}%
  \BibitemOpen
  \bibfield  {author} {\bibinfo {author} {\bibfnamefont {Jingke}\ \bibnamefont
  {Xu}} \emph {et~al.},\ }\bibfield  {title} {\enquote {\bibinfo {title}
  {{Scintillation Efficiency Measurement of Na Recoils in NaI(Tl) Below the
  DAMA/LIBRA Energy Threshold}},}\ }\href {\doibase 10.1103/PhysRevC.92.015807}
  {\bibfield  {journal} {\bibinfo  {journal} {Phys. Rev.}\ }\textbf {\bibinfo
  {volume} {C92}},\ \bibinfo {pages} {015807} (\bibinfo {year} {2015})},\
  \Eprint {http://arxiv.org/abs/1503.07212} {arXiv:1503.07212
  [physics.ins-det]} \BibitemShut {NoStop}%
\bibitem [{\citenamefont {Joo}\ \emph {et~al.}(2019)\citenamefont {Joo} \emph
  {et~al.}}]{joo-quench}%
  \BibitemOpen
  \bibfield  {author} {\bibinfo {author} {\bibfnamefont {H.W.}\ \bibnamefont
  {Joo}} \emph {et~al.},\ }\bibfield  {title} {\enquote {\bibinfo {title}
  {{Quenching factor measurement for NaI(Tl) scintillation crystal}},}\ }\href
  {\doibase https://doi.org/10.1016/j.astropartphys.2019.01.001} {\bibfield
  {journal} {\bibinfo  {journal} {Astropart. Phys.}\ }\textbf {\bibinfo
  {volume} {108}},\ \bibinfo {pages} {50 -- 56} (\bibinfo {year}
  {2019})}\BibitemShut {NoStop}%
\bibitem [{\citenamefont {Ko}\ \emph {et~al.}(2019)\citenamefont {Ko} \emph
  {et~al.}}]{cosine-quench}%
  \BibitemOpen
  \bibfield  {author} {\bibinfo {author} {\bibfnamefont {Y.J.}\ \bibnamefont
  {Ko}} \emph {et~al.} (\bibinfo {collaboration} {COSINE-100 Collaboration}),\
  }\bibfield  {title} {\enquote {\bibinfo {title} {{Comparison between
  {DAMA/LIBRA} and {COSINE-100} in the light of quenching factors}},}\ }\href
  {\doibase 10.1088/1475-7516/2019/11/008} {\bibfield  {journal} {\bibinfo
  {journal} {J. Cosmol. Astropart. Phys.}\ }\textbf {\bibinfo {volume}
  {2019}},\ \bibinfo {pages} {008–008} (\bibinfo {year} {2019})}\BibitemShut
  {NoStop}%
\bibitem [{\citenamefont {Bignell}\ \emph {et~al.}(2021)\citenamefont {Bignell}
  \emph {et~al.}}]{Bignell:2021bjx}%
  \BibitemOpen
  \bibfield  {author} {\bibinfo {author} {\bibfnamefont {L.~J.}\ \bibnamefont
  {Bignell}} \emph {et~al.},\ }\bibfield  {title} {\enquote {\bibinfo {title}
  {{Quenching factor measurements of sodium nuclear recoils in NaI:Tl
  determined by spectrum fitting}},}\ }\href@noop {} {\  (\bibinfo {year}
  {2021})},\ \Eprint {http://arxiv.org/abs/2102.02833} {arXiv:2102.02833
  [physics.ins-det]} \BibitemShut {NoStop}%
\bibitem [{\citenamefont {Sarsa}\ \emph {et~al.}(1996)\citenamefont {Sarsa}
  \emph {et~al.}}]{Sarsa:1996pa}%
  \BibitemOpen
  \bibfield  {author} {\bibinfo {author} {\bibfnamefont {M.~L.}\ \bibnamefont
  {Sarsa}} \emph {et~al.},\ }\bibfield  {title} {\enquote {\bibinfo {title}
  {{Searching for annual modulations of WIMPs with NaI scintillators}},}\
  }\href {\doibase 10.1016/0370-2693(96)01070-2} {\bibfield  {journal}
  {\bibinfo  {journal} {Phys. Lett. B}\ }\textbf {\bibinfo {volume} {386}},\
  \bibinfo {pages} {458--462} (\bibinfo {year} {1996})}\BibitemShut {NoStop}%
\bibitem [{\citenamefont {Sarsa}\ \emph {et~al.}(1997)\citenamefont {Sarsa}
  \emph {et~al.}}]{Sarsa:1997hb}%
  \BibitemOpen
  \bibfield  {author} {\bibinfo {author} {\bibfnamefont {M.~L.}\ \bibnamefont
  {Sarsa}} \emph {et~al.},\ }\bibfield  {title} {\enquote {\bibinfo {title}
  {{Results of a search for annual modulation of WIMP signals}},}\ }\href
  {\doibase 10.1103/PhysRevD.56.1856} {\bibfield  {journal} {\bibinfo
  {journal} {Phys. Rev. D}\ }\textbf {\bibinfo {volume} {56}},\ \bibinfo
  {pages} {1856--1862} (\bibinfo {year} {1997})}\BibitemShut {NoStop}%
\bibitem [{\citenamefont {Amar{\'e}}\ \emph
  {et~al.}(2019{\natexlab{b}})\citenamefont {Amar{\'e}} \emph
  {et~al.}}]{Amare:2018ndh}%
  \BibitemOpen
  \bibfield  {author} {\bibinfo {author} {\bibfnamefont {J.}~\bibnamefont
  {Amar{\'e}}} \emph {et~al.},\ }\bibfield  {title} {\enquote {\bibinfo {title}
  {{Analysis of backgrounds for the ANAIS-112 dark matter experiment}},}\
  }\href {\doibase 10.1140/epjc/s10052-019-6911-4} {\bibfield  {journal}
  {\bibinfo  {journal} {Eur. Phys. J. C}\ }\textbf {\bibinfo {volume} {79}},\
  \bibinfo {pages} {412} (\bibinfo {year} {2019}{\natexlab{b}})},\ \Eprint
  {http://arxiv.org/abs/1812.01377} {arXiv:1812.01377 [astro-ph.GA]}
  \BibitemShut {NoStop}%
\bibitem [{\citenamefont {Amar\'e}\ \emph {et~al.}(2020)\citenamefont {Amar\'e}
  \emph {et~al.}}]{Amare:2019ncj}%
  \BibitemOpen
  \bibfield  {author} {\bibinfo {author} {\bibfnamefont {J.}~\bibnamefont
  {Amar\'e}} \emph {et~al.},\ }\bibfield  {title} {\enquote {\bibinfo {title}
  {{ANAIS-112 status: two years results on annual modulation}},}\ }\href
  {\doibase 10.1088/1742-6596/1468/1/012014} {\bibfield  {journal} {\bibinfo
  {journal} {J. Phys. Conf. Ser.}\ }\textbf {\bibinfo {volume} {1468}},\
  \bibinfo {pages} {012014} (\bibinfo {year} {2020})},\ \Eprint
  {http://arxiv.org/abs/1910.13365} {arXiv:1910.13365 [astro-ph.IM]}
  \BibitemShut {NoStop}%
\bibitem [{\citenamefont {Oliván}\ \emph {et~al.}(2017)\citenamefont {Oliván}
  \emph {et~al.}}]{Olivan:2017akd}%
  \BibitemOpen
  \bibfield  {author} {\bibinfo {author} {\bibfnamefont {M.~A.}\ \bibnamefont
  {Oliván}} \emph {et~al.},\ }\bibfield  {title} {\enquote {\bibinfo {title}
  {{Light yield determination in large sodium iodide detectors applied in the
  search for dark matter}},}\ }\href {\doibase
  10.1016/j.astropartphys.2017.06.005} {\bibfield  {journal} {\bibinfo
  {journal} {Astropart. Phys.}\ }\textbf {\bibinfo {volume} {93}},\ \bibinfo
  {pages} {86--95} (\bibinfo {year} {2017})},\ \Eprint
  {http://arxiv.org/abs/1703.01262} {arXiv:1703.01262 [physics.ins-det]}
  \BibitemShut {NoStop}%
\bibitem [{\citenamefont {Bernabei}\ \emph {et~al.}(2012)\citenamefont
  {Bernabei} \emph {et~al.}}]{Bernabei:2012zzb}%
  \BibitemOpen
  \bibfield  {author} {\bibinfo {author} {\bibfnamefont {R.}~\bibnamefont
  {Bernabei}} \emph {et~al.},\ }\bibfield  {title} {\enquote {\bibinfo {title}
  {{Performances of the new high quantum efficiency PMTs in DAMA/LIBRA}},}\
  }\href {\doibase 10.1088/1748-0221/7/03/P03009} {\bibfield  {journal}
  {\bibinfo  {journal} {JINST}\ }\textbf {\bibinfo {volume} {7}},\ \bibinfo
  {pages} {P03009} (\bibinfo {year} {2012})}\BibitemShut {NoStop}%
\bibitem [{\citenamefont {Oliv{\'a}n}(2016)}]{MAThesis}%
  \BibitemOpen
  \bibfield  {author} {\bibinfo {author} {\bibfnamefont {M.A.}\ \bibnamefont
  {Oliv{\'a}n}},\ }\href@noop {} {Ph.D. thesis},\ \bibinfo  {school}
  {Universidad de Zaragoza} (\bibinfo {year} {2016})\BibitemShut {NoStop}%
\bibitem [{\citenamefont {Trzaska}\ \emph {et~al.}(2019)\citenamefont {Trzaska}
  \emph {et~al.}}]{Trzaska:2019kuk}%
  \BibitemOpen
  \bibfield  {author} {\bibinfo {author} {\bibfnamefont {W.~H.}\ \bibnamefont
  {Trzaska}} \emph {et~al.},\ }\bibfield  {title} {\enquote {\bibinfo {title}
  {{Cosmic-ray muon flux at Canfranc Underground Laboratory}},}\ }\href
  {\doibase 10.1140/epjc/s10052-019-7239-9} {\bibfield  {journal} {\bibinfo
  {journal} {Eur. Phys. J. C}\ }\textbf {\bibinfo {volume} {79}},\ \bibinfo
  {pages} {721} (\bibinfo {year} {2019})},\ \Eprint
  {http://arxiv.org/abs/1902.00868} {arXiv:1902.00868 [physics.ins-det]}
  \BibitemShut {NoStop}%
\bibitem [{\citenamefont {Jordan}\ \emph {et~al.}(2013)\citenamefont {Jordan}
  \emph {et~al.}}]{Jordan:2013exa}%
  \BibitemOpen
  \bibfield  {author} {\bibinfo {author} {\bibfnamefont {D.}~\bibnamefont
  {Jordan}} \emph {et~al.},\ }\bibfield  {title} {\enquote {\bibinfo {title}
  {{Measurement of the neutron background at the Canfranc Underground
  Laboratory LSC}},}\ }\href {\doibase 10.1016/j.astropartphys.2012.11.007}
  {\bibfield  {journal} {\bibinfo  {journal} {Astropart. Phys.}\ }\textbf
  {\bibinfo {volume} {42}},\ \bibinfo {pages} {1--6} (\bibinfo {year}
  {2013})},\ \bibinfo {note} {[Erratum: Astropart.Phys. 118, 102372
  (2020)]}\BibitemShut {NoStop}%
\bibitem [{\citenamefont {Bernabei}\ \emph
  {et~al.}(2008{\natexlab{a}})\citenamefont {Bernabei} \emph
  {et~al.}}]{2008NIMPA.592..297B}%
  \BibitemOpen
  \bibfield  {author} {\bibinfo {author} {\bibfnamefont {R.}~\bibnamefont
  {Bernabei}} \emph {et~al.},\ }\bibfield  {title} {\enquote {\bibinfo {title}
  {{The DAMA/LIBRA apparatus}},}\ }\href {\doibase 10.1016/j.nima.2008.04.082}
  {\bibfield  {journal} {\bibinfo  {journal} {Nuclear Instruments and Methods
  in Physics Research A}\ }\textbf {\bibinfo {volume} {592}},\ \bibinfo {pages}
  {297--315} (\bibinfo {year} {2008}{\natexlab{a}})},\ \Eprint
  {http://arxiv.org/abs/0804.2738} {arXiv:0804.2738} \BibitemShut {NoStop}%
\bibitem [{\citenamefont {Kim}\ \emph {et~al.}(2019)\citenamefont {Kim} \emph
  {et~al.}}]{Kim:2018wcl}%
  \BibitemOpen
  \bibfield  {author} {\bibinfo {author} {\bibfnamefont {K.~W.}\ \bibnamefont
  {Kim}} \emph {et~al.} (\bibinfo {collaboration} {KIMS}),\ }\bibfield  {title}
  {\enquote {\bibinfo {title} {{Limits on Interactions between Weakly
  Interacting Massive Particles and Nucleons Obtained with NaI(Tl) crystal
  Detectors}},}\ }\href {\doibase 10.1007/JHEP03(2019)194} {\bibfield
  {journal} {\bibinfo  {journal} {J. High Energy Phys.}\ }\textbf {\bibinfo
  {volume} {03}},\ \bibinfo {pages} {194} (\bibinfo {year} {2019})},\ \Eprint
  {http://arxiv.org/abs/1806.06499} {arXiv:1806.06499 [hep-ex]} \BibitemShut
  {NoStop}%
\bibitem [{\citenamefont {Amar{\'e}}\ \emph {et~al.}(2015)\citenamefont
  {Amar{\'e}} \emph {et~al.}}]{Amare:2014bea}%
  \BibitemOpen
  \bibfield  {author} {\bibinfo {author} {\bibfnamefont {J.}~\bibnamefont
  {Amar{\'e}}} \emph {et~al.},\ }\bibfield  {title} {\enquote {\bibinfo {title}
  {{Cosmogenic radionuclide production in NaI(Tl) crystals}},}\ }\href
  {\doibase 10.1088/1475-7516/2015/02/046} {\bibfield  {journal} {\bibinfo
  {journal} {JCAP}\ }\textbf {\bibinfo {volume} {1502}},\ \bibinfo {pages}
  {046} (\bibinfo {year} {2015})},\ \Eprint {http://arxiv.org/abs/1411.0106}
  {arXiv:1411.0106 [physics.ins-det]} \BibitemShut {NoStop}%
\bibitem [{\citenamefont {Villar}\ \emph {et~al.}(2018)\citenamefont {Villar}
  \emph {et~al.}}]{Villar:2018ymt}%
  \BibitemOpen
  \bibfield  {author} {\bibinfo {author} {\bibfnamefont {P.}~\bibnamefont
  {Villar}} \emph {et~al.},\ }\bibfield  {title} {\enquote {\bibinfo {title}
  {{Study of the cosmogenic activation in NaI(Tl) crystals within the ANAIS
  experiment}},}\ }\href {\doibase 10.1142/S0217751X18430066} {\bibfield
  {journal} {\bibinfo  {journal} {Int. J. Mod. Phys.}\ }\textbf {\bibinfo
  {volume} {A33}},\ \bibinfo {pages} {1843006} (\bibinfo {year}
  {2018})}\BibitemShut {NoStop}%
\bibitem [{\citenamefont {Bernabei}\ \emph
  {et~al.}(2008{\natexlab{b}})\citenamefont {Bernabei} \emph
  {et~al.}}]{Bernabei:2008yi}%
  \BibitemOpen
  \bibfield  {author} {\bibinfo {author} {\bibfnamefont {R.}~\bibnamefont
  {Bernabei}} \emph {et~al.} (\bibinfo {collaboration} {DAMA}),\ }\bibfield
  {title} {\enquote {\bibinfo {title} {{First results from DAMA/LIBRA and the
  combined results with DAMA/NaI}},}\ }\href {\doibase
  10.1140/epjc/s10052-008-0662-y} {\bibfield  {journal} {\bibinfo  {journal}
  {Eur. Phys. J.}\ }\textbf {\bibinfo {volume} {C56}},\ \bibinfo {pages}
  {333--355} (\bibinfo {year} {2008}{\natexlab{b}})},\ \Eprint
  {http://arxiv.org/abs/0804.2741} {arXiv:0804.2741 [astro-ph]} \BibitemShut
  {NoStop}%
\bibitem [{\citenamefont {Buttazzo}\ \emph {et~al.}(2020)\citenamefont
  {Buttazzo}, \citenamefont {Panci}, \citenamefont {Rossi},\ and\ \citenamefont
  {Strumia}}]{Buttazzo:2020bto}%
  \BibitemOpen
  \bibfield  {author} {\bibinfo {author} {\bibfnamefont {Dario}\ \bibnamefont
  {Buttazzo}}, \bibinfo {author} {\bibfnamefont {Paolo}\ \bibnamefont {Panci}},
  \bibinfo {author} {\bibfnamefont {Nicola}\ \bibnamefont {Rossi}}, \ and\
  \bibinfo {author} {\bibfnamefont {Alessandro}\ \bibnamefont {Strumia}},\
  }\bibfield  {title} {\enquote {\bibinfo {title} {{Annual modulations from
  secular variations: relaxing DAMA?}}}\ }\href {\doibase
  10.1007/JHEP04(2020)137} {\bibfield  {journal} {\bibinfo  {journal} {JHEP}\
  }\textbf {\bibinfo {volume} {04}},\ \bibinfo {pages} {137} (\bibinfo {year}
  {2020})},\ \Eprint {http://arxiv.org/abs/2002.00459} {arXiv:2002.00459
  [hep-ex]} \BibitemShut {NoStop}%
\bibitem [{\citenamefont {Messina}\ \emph {et~al.}(2020)\citenamefont
  {Messina}, \citenamefont {Nardecchia},\ and\ \citenamefont
  {Piacentini}}]{Messina_2020}%
  \BibitemOpen
  \bibfield  {author} {\bibinfo {author} {\bibfnamefont {Andrea}\ \bibnamefont
  {Messina}}, \bibinfo {author} {\bibfnamefont {Marco}\ \bibnamefont
  {Nardecchia}}, \ and\ \bibinfo {author} {\bibfnamefont {Stefano}\
  \bibnamefont {Piacentini}},\ }\bibfield  {title} {\enquote {\bibinfo {title}
  {Annual modulations from secular variations: not relaxing {DAMA}?}}\ }\href
  {\doibase 10.1088/1475-7516/2020/04/037} {\bibfield  {journal} {\bibinfo
  {journal} {Journal of Cosmology and Astroparticle Physics}\ }\textbf
  {\bibinfo {volume} {2020}},\ \bibinfo {pages} {037--037} (\bibinfo {year}
  {2020})}\BibitemShut {NoStop}%
\bibitem [{\citenamefont {Bernabei}\ \emph {et~al.}(2018)\citenamefont
  {Bernabei} \emph {et~al.}}]{Bernabei:2018yyw}%
  \BibitemOpen
  \bibfield  {author} {\bibinfo {author} {\bibfnamefont {R.}~\bibnamefont
  {Bernabei}} \emph {et~al.},\ }\bibfield  {title} {\enquote {\bibinfo {title}
  {{First Model Independent Results from DAMA/LIBRA-Phase2}},}\ }\bibfield
  {booktitle} {\emph {\bibinfo {booktitle} {{Proceedings, 7th International
  Conference on New Frontiers in Physics (ICNFP 2018): Kolymbari, Crete,
  Greece, July 4-12, 2018}}},\ }\href {\doibase 10.3390/universe4110116,
  10.15407/jnpae2018.04.307} {\bibfield  {journal} {\bibinfo  {journal}
  {Universe}\ }\textbf {\bibinfo {volume} {4}},\ \bibinfo {pages} {116}
  (\bibinfo {year} {2018})},\ \bibinfo {note} {[At. Energ.19,307(2018)]},\
  \Eprint {http://arxiv.org/abs/arXiv:1805.10486} {arXiv:arXiv:1805.10486
  [hep-ex]} \BibitemShut {NoStop}%
\bibitem [{\citenamefont {{Alegria}}(2009)}]{2009Meas...42..748A}%
  \BibitemOpen
  \bibfield  {author} {\bibinfo {author} {\bibfnamefont {F.~Corr{\^e}a}\
  \bibnamefont {{Alegria}}},\ }\bibfield  {title} {\enquote {\bibinfo {title}
  {{Bias of amplitude estimation using three-parameter sine fitting in the
  presence of additive noise}},}\ }\href {\doibase
  10.1016/j.measurement.2008.12.006} {\bibfield  {journal} {\bibinfo  {journal}
  {Measurements}\ }\textbf {\bibinfo {volume} {42}},\ \bibinfo {pages}
  {748--756} (\bibinfo {year} {2009})}\BibitemShut {NoStop}%
\bibitem [{\citenamefont {Ahmed}\ \emph {et~al.}(2012)\citenamefont {Ahmed}
  \emph {et~al.}}]{Ahmed:2012vq}%
  \BibitemOpen
  \bibfield  {author} {\bibinfo {author} {\bibfnamefont {Z.}~\bibnamefont
  {Ahmed}} \emph {et~al.} (\bibinfo {collaboration} {CDMS-II}),\ }\bibfield
  {title} {\enquote {\bibinfo {title} {{Search for annual modulation in
  low-energy CDMS-II data}},}\ }\href@noop {} {\  (\bibinfo {year} {2012})},\
  \Eprint {http://arxiv.org/abs/1203.1309} {arXiv:1203.1309 [astro-ph.CO]}
  \BibitemShut {NoStop}%
\bibitem [{\citenamefont {Fox}\ \emph {et~al.}(2012)\citenamefont {Fox},
  \citenamefont {Kopp}, \citenamefont {Lisanti},\ and\ \citenamefont
  {Weiner}}]{Fox:2011px}%
  \BibitemOpen
  \bibfield  {author} {\bibinfo {author} {\bibfnamefont {Patrick~J.}\
  \bibnamefont {Fox}}, \bibinfo {author} {\bibfnamefont {Joachim}\ \bibnamefont
  {Kopp}}, \bibinfo {author} {\bibfnamefont {Mariangela}\ \bibnamefont
  {Lisanti}}, \ and\ \bibinfo {author} {\bibfnamefont {Neal}\ \bibnamefont
  {Weiner}},\ }\bibfield  {title} {\enquote {\bibinfo {title} {{A CoGeNT
  Modulation Analysis}},}\ }\href {\doibase 10.1103/PhysRevD.85.036008}
  {\bibfield  {journal} {\bibinfo  {journal} {Phys. Rev. D}\ }\textbf {\bibinfo
  {volume} {85}},\ \bibinfo {pages} {036008} (\bibinfo {year} {2012})},\
  \Eprint {http://arxiv.org/abs/1107.0717} {arXiv:1107.0717 [hep-ph]}
  \BibitemShut {NoStop}%
\bibitem [{\citenamefont {Yang}\ \emph {et~al.}(2019)\citenamefont {Yang} \emph
  {et~al.}}]{Yang:2019lao}%
  \BibitemOpen
  \bibfield  {author} {\bibinfo {author} {\bibfnamefont {L.~T.}\ \bibnamefont
  {Yang}} \emph {et~al.} (\bibinfo {collaboration} {CDEX}),\ }\bibfield
  {title} {\enquote {\bibinfo {title} {{Search for Light
  Weakly-Interacting-Massive-Particle Dark Matter by Annual Modulation Analysis
  with a Point-Contact Germanium Detector at the China Jinping Underground
  Laboratory}},}\ }\href {\doibase 10.1103/PhysRevLett.123.221301} {\bibfield
  {journal} {\bibinfo  {journal} {Phys. Rev. Lett.}\ }\textbf {\bibinfo
  {volume} {123}},\ \bibinfo {pages} {221301} (\bibinfo {year} {2019})},\
  \Eprint {http://arxiv.org/abs/1904.12889} {arXiv:1904.12889 [hep-ex]}
  \BibitemShut {NoStop}%
\bibitem [{\citenamefont {Aprile}\ \emph
  {et~al.}(2015{\natexlab{b}})\citenamefont {Aprile} \emph
  {et~al.}}]{Aprile:2015ibr}%
  \BibitemOpen
  \bibfield  {author} {\bibinfo {author} {\bibfnamefont {E.}~\bibnamefont
  {Aprile}} \emph {et~al.} (\bibinfo {collaboration} {XENON100}),\ }\bibfield
  {title} {\enquote {\bibinfo {title} {{Search for Event Rate Modulation in
  XENON100 Electronic Recoil Data}},}\ }\href {\doibase
  10.1103/PhysRevLett.115.091302} {\bibfield  {journal} {\bibinfo  {journal}
  {Phys. Rev. Lett.}\ }\textbf {\bibinfo {volume} {115}},\ \bibinfo {pages}
  {091302} (\bibinfo {year} {2015}{\natexlab{b}})},\ \Eprint
  {http://arxiv.org/abs/1507.07748} {arXiv:1507.07748 [astro-ph.CO]}
  \BibitemShut {NoStop}%
\bibitem [{\citenamefont {{VanderPlas}}(2018)}]{2018ApJS..236...16V}%
  \BibitemOpen
  \bibfield  {author} {\bibinfo {author} {\bibfnamefont {Jacob~T.}\
  \bibnamefont {{VanderPlas}}},\ }\bibfield  {title} {\enquote {\bibinfo
  {title} {{Understanding the Lomb-Scargle Periodogram}},}\ }\href {\doibase
  10.3847/1538-4365/aab766} {\bibfield  {journal} {\bibinfo  {journal} {ApJS}\
  }\textbf {\bibinfo {volume} {236}},\ \bibinfo {eid} {16} (\bibinfo {year}
  {2018})},\ \Eprint {http://arxiv.org/abs/1703.09824} {arXiv:1703.09824
  [astro-ph.IM]} \BibitemShut {NoStop}%
\bibitem [{\citenamefont {Lista}(2017)}]{Lista:2016chp}%
  \BibitemOpen
  \bibfield  {author} {\bibinfo {author} {\bibfnamefont {L.}~\bibnamefont
  {Lista}},\ }\bibfield  {title} {\enquote {\bibinfo {title} {{Practical
  Statistics for Particle Physicists}},}\ }in\ \href {\doibase
  10.23730/CYRSP-2017-005.213} {\emph {\bibinfo {booktitle} {{2016 European
  School of High-Energy Physics}}}}\ (\bibinfo {year} {2017})\ pp.\ \bibinfo
  {pages} {213--258},\ \Eprint {http://arxiv.org/abs/1609.04150}
  {arXiv:1609.04150 [physics.data-an]} \BibitemShut {NoStop}%
\bibitem [{\citenamefont {Eadie}\ \emph {et~al.}(1971)\citenamefont {Eadie}
  \emph {et~al.}}]{Eadie_1971}%
  \BibitemOpen
  \bibfield  {author} {\bibinfo {author} {\bibfnamefont {W.~T.}\ \bibnamefont
  {Eadie}} \emph {et~al.},\ }\href@noop {} {\emph {\bibinfo {title}
  {Statistical methods in experimental physics}}}\ (\bibinfo  {publisher}
  {Amsterdam: North--Holland},\ \bibinfo {year} {1971})\ \bibinfo {note} {ch.
  8, pp. 163--165}\BibitemShut {NoStop}%
\end{thebibliography}
%

\end{document}